\pdfoutput=1  

\documentclass[aps,prd,superscriptaddress,floatfix,nofootinbib,preprintnumbers,eqsecnum,twocolumn]{revtex4-1}

\usepackage[utf8]{inputenc}
\usepackage[dvips]{graphicx}
\usepackage[dvipsnames]{xcolor}

\usepackage{amsmath,float,graphicx,placeins,amssymb,multirow,pgfplots,pgfplotstable}
\usepackage{empheq}
\usepackage{tikz}
\usepackage{comment}
\usepackage{enumerate,slashed,cancel,comment,bbm,rotating,placeins,mathtools,multirow,hhline}
\usepackage[normalem]{ulem}
\usepackage{array}

\usepackage{epstopdf}
\usepackage{hyperref}
\usepackage{mathrsfs}
\usepackage{amssymb}
\usepackage{physics}
\usepackage{booktabs}

\usepackage{color}
\usepackage{relsize}
\usepackage{graphics}
\usepackage{epstopdf}
\usepackage{hyperref}
\usepackage{mathrsfs}
\usepackage{amssymb}
\usepackage{physics}
\usepackage{booktabs}

\usepackage[normalem]{ ulem }
\usepackage{amsthm}
\usepackage{amsmath}
\usepackage{cancel}
\usepackage{float}
\usepackage{fontawesome} 
\usepackage[caption=false]{subfig}
\usepackage{tabularx,ragged2e} 
 \newcolumntype{L}{>{\RaggedRight\arraybackslash}X}
 
\colorlet{mylinkcolor}{ForestGreen}
\colorlet{mycitecolor}{Red}
\colorlet{myurlcolor}{violet}

\usepackage{lipsum}
\usepackage{diagbox}
\usetikzlibrary{shapes.geometric,arrows}
\graphicspath{}

\def\beq{\begin{equation}}
\def\eeq{\end{equation}}
\def\beqn{\begin{eqnarray}}
\def\eeqn{\end{eqnarray}}

\hypersetup{
 linkcolor = {blue!80!black},
 citecolor = {blue!70!black},
 urlcolor = {blue!80!black},
 colorlinks = true
}
\def\barOmega{{\overline{\Omega}}}

\def\ie{{\it i.e.}\/}
\def\eg{{\it e.g.}\/}
\def\etc{{\it etc}.\/}

\def\IR{\relax{\rm I\kern-.18em R}}
 \font\cmss=cmss10 \font\cmsss=cmss10 at 7pt
\def\IQ{\relax{\rm I\kern-.18em Q}}
\def\IZ{\relax\ifmmode\mathchoice
 {\hbox{\cmss Z\kern-.4em Z}}{\hbox{\cmss Z\kern-.4em Z}}
 {\lower.9pt\hbox{\cmsss Z\kern-.4em Z}}
 {\lower1.2pt\hbox{\cmsss Z\kern-.4em Z}}\else{\cmss Z\kern-.4em Z}\fi}

\def\fBH{f_{\rm BH}}
\def\rhoBH{\rho_{\rm BH}}

\def\TBH{T_{\rm BH}}

\def\M0{M^{(0)}}
\def\f0{f^{(0)}}
\def\t0{t^{(0)}}
\def\X0{X^{(0)}}
\def\a0{a^{(0)}}
\def\H0{H^{(0)}}

\def\nn{\nonumber}

\def\OmegaBar{\overline{\Omega}_{\rm BH}}
\def\OmegaBHAvg{\langle\Omega_{\rm BH}\rangle}

\pgfplotsset{compat=1.17}
\begin{document}


\title{Primordial Black Holes Place the Universe in Stasis}

\def\andname{\hspace*{-0.5em}} 

\author{Keith R. Dienes}
\email[Email address: ]{dienes@arizona.edu}
\affiliation{Department of Physics, University of Arizona, Tucson, AZ 85721 USA}
\affiliation{Department of Physics, University of Maryland, College Park, MD 20742 USA}
\author{Lucien Heurtier}
\email[Email address: ]{lucien.heurtier@durham.ac.uk}
\affiliation{IPPP, Durham University, Durham, DH1 3LE, United Kingdom}
\author{Fei Huang}
\email[Email address: ]{fei.huang@weizmann.ac.il}
\affiliation{CAS Key Laboratory of Theoretical Physics, Institute of Theoretical Physics,\\ 
Chinese Academy of Sciences, Beijing 100190, China}
\affiliation{Department of Physics and Astronomy, University of California, Irvine, CA  92697  USA}
\affiliation{Department of Particle Physics and Astrophysics, Weizmann Institute of Science, Rehovot 7610001, Israel}
\author{Doojin Kim}
\email[Email address: ]{doojin.kim@tamu.edu}
\affiliation{Mitchell Institute for Fundamental Physics and Astronomy,\\
Department of Physics and Astronomy, Texas A\&M University, College Station, TX  77843 USA}
\author{\\ Tim M.P. Tait}
\email[Email address: ]{ttait@uci.edu}
\affiliation{Department of Physics and Astronomy, University of California, Irvine, CA  92697  USA}
\author{Brooks Thomas}
\email[Email address: ]{thomasbd@lafayette.edu}
\affiliation{Department of Physics, Lafayette College, Easton, PA  18042 USA}

\begin{abstract}
A variety of scenarios for early-universe cosmology give rise to a population of primordial
black holes (PBHs) with a broad spectrum of masses.   
The evaporation of PBHs in such scenarios has the 
potential to place the universe into an extended period of ``stasis'' during which the 
abundances of matter and radiation --- \ie, their energy densities divided by the
critical density --- remain absolutely constant despite cosmological 
expansion.   This surprising phenomenon can give rise to new possibilities for 
early-universe dynamics and lead to distinctive signatures of the evaporation of such PBHs.
In this paper, we discuss how this stasis epoch arises and explore a number of its 
phenomenological consequences, including implications for inflationary observables,
the stochastic gravitational-wave background, baryogenesis, and the 
production of dark matter and dark radiation.
\end{abstract}
\maketitle

\tableofcontents


\section{Introduction}


In a broad class of inflationary scenarios, a population of black holes (BHs) is 
generated shortly after inflation as a consequence of the gravitational 
collapse of primordial density fluctuations.  Such primordial black holes (PBHs)
have received a significant amount of recent attention, in part because
they can potentially provide a solution to the dark-matter problem.  Indeed,  
while black holes evaporate over time as a consequence of Hawking 
radiation~\cite{Hawking:1974rv,Hawking:1975vcx}, PBHs with masses 
$M \gtrsim 10^{15}$~g would nevertheless have lifetimes longer than 
the age of the universe.  Indeed, a population of PBHs with masses within
the range $10^{17}~{\rm g} \lesssim M \lesssim 10^{23}~{\rm g}$ can 
potentially account for the entirety of the present-day dark-matter 
abundance, even when the spectrum of PBHs is approximately 
monochromatic (for reviews, see, \eg, 
Refs.~\cite{Escriva:2022duf, Carr:2020gox,Green:2020jor,Villanueva-Domingo:2021spv}).  
PBHs with lower masses can also have implications for cosmology.  Indeed, 
PBHs with masses in the range 
$10^9~{\rm g} \lesssim M \lesssim 10^{14}~{\rm g}$ evaporate 
at a significant rate during or after Big-Bang nucleosynthesis (BBN), generating 
energetic particles which can modify the primordial abundances of light nuclei. 
As a result, the abundance of PBHs with masses in this range is 
tightly constrained~\cite{Carr:2009jm,Carr:2020gox,Keith:2020jww}.

By contrast, PBHs with masses $M \lesssim 10^9~{\rm g}$  
evaporate completely prior to the BBN epoch and are therefore 
essentially unconstrained by these considerations. 
For this reason, far less attention has been
focused on PBHs within this mass range.  Nevertheless, such light PBHs can
potentially have an impact on early-universe cosmology.  For example, 
their evaporation can serve as a source for dark matter or 
dark radiation~\cite{Hooper:2019gtx,Masina:2020xhk,Baldes:2020nuv,Gondolo:2020uqv,
Bernal:2020kse,Bernal:2020ili,Bernal:2020bjf,Masina:2021zpu,Arbey:2021ysg,
JyotiDas:2021shi,Cheek:2021odj,Cheek:2021cfe,Sandick:2021gew,Schiavone:2021imu,
Bernal:2021yyb,Cheek:2022dbx,Bernal:2022oha,Morrison:2018xla,Auffinger:2020afu,
Khlopov:2004tn,Allahverdi:2017sks,Lennon:2017tqq,Kitabayashi:2021hox}.  Such 
PBHs can also trigger baryogenesis~\cite{Barrow:1990he,Hamada:2016jnq,Hooper:2020otu,
Perez-Gonzalez:2020vnz,Datta:2020bht,Gehrman:2022imk} or lead to an epoch of early 
matter domination~\cite{Bernal:2021bbv}. 

In this paper,  we point out that a population of PBHs with masses in this 
range can also have another important impact on early-universe cosmology.
In particular, as first noted in Ref.~\cite{Barrow:1991dn}, such a population 
of PBHs can give rise to an extended period during which the abundances of PBHs 
and radiation can remain approximately constant.  Indeed, this is an example of  
{\it cosmic stasis}\/~\cite{Dienes:2021woi}, a general phenomenon wherein multiple 
cosmological components with different equations of state --- in this case, matter in 
the form of black holes and the  radiation generated by their evaporation  --- have 
abundances which remain constant despite cosmological expansion.
Since the energy density of radiation dilutes faster as a result of cosmic expansion 
than does that of matter, the abundance of matter typically increases over time, 
while the abundance of radiation decreases.  Thus, in order for a stasis epoch 
involving these two cosmological components to arise, a mechanism must exist which
counteracts this tendency by gradually transferring energy density from matter to 
radiation over an extended period.  In Ref.~\cite{Dienes:2021woi}, it was shown that
particle decay constitutes such a mechanism in the case in which the matter sector 
consists of a tower of unstable particles with a broad spectrum of lifetimes and 
individual abundances.   However, as shown in Ref.~\cite{Barrow:1991dn} and 
as we shall demonstrate here, Hawking radiation likewise provides such a mechanism in 
the case in which the matter sector comprises a population of PBHs.

This paper is devoted to a study of PBH-induced stasis and its phenomenological 
consequences for the early universe.  In Sect.~\ref{sec:PBH}, we begin by reviewing the 
critical ingredients which are needed for our subsequent analysis.  In particular, 
we consider the processes through which a population of PBHs can arise in the early 
universe and discuss the resulting PBH mass spectra.  We also review the dynamics 
of PBH evaporation and assess the impact of this dynamics on the overall energy density 
of this population. We then review how this dynamics can give rise to a period of  cosmic 
stasis --- one which is a global attractor within cosmologies of this sort.
Having thus set the stage for our work, we then proceed to the main purpose of this 
paper, namely to study some of the phenomenological implications of a period of cosmic 
stasis.  We begin in Sect.~\ref{sec:NotEternal} by demonstrating that cosmic stasis not 
only has a beginning {\it but also an end}\/, so that the universe can not only enter 
into stasis in a natural way but also exit from it equally naturally.   This property 
allows stasis to be the underpinning of a true cosmological {\it epoch}\/ from which the 
universe may subsequently transition into an epoch of another type, thereby allowing it 
to be ``spliced'' into more traditional cosmological timelines.   In 
Sect.~\ref{sec:Implications}, we then proceed to consider various phenomenological 
implications of a PBH-induced period of stasis.  In particular, in 
Sect.~\ref{sec:ExpansionHistory} we consider the effects of a stasis epoch on the 
cosmic expansion history, while in Sect.~\ref{sec:Inflation} we consider the effects 
of such an epoch on inflationary observables.  In Sect.~\ref{sec:GravitationalWaves} 
we consider the implications for gravitational waves.  In Sects.~\ref{sec:DarkRadiation}
and~\ref{sec:DarkMatter}, we consider the effects that a PBH-induced stasis 
epoch can potentially have on dark-radiation and dark-matter production, respectively.   
In Sect.~\ref{sec:Baryogenesis} we consider the implications of such an epoch for 
baryogenesis.  Finally, in Sect.~\ref{sec:conclusions} we conclude with a summary of our 
main results and highlight possible directions for future work.


\section{The Emergence of a Stasis Epoch}\label{sec:PBH}


We begin by reviewing the physics that will be needed for our 
subsequent work.  This includes the dynamics of PBH formation and evaporation, 
and the manner in which these features can give rise to cosmic stasis.

\subsection{PBH Formation}

Quantum fluctuations during cosmic inflation give rise to a spectrum of 
primordial density fluctuations.  After inflation ends, the size of a 
causally connected region grows as the universe expands, as does the 
range of gravitational interactions.  More specifically, the comoving radius 
of such a region is given by the comoving Hubble horizon $(aH)^{-1}$, where $a$ 
is the scale factor and $H\equiv \dot a/a$ is the Hubble parameter.  Primordial 
density fluctuations with a given comoving wavenumber $k$ remain frozen until 
this comoving radius becomes comparable to $2\pi/k$ and gravity comes into
play.  Any fluctuation sufficiently overdense that its radius lies below its 
Schwarzschild radius is expected to collapse and form a black hole.  
The characteristic initial mass $M_i$ of a black hole 
formed in this way is (for reviews, see, \eg, Ref.~\cite{Byrnes:2021jka})
\beq
  M_i ~=~ \frac{4}{3}\pi\gamma \rho \left(\frac{1}{H}\right)^3 
    ~=~ \frac{\gamma M_P^2}{2H}~,
  \label{eq:InitialMass}
\eeq
where $\gamma$ is an $\mathcal{O}(1)$ proportionality factor, where 
$\rho$ is the average energy density of the universe at the time the
fluctuation enters the horizon, and where $M_P = G^{-1/2}$ is the Planck mass,
which we define in the usual way in terms of the gravitational constant $G$.

As the universe expands, density fluctuations with increasingly long comoving 
wavelengths enter the horizon and form PBHs with different masses.  The mass 
spectrum of these PBHs --- \ie, the number density $f_{\rm BH}(M,t)$ of PBHs 
per unit mass --- can therefore be viewed as resulting from the interplay 
between two factors.  The first of these is 
the power spectrum of primordial density perturbations present in the early 
universe.  The second is the equation-of-state parameter of the universe 
$w_c$ during the period when these perturbations collapse.  
We note that contributions to $f_{\rm BH}(M,t)$ can also be generated 
through other mechanisms at later times as well (for a review, see, \eg,
Ref.~\cite{Carr:2020xqk}).  However, we shall not consider such additional 
contributions in this paper. 

Different scenarios for PBH production in the early universe give rise to 
different forms for $f_{\rm BH}(M,t)$.  Examples of such production scenarios 
include the collapse of Gaussian primordial 
inhomogeneities~\cite{Carr:1975qj}, inflation-induced power
spectra~\cite{Clesse:2015wea,Dolgov:1992pu,Green:2016xgy,Carr:1993aq,
Ivanov:1994pa,Garcia-Bellido:1996mdl,Randall:1995dj, Heurtier:2022rhf, 
Dimopoulos:2017ged, Ballesteros:2017fsr, Karam:2022nym, Dalianis:2018frf, 
Kannike:2017bxn}, the collapse of density perturbations during the QCD phase
transition~\cite{Crawford:1982yz,Jedamzik:1996mr,Schmid:1998mx,Widerin:1998my}, 
bubble collisions~\cite{Crawford:1982yz,Hawking:1982ga,
Kodama:1982sf,Leach:2000ea,
Moss:1994iq,Kitajima:2020kig,Kodama:1981gu,Maeda:1985gz,Khlopov:1998nm,
Konoplich:1998ugi,Pavsic:1995jt,Khlopov:2000js}, 
or the collapse of domain walls~\cite{Khlopov:1998nm,Konoplich:1998ugi,
Pavsic:1995jt,Khlopov:2000js,
Dokuchaev:2004kr,Rubin:2001yw,Garriga:2015fdk} 
and cosmic loops~\cite{Hawking:1987bn,Polnarev:1988dh,
Garriga:1993gj,Caldwell:1995fu,MacGibbon:1997pu,Jenkins:2020ctp}. 
 Interestingly, light PBHs can also be produced by the resonant 
amplification of scalar perturbations shortly after the end of cosmic 
inflation~\cite{Martin:2019nuw,Martin:2020fgl}.
Some of these BH-production scenarios give rise to a sharply peaked, 
nearly monochromatic PBH mass spectrum.  By contrast, others give rise to a 
spectrum which is well-modeled by a power law over a broad range of 
masses.  

In what follows, we consider PBH spectra of this latter sort.  In particular, 
we shall focus on a broad class of PBH mass spectra for which the initial
number density $f_{\rm BH}(M_i,t_i)$ of PBHs per unit mass $M_i$ at 
the initial time $t_i$ at which the spectrum has effectively been established
takes the form
\beq
  f_{\rm BH}(M_i,t_i) ~=~ 
  \begin{cases}
    C M_i^{\alpha-1} & {\rm for}~ M_{\rm min}\leq M_i\leq M_{\rm max}\\
    0 & \text{otherwise}~
  \end{cases}
  \label{eq:dist}
\eeq
for some minimum and maximum PBH masses $M_{\rm min}$ and $M_{\rm max}$,
where $\alpha$ is a power-law exponent and where $C$ is an overall normalization 
coefficient.  The corresponding energy density of PBHs per unit mass is 
therefore given by $C M_i^{\alpha}$.

PBH spectra of the form given in Eq.~(\ref{eq:dist}) arise naturally, for example, 
in scenarios in which the PBHs form via the collapse of a scale-invariant power 
spectrum~\cite{Green:1997sz,Bringmann:2001yp,Kim:1999iv,Carr:1975qj,Carr:2017jsz}.
In scenarios of this sort, the power-law exponent $\alpha$ is related to the 
value of the equation-of-state parameter $w_c$ for the universe as a 
whole during the epoch immediately following inflation, during which primordial 
density perturbations collapse to form PBHs.  In particular, one finds that
\beq
  \alpha~\equiv~ -\frac{3w_c+1}{w_c+1}~.
  \label{eq:w_c}
\eeq 
Since the collapse of PBHs occurs after (rather than during) inflation, the
physically motivated range for $w_c$ is $-1/3<w_c\leq 1$. 
Such values correspond to values of the scaling exponent 
\beq
  -2~\leq~ \alpha ~<~ 0~.
  \label{eq:range}
\eeq

Phenomenological considerations likewise place constraints on
the values of $M_{\rm min}$ and $M_{\rm max}$, although these constraints 
are considerably more model-dependent than those on $\alpha$.
As discussed in the Introduction, BBN constraints on the production of energetic 
particles from black-hole evaporation place stringent bounds on
$f_{\rm BH}(M_i,t_i)$.  However, PBHs with initial masses 
$M_i\lesssim 10^9 ~\mathrm{g}$ evaporate before the BBN epoch begins and 
consequently evade these constraints~\cite{Carr:2009jm,Carr:2020gox,Keith:2020jww}.
Of course, BBN constraints on the continuous injection of energy density over an
extended period depend on both on the overall energy density injected and
on the time-dependent manner in which that injection occurs~\cite{Dienes:2018yoq}.
Since the energy density of PBHs per unit mass falls rapidly with 
$M_i$ for values of $\alpha$ at the lower end of the allowed range in 
Eq.~(\ref{eq:range}), some fraction of the PBH population may 
evaporate after the onset of the BBN epoch without running afoul of 
these constraints.  However, in what follows, we adopt a conservative 
approach and focus on the regime in which $M_{\rm max}\lesssim 10^9 ~\mathrm{g}$.

A lower bound on $M_{\rm min}$ arises from CMB data, which place an upper bound 
$H_\star < 2.5 \times 10^{-5} M_P$ on the Hubble parameter when the mode 
$k_\star=0.002\,\mathrm{Mpc}^{-1}$ exits the horizon during 
inflation~\cite{Akrami:2018odb}.
Since the Hubble parameter $H_{\rm end}$ at the end of inflation must be smaller 
than $H_\star$, this bound in turn translates into a lower bound on $M_{\rm min}$.
The precise value of this bound depends on the details of both the 
inflationary model and the mechanics of gravitational collapse. 
For concreteness, we take the bound $M_{\rm min}\gtrsim 0.1$~g~\cite{Carr:2020gox} 
obtained for the case of slow-roll inflation as a rough benchmark.
Note that, in practice, the power spectrum produced from 
inflation may need to grow smoothly as a function of 
frequency before it plateaus.  This may effectively lead to an increase of the $M_{\rm min}$ 
value (see, \eg, Ref.~\cite{Cole:2022xqc}).  In this paper, we remain agnostic about the 
precise origin of this power spectrum and simply consider this lower bound on $M_{\rm min}$ to 
be a rough benchmark.  We emphasize, however, that bounds of this sort apply only in situations in 
which the population of PBHs is generated entirely from primordial perturbations which 
collapse upon re-entering the horizon after inflation.  Additional mechanisms for 
PBH production which involve the gravitational collapse of overdense regions with
comoving wavelengths well below $(aH)^{-1}$ can result in PBHs with far lower 
masses.

Thus, in summary, we shall primarily focus on PBH mass spectra wherein $\alpha$
falls within the range specified in Eq.~(\ref{eq:range}) and wherein
\begin{equation}
  0.1~\mathrm{g}~\lesssim~ M_{\rm min} ~<~ M_{\rm max}
    ~\lesssim~ 10^{9}~\mathrm{g} 
  \label{eq:MminMmaxBounds}
\end{equation}
in what follows.

\subsection{PBH Evaporation}

A Schwarzschild black hole of mass $M$ evaporates by emitting a 
thermal spectrum of particles at the Hawking
temperature~\cite{Hawking:1974rv,Hawking:1975vcx} 
\begin{align}
 T_{\rm BH} ~=~ \frac{1}{8\pi G M}
   ~\sim~ 1.06~{\rm GeV}\left(\frac{10^{13}\,{\rm g}}{M}\right)~.
\end{align}
The efficiency of the evaporation 
process depends both on $\TBH$ and on the mass 
spectrum and quantum numbers of the particles that can be produced.  
When one accounts for all the 
corresponding graybody factors, one finds that the rate of change of 
$M$ due to evaporation is~\cite{PhysRevD.41.3052,PhysRevD.44.376}
\begin{equation}
  \frac{d M}{d t}~\equiv~ -\varepsilon(M)\frac{M_P^4}{M^2}~,
  \label{eq:dMBHdt}
\end{equation}
where the function $\varepsilon(M)$ encodes how this rate varies with 
the BH mass and temperature. 

A detailed description of how this function is computed can be found in 
Ref.~\cite{Cheek:2021cfe} and references therein.
As discussed in the Introduction, PBHs which evaporate before BBN --- and 
which therefore evade current observational constraints --- have masses 
$M\lesssim 10^9 ~\mathrm{g}$~\cite{Carr:2009jm,Carr:2020gox,Keith:2020jww}. 
The corresponding Hawking temperatures $\TBH \gtrsim 10^4\mathrm{~GeV}$ for
PBHs in this mass range are sufficiently high that the average momentum 
of any Standard-Model particles produced as Hawking radiation is 
$\langle p \rangle\sim \TBH$.  As a result, provided that any additional 
particle species that might happen to be present in the theory likewise have 
masses that lie well below this value of $\TBH$, one finds that 
$\varepsilon(M)$ can be approximated as effectively 
constant for $M\lesssim 10^9 ~\mathrm{g}$.  Thus, for $M$ in this 
regime, we may take $\varepsilon(M) \approx \varepsilon$.
It therefore follows from Eq.~\eqref{eq:dMBHdt} that at time $t$, the mass 
$M$ of a PBH produced at time $t_i$ with initial mass $M_i$ is given by
\beq
  M ~=~ M_i\left[1-\frac{t-t_i}{\tau(M_i)} \right]^{1/3}\!
  \Theta\Big(\tau(M_i) - (t-t_i)\Big)~,
  \label{eq:Mt}
\eeq
where $\Theta(x)$ denotes the Heaviside function of $x$ and where
\beq
  \tau(M_i) ~\equiv~ \frac{{M_i}^3}{3\varepsilon M_P^4}
  \label{eq:tau}
\eeq
is the lifetime $\tau$ of the black hole --- \ie, the time after $t_i$ at which 
$M$ reaches zero.  


\subsection{Evolution of the Energy Density}

Once an initial spectrum $\fBH(M_i,t_i)$ of PBHs is established at time
$t_i$, this distribution evolves with time as a result of both evaporation 
and cosmic expansion.  However, the comoving number density of PBHs with initial
masses in the infinitesimal range from $M_i$ to $M_i + dM_i$ remains constant
under the influence of these two processes
until $t = \tau(M_i) + t_i$, at which point the PBHs evaporate completely and the
number density drops to zero.  This implies that
\begin{eqnarray}
  a^3\fBH(M,t)dM &=& a_i^3 \Theta\Big(\tau(M_i)-(t-t_i)\Big)\nonumber \\ & & ~~~
      \times \, \fBH(M_i,t_i) \, dM_i~, ~~
   \label{eq:fBHConserve}
\end{eqnarray}
where $a_i$ is the scale factor at time $t_i$.  

Additional processes beyond evaporation and cosmic
expansion can in principle affect the differential number density $\fBH(M,t)$ 
of PBHs in the universe per unit mass after their initial spectrum is established.
For example, PBHs can acquire additional mass through the accretion of 
particles present in the radiation bath or undergo mergers with other PBHs.  
However, it turns out that for much of the parameter space relevant for cosmic 
stasis, these processes have a negligible affect on $\fBH(M,t)$.  We shall 
therefore focus in what follows on the regime in which the effects of accretion 
and mergers can be neglected.  We shall return to discuss these processes and the 
conditions under which they can potentially have non-negligible
effects on $\fBH(M,t)$ in Sect.~\ref{sec:mergers}.

The total energy density 
\begin{equation}
  \rhoBH(t) ~=~ \int_0^{\infty}  d M~\fBH(M,t) M
\end{equation}
of the PBH population evolves with $t$ according to the Boltzmann 
equation\footnote{For a complete description 
of the PBH spectrum and its dynamical evolution which incorporates black-hole 
spins as well as masses, see Ref.~\cite{Cheek:2022mmy}.} 
\begin{equation}
  \frac{d\rhoBH}{d t}+3H\rhoBH ~=~ 
    \int_0^\infty dM~\fBH(M,t) \frac{d M}{d t}~,
      \label{eq:drhoBHdt}
\end{equation}
where $dM/dt$ is given by Eq.~(\ref{eq:dMBHdt}).  The conservation 
relation in Eq.~\eqref{eq:fBHConserve} permits us to recast the integral 
over PBH masses $M$ at time $t$ on the right side of this equation as an 
integral over the corresponding initial masses $M_i$.  In particular,
we find that
\begin{equation}
  \frac{d\rhoBH}{d t}+3H\rhoBH ~=~ \left(\frac{a_i}{a}\right)^3
    \int_{\widetilde{M}_i(t)}^\infty d M_i \,\fBH(M_i,t_i) \frac{d M}{d t}
  \label{eq:drhoBHdtwLim}~,
\end{equation}
where $dM/dt$ is understood to be an implicit function of $M_i$ by virtue
of the functional map between $M$ and $M_i$ in Eq.~\eqref{eq:Mt} and where 
\begin{equation}
  \widetilde{M}_i(t)~\equiv~ \Big[3 \varepsilon M_P^4(t-t_i)\Big]^{1/3}~.
  \label{eq:Miev}
\end{equation}
This lower limit of integration arises from the Heaviside function in 
Eq.~(\ref{eq:fBHConserve}) and accounts for the fact that 
PBHs with masses $M_i < \widetilde{M}_i(t)$ evaporate completely by time $t$
and therefore no longer contribute to the energy density.

\subsection{Cosmological Dynamics and the Emergence of Stasis\label{sec:stasis}}

We are now ready to examine the cosmological implications of a population 
of evaporating PBHs.  In so doing, we shall follow as closely as possible the 
presentation in Ref.~\cite{Dienes:2021woi} concerning the general possibility 
of stasis between matter and radiation.  The special case in which the matter 
comprises a population of PBHs and the radiation is produced through their 
evaporation was also previously discussed in Ref.~\cite{Barrow:1991dn}.

For simplicity, we consider a flat 
Friedmann-Robertson-Walker (FRW) universe consisting of 
two cosmological components.  The first is a population of PBHs characterized by a total 
energy density $\rho_{\rm BH}$ and corresponding total abundance 
\begin{equation}
  \Omega_{\rm BH} ~\equiv~ \frac{8\pi G}{3H^2} \,\rho_{\rm BH} ~.
\end{equation}
The second is a radiation component characterized by an energy density $\rho_\gamma$ 
and a corresponding abundance $\Omega_\gamma = 1-\Omega_{\rm BH}$.
The time-evolution of the PBH abundance is given by 
\begin{equation}
    \frac{d\Omega_{\rm BH}}{dt} ~=~ \frac{8\pi G}{3}\left(
      \frac{1}{H^2}\frac{d\rho_{\rm BH}}{dt} -2\frac{\rho_{\rm BH}}{H^3} 
      \frac{dH}{dt} \right)~,
  \label{eq:dOmegaBHdt}
\end{equation}
where $d\rho_{\rm BH}/dt$ is given by Eq.~(\ref{eq:drhoBHdt}) and where
$dH/dt$ is given by the Friedmann acceleration equation, which in this
case takes the form
\begin{equation}
  \frac{dH}{dt} ~=~ -H^2 - \frac{4\pi G}{3}\Big[\rho_{\rm BH}(1+3w_{\rm BH}) + 
    \rho_\gamma(1 + 3 w_\gamma)\Big]~,
\end{equation}
where $w_{\rm BH}$ and $w_\gamma$ are the equation-of-state parameters for 
black holes and radiation, respectively. Since $w_{\rm BH} = 0$ and 
$w_\gamma = 1/3$, this equation reduces 
to~\cite{Dienes:2021woi}
\begin{eqnarray}
  \frac{dH}{dt} &=& -\frac{1}{2}H^2(4 - \Omega_{\rm BH})~.
  \label{eq:FriedmannAccelEq}
\end{eqnarray}
Substituting this result for $dH/dt$ into Eq.~(\ref{eq:dOmegaBHdt}),
we obtain
\begin{equation}
   \frac{d\Omega_{\rm BH}}{dt} ~=~ -\Gamma_{\rm BH}(t)\,\Omega_{\rm BH}
     +H\left(\Omega_{\rm BH} - \Omega_{\rm BH}^2\right)~,
  \label{eq:OmegaBHEvolEq}
\end{equation}
where we have defined the effective BH evaporation rate
\begin{eqnarray}
  \Gamma_{\rm BH}(t) & \equiv & - \frac{
      \int_0^\infty \fBH(M, t)
      \frac{dM}{dt}\,dM}{
    \int_0^\infty \fBH(M, t) M \,dM} \nonumber \\
    & = & -\frac{d}{dt}\log \left(\int_{\widetilde{M}_i(t)}^\infty 
        \fBH(M_i, t_i)M(t)\,dM_i \right)~.\nonumber \\
\end{eqnarray}
For an initial PBH spectrum $\fBH(M_i,t_i)$ of the form given in 
Eq.~(\ref{eq:dist}), this expression becomes
\begin{equation}
  \Gamma_{\rm BH}(t)  \equiv  
     -\frac{d}{dt}\log\!
     \left(\int_{\mu_i(t)}^{M_{\rm max}}\!\!\!\! 
      M_i^{\alpha-1}\Big[M_i^3 - \widetilde{M}_i^3(t)\Big]^{1/3} dM_i\right)~,
  \label{eq:GammaSimpModelExact}
\end{equation}
where $\mu_i(t) \equiv {\rm max}\{M_{\rm min},\widetilde{M}_i(t)\}$.
While Eq.~(\ref{eq:GammaSimpModelExact}) is exact, we note that 
once $\widetilde{M}_i(t) > M_{\rm min}$ and the lightest PBHs begin to 
evaporate, this expression reduces to a particularly simple form in the 
regime in which $\widetilde{M}_i(t) \ll M_{\rm max}$.  In this regime,
the value of the integral over $M_i$ in Eq.~(\ref{eq:GammaSimpModelExact}) 
will not be significantly impacted by taking $M_{\rm max} \rightarrow \infty$.
Thus, in this regime $\Gamma_{\rm BH}(t)$ can be approximated as 
\begin{eqnarray}
  \Gamma_{\rm BH}(t) \!& \approx &\! 
     -\frac{d}{dt}\log\!\left(\int_{\widetilde{M}_i(t)}^\infty\!\!\!\! 
      M_i^{\alpha-1}\Big[M_i^3 - \widetilde{M}_i^3(t)\Big]^{1/3} \,dM_i\right) 
      \nonumber \\
   & \approx &\! -\frac{\alpha+1}{3(t-t_i)}~.
   \label{eq:GammaSimpModel}
\end{eqnarray}

Taken together, the differential equations in Eqs.~(\ref{eq:FriedmannAccelEq}) 
and (\ref{eq:OmegaBHEvolEq}) with $\Gamma_{\rm BH}(t)$ given in 
Eq.~(\ref{eq:GammaSimpModelExact}) describe the evolution of the
PBH abundance.  However, it is possible (and indeed particularly useful) to 
express this dynamics in terms of the dynamical variables $\Omega_{\rm BH}$ and 
its time-averaged value 
\begin{equation}
  \OmegaBHAvg ~\equiv~ \frac{1}{t-t_i} 
    \int_{ t_i }^t dt' \, \Omega_{\rm BH}(t')
\label{eq:AvgHubble}    
\end{equation}
where the time averaging extends over the interval $(t_i,t)$ where $t_i$ is 
the initial time $t_i$ at which the PBHs were produced and their mass spectrum 
implicitly established.  In particular, integrating Eq.~(\ref{eq:FriedmannAccelEq})
yields a relation between these two variables of the form
\begin{equation}
  \frac{1}{H} - \frac{1}{H_i} ~=~  \frac{1}{2}(t-t_i)(4 - \OmegaBHAvg)~,
  \label{eq:IntegrateHEqn}
\end{equation}
where $H_i$ is the initial value of $H$ at $t=t_i$.  At times $t \gg t_i$,
when $H \ll H_i$, this relation reduces to 
\begin{equation}
  H ~\approx~ \frac{2}{4-\OmegaBHAvg}\frac{1}{t}~.
  \label{eq:HIntegrateHEqnApprox}
\end{equation}
Recasting the coupled differential equations which govern the evolution of the
PBH abundance in terms of $\Omega_{\rm BH}$ and $\OmegaBHAvg$ is then simply 
a matter of substituting this result for $H$ into Eqs.~(\ref{eq:OmegaBHEvolEq}) 
and~(\ref{eq:HinStasis}).  We thus obtain
\begin{eqnarray}
  \displaystyle  \frac{d \Omega_{\rm BH}}{d t} 
    & ~=~ & \frac{1}{t} \, 
    \Omega_{\rm BH} \left[\frac{ \alpha+1}{3} +  
    \frac{2(1-\Omega_{\rm BH})}{4-\OmegaBHAvg}\right] 
    \nonumber \\
  \displaystyle \frac{d \OmegaBHAvg}{d t} 
    &~=~& \frac{1}{t} \, 
    \Bigl( \Omega_{\rm BH} - \OmegaBHAvg \Bigr)~.
  \label{eq:2Dsystem}
\end{eqnarray}

One advantage of expressing these coupled equations in this form is that
it is more readily apparent that they are autonomous.  
Indeed, we observe that by performing a change of variables from $t$ to 
$\log t$, we can remove all explicit time-dependence from the right 
sides of both equations in Eq.~(\ref{eq:2Dsystem}).

A more important observation that follows from the dynamical equations 
in Eq.~(\ref{eq:2Dsystem}) is that this dynamical system has a 
{\it fixed-point}\/ solution in which
\beq
  \Omega_{\rm BH} ~=~ \langle \Omega_{\rm BH}
    \rangle~=~\barOmega_{\rm BH}~
  \label{fixedpoint}
\eeq
for all time, where
\begin{equation}
  \overline{\Omega}_{\rm BH} ~=~ \frac{4\alpha+10}{\alpha+7}~.
  \label{eq:omegabar}
\end{equation}
Indeed, any configuration in which Eq.~(\ref{fixedpoint}) is satisfied for all 
time is a solution to the dynamical equations in Eq.~(\ref{eq:2Dsystem}).     
Within such a system, the abundance $\Omega_{\rm BH}$ remains fixed even though 
the universe continues to expand!   The abundance of radiation within this universe 
must therefore remain fixed as well.  This solution, originally observed for 
black-hole dynamics in Ref.~\cite{Barrow:1991dn}, is an example of a general 
phenomenon called {\it cosmic stasis}\/~\cite{Dienes:2021woi} in which the universe 
contains non-trivial mixtures of different energy components but in which the 
abundances of these components remains fixed despite cosmological expansion.
Indeed, the solution in Eq.~(\ref{fixedpoint}) is a particular example of cosmic 
stasis between matter and radiation when the matter is comprised of PBHs and the 
radiation is the product of their Hawking evaporation. Within 
Ref.~\cite{Dienes:2021woi}, cosmic stasis was identified in the context of
cosmologies involving towers of decaying particles, such as arise in many models 
of BSM physics.  Indeed, it was shown in Ref.~\cite{Dienes:2021woi} that stasis 
will emerge regardless of the kinds of particles considered so long as their 
decay widths obey certain scaling relations.  The above analysis demonstrates 
that this remains true even for cosmologies involving decaying towers
of primordial black holes!

It is straightforward to determine the properties of the universe corresponding to 
the stasis solution.  For constant $\Omega_{\rm BH} = \overline{\Omega}_{\rm BH}$, 
we may integrate Eq.~(\ref{eq:FriedmannAccelEq}) directly in order to obtain the 
functional form of $H$ during stasis.  In particular, we find that the Hubble 
parameter is given by
\begin{equation}
    H ~=~ \left(\frac{2}{4-\overline{\Omega}_{\rm BH}}\right)\frac{1}{t}~.
    \label{eq:HinStasis}
\end{equation}
Likewise, since $\overline{\Omega}_{\rm BH}$ is only sensibly defined 
within the range $0 \leq \overline{\Omega}_{\rm BH} \leq 1$, it follows from 
Eq.~(\ref{eq:omegabar}) that a value of $\alpha$ within the range  
$-5/2 \leq \alpha \leq -1$ is required in order to obtain stasis.  Combining 
this constraint with the one in Eq.~(\ref{eq:range}), we find that 
\begin{equation}
   -2 ~\leq~ \alpha ~\leq~ -1~,
  \label{eq:AlphaRangeCombined}
\end{equation}
which corresponds to the range $0 \leq w_c \leq 1$, where $w_c$ is the 
equation-of-state parameter during the epoch within which the population of 
PBHs is initially established.  Finally, the effective equation-of-state 
parameter $\overline{w}$ during such a stasis can also be obtained directly by 
substituting this value of $\overline{\Omega}_{\rm BH}$ from Eq.~(\ref{eq:omegabar})
into Eq.~(\ref{eq:HinStasis}), yielding
\begin{equation}
  \overline{w} ~=~ -\frac{\alpha + 1}{\alpha + 7}~.
  \label{eq:wBar}
\end{equation}
Thus, if we demand that $\alpha$ satisfy the criterion in 
Eq.~(\ref{eq:AlphaRangeCombined}), we find that our stasis solution requires
$0 \leq \overline{w} \leq 1/5$.

Our final observation from the dynamical equations in Eq.~(\ref{eq:2Dsystem}) is 
that the fixed-point solution in Eq.~(\ref{fixedpoint}) is actually a 
{\it global attractor}~\cite{Barrow:1991dn,Dienes:2021woi}.  As we have indicated, 
the stasis solution in Eq.~(\ref{fixedpoint}) is valid for all time, and therefore 
assumes initial conditions for the dynamical equations in Eq.~(\ref{eq:2Dsystem}) 
that also happen to satisfy Eq.~(\ref{fixedpoint}).
Indeed, this is a fine-tuned initial condition.
However, in most actual situations, our initial conditions are highly unlikely 
to be fine-tuned in this way.  It is nevertheless straightforward to demonstrate 
that the dynamical system will generally flow toward the stasis solution even if 
it does not begin there. 

In order to see this, we solve the system of equations in Eq.~(\ref{eq:2Dsystem}) 
numerically.  Towards this end, we shall consider all possible combinations of 
initial conditions $\Omega_{\rm BH}(t_i)$ and $\OmegaBHAvg(t_i)$ within the 
domains $0 \leq \{\Omega_{\rm BH}(t_\ast),\OmegaBHAvg(t_\ast)\}\leq 1$.
In order to ensure that PBH evaporation persists until
arbitrarily late times, we take $M_{\rm max} \rightarrow \infty$.
We emphasize that the manner in which the system evolves in this limit
for any particular choice of $\alpha$ is effectively identical to the 
way in which it evolves for finite $M_{\rm max}$ at times 
$t \lesssim \tau(M_{\rm max})$.  Moreover, for simplicity, we also assume 
that $M_{\rm min}$ is such that the lightest PBHs have already begun evaporating 
by the time $t = t_\ast$.

\begin{figure}[t]
 \centering
\includegraphics[width=\linewidth]{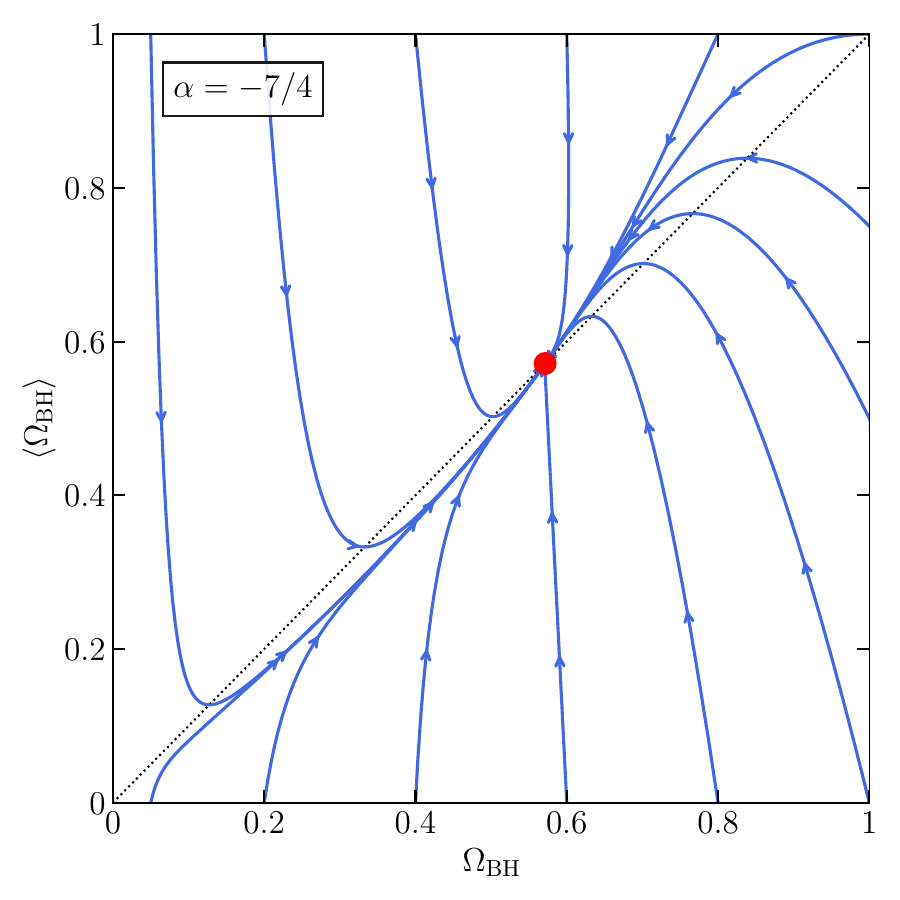}
\caption{Trajectories (blue curves) in the 
  $(\Omega_{\rm BH},\OmegaBHAvg)$-plane for the system of 
  differential equations in Eq.~(\ref{eq:2Dsystem}) with $\alpha = -7/4$.  
  The red dot indicates the point at which 
  $\Omega_{\rm BH} = \OmegaBHAvg = \overline{\Omega}_{\rm BH}$, 
  where in this case $\OmegaBar \approx 0.57$. 
  \label{fig:attractor}}
 \end{figure}

The results of this analysis are shown in Fig.~\ref{fig:attractor} 
for the case in which $\alpha=-7/4$.  The blue curves represent the trajectories 
along which the system evolves in the $(\Omega_{\rm BH},\OmegaBHAvg)$-plane 
for $\Omega_{\rm BH}$ and $\OmegaBHAvg$ anywhere within their domains 
$0 \leq \{\Omega_{\rm BH},\OmegaBHAvg\}\leq 1$.  The red dot in the figure 
indicates the fixed-point solution in Eq.~(\ref{fixedpoint}), 
where in this case $\OmegaBar \approx 0.57$.
We observe that throughout the entirety of the plane, the system 
ultimately flows toward the stasis solution.  Thus, we see that stasis is
indeed a global attractor.  Similar behavior arises for other choices of 
$\alpha$ within the range given in Eq.~(\ref{eq:AlphaRangeCombined}).

At first glance, it might seem that this analysis is misleading because the 
dynamical equations in Eq.~(\ref{eq:2Dsystem}) are valid only at times 
$t \gg t_i$.  It might therefore seem that we can only use these equations to 
study a possible {\it late-time}\/ approach to stasis --- \ie, to study the 
dynamics starting at some fiducial time $t_\ast \gg t_i$.   This in turn would 
imply that we are ignoring the possible non-trivial dynamics that occurred 
during the time interval from $t_i$ to $t_\ast$.  However, {\it regardless}\/ of 
whatever non-trivial dynamics might have transpired between $t_i$ and $t_\ast$, 
our system at $t=t_\ast$ must nevertheless still lie somewhere within the 
$(\Omega_{\rm BH}, \langle \barOmega_{\rm BH}\rangle)$ plane by the time our 
fiducial time $t_\ast$ is reached, since this plane includes all possible values 
for these parameters.  Our analysis then ensures that our system will subsequently 
be drawn to the stasis point.

We conclude, then, that the stasis solution is actually a global attractor within 
a cosmology of this type~\cite{Barrow:1991dn,Dienes:2021woi}.  
Indeed, the primary message of Fig.~\ref{fig:attractor} is not that it is merely 
{\it possible}\/ for stasis to arise in cosmological scenarios involving PBHs,
but that stasis is in fact a {\it generic feature}\/ of cosmologies with 
certain physically motivated PBH mass spectra.  Indeed, we have seen that stasis 
generically arises in {\it any}\/ cosmological scenario in which a population of 
PBHs with a mass spectrum of the form given in Eq.~(\ref{eq:dist}) with 
$-2 \leq \alpha \leq -1$ is generated with a sizable initial abundance after 
inflation.


\section{Stasis in the Early Universe: Stasis is Not Eternal\label{sec:NotEternal}}


Given the emergence of cosmological stasis arising from PBH decay, as discussed in
Sect.~\ref{sec:stasis}, the primary goal of this paper is to understand the numerous
phenomenological implications which follow.  The first of our 
observations --- indeed, the subject of this section --- is that stasis itself 
does not last forever.  Even though we have seen that stasis is an attractor, stasis 
ultimately persists for only a finite duration.  As stressed in 
Ref.~\cite{Dienes:2021woi}, this is a critical feature associated with stasis, 
allowing the universe to {\it exit}\/ stasis as naturally as it enters it.
{\it Indeed, without the ability to exit stasis in a natural way, the universe 
would have been trapped in an eternal stasis which would clearly be in conflict 
with direct observations of the present-day universe.}\/   In this section we 
therefore discuss the entrance into {\it and exit from}\/ stasis, with the goal 
of ultimately obtaining an expression for the number of {\it e}\/-folds across 
which a stasis epoch will last.

\begin{figure}
  \includegraphics[width=\linewidth]{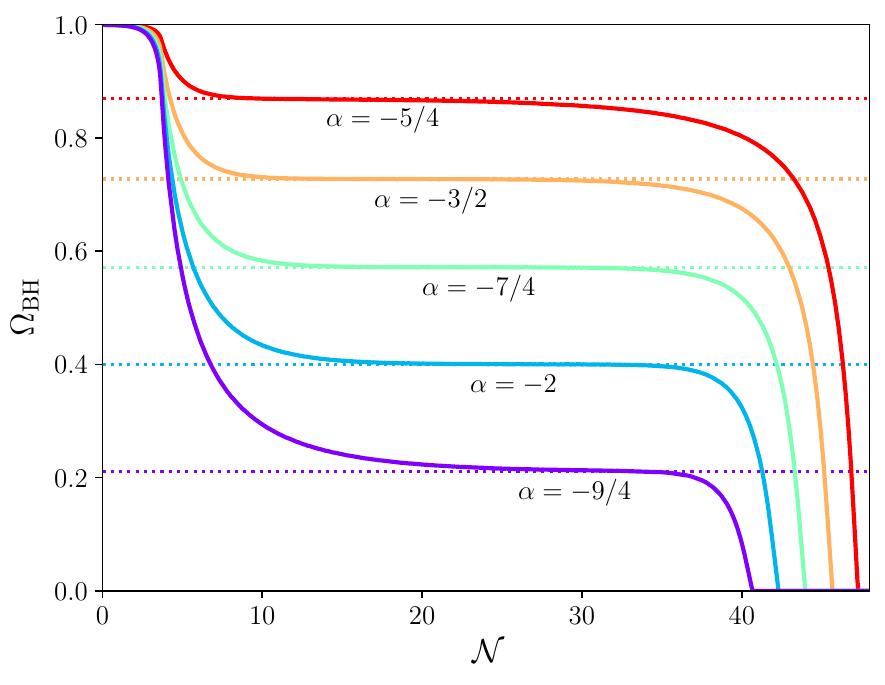}
  \caption{
    The total abundance $\Omega_{\rm BH}$ of a population of
    PBHs with an initial mass spectrum of the form given in 
    Eq.~(\protect\ref{eq:dist}), plotted as a function of the number of $e$-folds 
    $\mathcal{N}$ of cosmic expansion since $t_i$ for several different 
    choices of the power-law exponent $\alpha$ (solid curves).  For all of the 
    curves shown, we have taken $M_{\rm min}=10^{-1}$~g and 
    $M_{\rm max}=10^{9}$~g.  In each case, we see that the universe experiences
    a period of stasis stretching over many $e$-folds.   Indeed, for each 
    curve, the dotted line of the corresponding color represents the theoretical 
    prediction in Eq.~(\protect\ref{eq:omegabar}) for the 
    PBH abundance $\overline{\Omega}_{\rm BH}$ during stasis.
\label{fig:results}}
\end{figure}

In order to set the stage for our analysis, we begin by considering the behavior of 
the PBH abundance $\Omega_{\rm BH}$ as a function of time.  In Fig.~\ref{fig:results}
we plot  $\Omega_{\rm BH}$ (solid curves) as functions of the number $\mathcal{N}$ 
of {\it e}\/-folds for the PBH spectra corresponding to different choices of 
$\alpha$.  In each case we have taken 
$M_{\rm min}=10^{-1}$~g and $M_{\rm max}=10^{9}$~g.  We note that all of these 
choices lie within the range of physically meaningful values specified in 
Eq.~(\ref{eq:range}) except for $\alpha = -9/4$, which we have included for purposes 
of illustration.  In each case, we evaluate $\Omega_{\rm BH}(t)$ using the 
publicly available code {\tt FRISBHEE}~\cite{frisbhee}.  For simplicity, we follow 
Ref.~\cite{Dienes:2021woi} in what follows by focusing on the case in which 
$\Omega_{\rm BH}(t_i) = 1$ --- \ie, the case in which the universe is initially 
dominated by PBHs.  For concreteness, we also evaluate $\varepsilon(M)$ by taking 
the particle content of the theory to be that of the Standard Model (SM).~ 
However, we note that the presence of a small number of additional 
light particle species in the theory would not have a significant effect on 
our results. 

Regardless of the value of $\alpha$, we see that all of these curves share 
certain common characteristics.  First, although they each begin at 
$\Omega_{\rm BH}(t_i)=1$, they each begin to fall below unity at a rate which 
initially increases with decreasing $\alpha$.  In other words, the more negative 
$\alpha$ becomes, the more steeply $\Omega_{\rm BH}(t)$ initially begins to fall.

Second, we see that in each case this falling behavior ultimately begins to level 
out, with each curve approaching a period of stasis.  This is consistent with the 
idea that stasis is an attractor.  Indeed, we see that the number of $e$-folds until 
stasis is reached also depends on $\alpha$, with more negative values of $\alpha$ 
corresponding to more $e$-folds until stasis is achieved.

Third, we see that in each case our curve for $\Omega_{\rm BH}(t)$ then 
enters a period of stasis.  During this period $\Omega_{\rm BH}(t)$ --- and 
therefore also $\Omega_\gamma(t)$ --- remain effectively 
constant, despite the effects of cosmic expansion.   
Indeed, in each case the stasis abundance $\barOmega_{\rm BH}$ precisely 
matches the prediction in Eq.~(\ref{eq:omegabar}).   These latter predictions are 
indicated with dashed horizontal lines in Fig.~\ref{fig:results}.

But perhaps most importantly, we see that in each case this period of stasis is 
of only finite duration.  Indeed, stasis ends in each case once the heaviest PBHs 
ultimately evaporate, whereupon the conversion of matter to radiation ceases.
Given the curves in Fig.~\ref{fig:results}, we see that the number of 
$e$-folds across which stasis extends decreases with $\alpha$, such that 
more negative values of $\alpha$ lead to shorter durations of stasis.
We shall quantify this observation shortly.

Finally, once released from stasis, we find that in each case 
$\Omega_{\rm BH}(t)$ immediately begins to drop again, ultimately reaching 
$\Omega_{\rm BH}=0$ once all of the PBHs have evaporated.  Indeed, these curves 
do not asymptote to $\Omega_{\rm BH}=0$, as might have been expected, but 
instead hit zero with non-zero slope.

The basic dynamical considerations which give rise to the stasis 
behavior that we observe in Fig.~\ref{fig:results} are the same as those
which give rise to stasis in cosmologies involving 
decaying particles~\cite{Dienes:2021woi}.  In a FRW universe, the 
energy density of radiation is diluted more quickly 
than that of matter as the universe expands; thus, the abundance of matter 
typically tends to increase over time, while the abundance of radiation 
tends to decrease.  However, Hawking radiation counteracts this tendency
by transferring energy from matter to radiation --- a function analogous to 
the one the particle decays play within the realization of stasis 
presented  in Ref.~\cite{Dienes:2021woi}.  The interplay between these two
effects drives the universe toward a steady state in which the abundances of 
matter and radiation remain effectively constant in time.   Indeed, the stasis 
epoch which arises from a population of evaporating PBHs is of finite duration 
only because there exists a cutoff $M_{\rm max}$ in the initial PBH mass 
spectrum --- a cutoff which in turn implies a maximum lifetime $\tau(M_{\rm max})$ 
for any PBH within that population.

That said, we stress that the dynamics of PBH evaporation discussed 
in Sect.~\ref{sec:PBH} is fundamentally different from the dynamics of 
particle decay in a number of ways.  For one thing, the form of the collision term 
in the Boltzmann equation which describes the transfer of energy 
density from matter to radiation for a population of decaying particles of the 
same species $\phi$ differs from the form of the corresponding 
collision term for a population of PBHs with a common initial mass $M_i$.  In the 
decaying-particle case, this collision term takes the form 
$\mathcal{C}_\phi = -\Gamma_\phi \rho_\phi$, where $\rho_\phi$ is the energy 
density of the particles and where $\Gamma_\phi$ is their 
proper decay width.  By contrast, in the PBH case, the collision term is the product 
of the number density $n_{\rm BH}$ associated with this monochromatic 
spectrum of PBHs and the evaporation rate in Eq.~(\ref{eq:dMBHdt}).  Expressed in 
terms of the corresponding energy density $\rho_{\rm BH} \approx M n_{\rm BH}$, 
the collision term in this case takes the form
\begin{equation}
  \mathcal{C}_{\rm BH} ~=~ -\varepsilon(M)\frac{M_P^4}{M^3} \rho_{\rm BH}~.
\end{equation}
The factor $\varepsilon(M) M_P^4/M^3$ multiplying the energy density in this expression 
grows with time as the mass of each individual PBH decreases according to 
Eq.~(\ref{eq:Miev}) until $t - t_i = \tau(M_i)$, at which point the PBHs evaporate 
completely.  Since the corresponding factor $\Gamma_\phi$ in the decaying-particle 
case is constant, it follows that the energy density of PBHs falls more 
precipitously when $t - t_i \sim \tau(M_i)$ than does the energy density of decaying 
particles when $t-t_i \sim 1/\Gamma_\phi$.

Another significant difference between particle decay and PBH evaporation involves the 
sequence in which different states transfer their energy to radiation.  In general, 
heavier particles with the same quantum numbers tend to decay before lighter ones.  
This causes the conversion from matter to radiation to proceed {\it down}\/ the tower 
of states in the sense that progressively lighter states dominate the rate at which 
rest energy is transferred to radiation as time evolves.  By contrast, for PBHs the 
reverse is true: lighter PBHs lose their energy to radiation more rapidly than heavier 
ones.  As a result, the conversion from matter to radiation proceeds {\it up}\/ the 
PBH tower in the sense that states with progressively larger initial masses $M_i$ 
dominate the rate at which rest energy is transferred to radiation as time evolves.  

Finally, we may 
derive an approximate expression for the 
{\it duration}\/ of the stasis epoch. 
The transfer of energy density from matter to radiation on which
stasis relies only occurs at a significant rate once the lightest 
PBHs in the tower begin to evaporate completely.  Thus, the window
during which PBH evaporation gives rise to stasis is roughly 
$\tau(M_{\rm min}) \lesssim t \lesssim \tau(M_{\rm max})$. 
The corresponding number of $e$-folds is 
\beq
\mathcal{N}_s ~\approx ~ 
  \log\left[\frac{a(\tau(M_{\rm max}))}{a(\tau(M_{\rm min}))}\right]~.
  \label{eq:NefoldRaw}
\eeq
During stasis, the scale factor scales with time according to the relation
$a(t) \propto t^{2/(3\overline{w} +3)}$, with $\overline{w}$
given by Eq.~(\ref{eq:wBar}).  Thus, we find that the number of $e$-folds
of stasis to which a population of evaporating PBHs gives rise is
\beq
  \mathcal{N}_s ~\approx~
    \frac{\alpha+7}{3}\log\left(\frac{M_{\rm max}}{M_{\rm min}}\right)~.
  \label{eq:Nefold}
\eeq

We note that Eq.~(\ref{eq:Nefold}) can be used in order to derive a rough 
upper bound on the number of $e$-folds of cosmic stasis which can be 
achieved through PBH evaporation in scenarios in which the PBHs are 
produced by the collapse of primordial density fluctuations after inflation.  
As discussed in Sect.~\ref{sec:PBH}, both $M_{\rm min}$ and $M_{\rm max}$ are 
constrained in such scenarios to lie within the range specified in 
Eq.~(\ref{eq:MminMmaxBounds}).  Taking $M_{\rm min}$ at the lower limit 
of this range and $M_{\rm max}$ at the upper limit yields a rough upper
bound on $\mathcal{N}_s$ of the form
\beq
  \mathcal{N}_s ~\lesssim~ 23\left(\frac{\alpha+7}{3}\right)~.
\eeq
Thus, even under the conservative assumptions which led to 
Eq.~(\ref{eq:MminMmaxBounds}), we observe that a significant number of 
$e$-folds of stasis --- up to $\mathcal{N}_s \approx 38.3$ for 
$\alpha = -2$, and even more for larger values of $\alpha$ --- can be 
achieved through PBH evaporation.  Thus, in cosmologies involving
evaporating PBHs, stasis can extend over a potentially significant 
portion of the cosmological timeline.

As indicated above, the fact that a period of stasis has at most a finite duration 
implies that our universe is not only inevitably drawn into a period of stasis 
but also  naturally expelled from it.  Thus it makes sense to speak of a stasis 
``epoch'', one which might resemble a matter- or radiation-dominated epoch in having 
a constant equation of state except that this epoch comprises non-trivial amounts of 
matter and radiation simultaneously and therefore has  a non-traditional equation 
of state.  This is therefore a new kind of cosmological epoch --- one which is an 
intrinsic feature of universes that give rise to an appropriate spectrum of PBHs 
at early times.   It is therefore important to understand the phenomenological 
implications of such a stasis epoch.


\section{Phenomenological Implications of PBH-Induced Stasis \label{sec:Implications}}


In this section we shall consider some of the phenomenological implications 
of a stasis epoch induced by PBH evaporation.  These will include the effects 
of such an epoch on the cosmic expansion history, on inflation, on gravitational waves, 
on dark radiation, on dark matter, and on baryogenesis.  Needless to say, 
many of these effects are interconnected.  We shall nevertheless attempt to 
separate the relevant physics into distinct discussions as much as possible.

\subsection{Cosmic Expansion History\label{sec:ExpansionHistory}}

In the previous sections, we have shown that the evaporation of a population
of PBHs with a mass spectrum described by the power-law distribution in
Eq.~(\ref{eq:dist}) can give rise to a stasis epoch lasting a significant number 
of $e$-folds.  Such a modification of the cosmic expansion history can have
an impact on variety of astrophysical quantities, including the spectrum of 
primordial density fluctuations and the stochastic gravitational-wave (GW) 
background.  

In this section, we examine the phenomenological implications of 
modified cosmologies involving a stasis epoch induced by PBH evaporation.
We begin by reviewing the expansion history of the universe that characterizes 
cosmologies of this sort from the end of inflation to the present time.  This 
expansion history includes not only a stasis epoch, but the entire sequence 
of epochs described below.
\begin{itemize}
  \item Immediately following inflation, the universe is dominated by a perfect 
    fluid with an equation-of-state parameter $0 \leq w_c\leq 1$. 
    PBHs form after inflation via the gravitational collapse of primordial
    density perturbations during this epoch.  The power spectrum
    of these perturbations is assumed to be scale-invariant across the entire 
    range of comoving wavenumbers $k$ which correspond to PBH masses within the
    range $M_{\rm min} \leq M_i \leq M_{\rm max}$.  As discussed in 
    Sect.~\ref{sec:PBH}, such a perturbation spectrum gives 
    rise to an initial PBH mass spectrum of the form given in Eq.~(\ref{eq:dist}).  
    We emphasize, however, that while we assume the power spectrum of these 
    perturbations to be scale-invariant {\it within}\/ this range of $k$, the 
    amplitude of these perturbations may potentially be significantly enhanced 
    relative to the amplitude of perturbations {\it outside}\/ this range.
  \item Prior to the time at which these PBHs begin evaporating at a non-negligible
    rate, their abundance increases for $w_c$ within this range.  This eventually 
    leads to a matter-dominated epoch during which the PBHs themselves dominate 
    the energy density of the universe.  The duration of this epoch --- which 
    we parametrize in terms of the number of $e$-folds of cosmic 
    expansion $\mathcal{N}_{\rm PBH}$ that take place therein --- 
    depends on the total abundance of PBHs generated over the course 
    of the preceding epoch.  This abundance in turn depends fundamentally on the 
    amplitude of the primordial density perturbations which collapse to form the 
    PBHs --- an amplitude which, as discussed above, can potentially be significantly
    enhanced within the region of $k$ relevant for PBH formation.  We therefore 
    take $\mathcal{N}_{\rm PBH}$ to be a free parameter in what follows.
  \item Once the PBHs begin evaporation, the universe evolves toward a stasis 
    due to the attractor behavior discussed in Sect.~\ref{sec:stasis}.~  The 
    effective equation-of-state parameter $\overline{w}$ for the universe during 
    the stasis epoch is given by Eq.~(\ref{eq:wBar}) and the duration of this 
    epoch is given by Eq.~(\ref{eq:Nefold}).
  \item After the heaviest PBHs have evaporated, the universe becomes 
    radiation-dominated.  The subsequent expansion history of the universe
    coincides with that of the standard cosmology.
\end{itemize}

\begin{figure}
  \includegraphics[width=\linewidth]{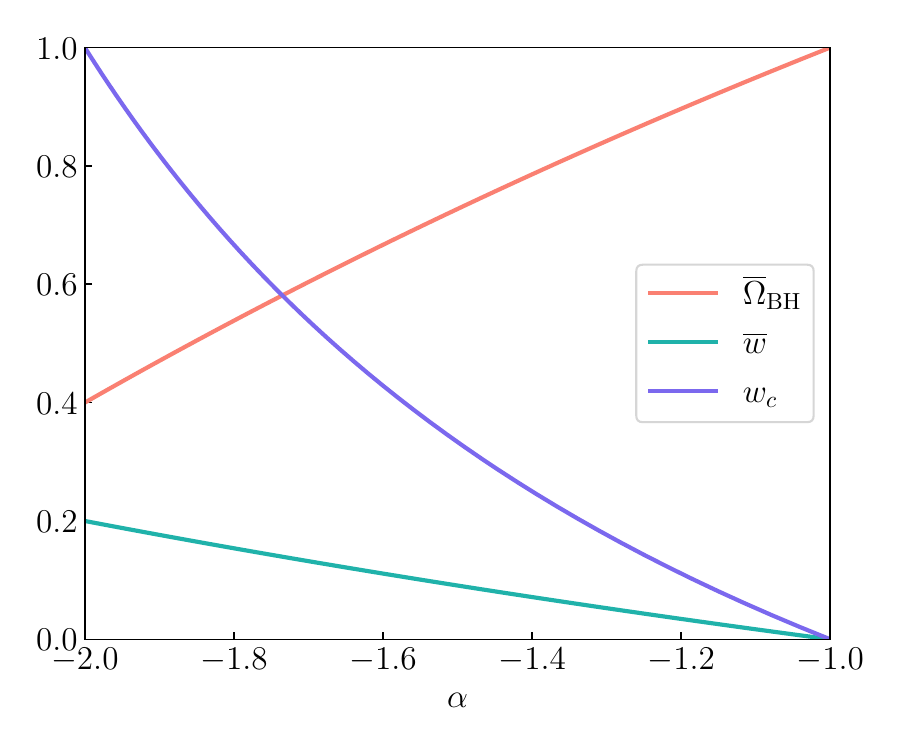}
  \caption{The PBH abundance $\barOmega_{\rm BH}$ during stasis, the equation-of-state 
  parameter $\overline w$ during stasis, and the equation-of-state parameter $w_c$  during the PBH-formation 
  epoch, each plotted as a function of $\alpha$.
  \label{fig:alpha_Omega_w}}
\end{figure}

We emphasize that the parameters $w_c$, $\overline{w}$, and $\alpha$ are 
not independent in cosmologies of this sort; rather, they are 
interrelated through Eqs.~\eqref{eq:w_c} and~\eqref{eq:wBar}.  The relationships 
between these three parameters are displayed graphically in Fig.~\ref{fig:alpha_Omega_w}.
Thus, since the duration of the stasis epoch depends not only
on $\alpha$, but also on the ratio $M_{\rm max}/M_{\rm min}$, the expansion 
history after inflation in these kinds of modified cosmologies is determined by 
two parameters, which we take to be $\alpha$ and $M_{\rm max}/M_{\rm min}$.

Since the equation-of-state parameter is constant during each of the 
individual cosmological epochs described above, it is straightforward to
obtain the total number of $e$-folds of expansion which take place between
the end of inflation and present time.  In particular, in the approximation 
that transitions between successive cosmological epochs may be considered 
effectively instantaneous, the number of $e$-folds of expansion which take 
place during an epoch with equation-of-state parameter $w_j$ is related to 
the values $H_j$ and $H_{j+1}$ of the Hubble parameter at the beginning of 
that epoch and at the beginning of the subsequent epoch, respectively, by
\begin{equation}
  \mathcal{N}_j ~=~ \frac{2}{3(1+w_j)}\log\left(\frac{H_j}{H_{j+1}}\right)~.
\end{equation}

Given a particular set of input parameters $\alpha$, $M_{\rm min}$, $M_{\rm max}$, and 
$\mathcal{N}_{\rm PBH}$ and a value of $H_{\rm end}$ at the end of inflation, this
relation is sufficient to allow us to reconstruct the entire expansion history. 

\begin{figure}
\centering
  \includegraphics[width=\linewidth]{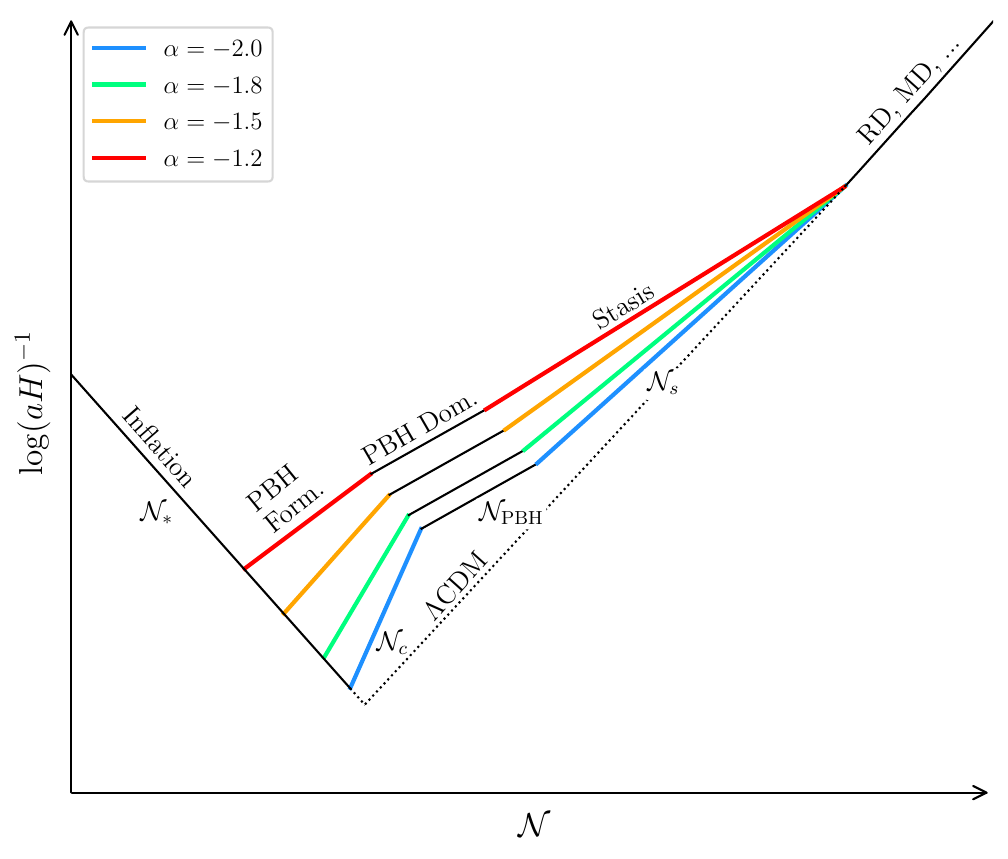}
  \caption{Qualitative sketch illustrating how the comoving Hubble horizon 
    evolves in cosmologies involving an epoch of PBH-induced stasis 
    as a function of the number of $e$-folds for several different values of 
    $\alpha$.  All curves shown correspond to the parameter choices 
    $M_{\rm max}/M_{\rm min} = 10^{7}$ and $\mathcal{N}_{\rm PBH} = 10$. 
    The evolution of $(aH)^{-1}$ in the standard cosmology
    is indicated by the dotted black curve. 
  \label{fig:cosmo}}  
\end{figure}

The manner in which the comoving Hubble horizon evolves in modified cosmologies 
of this sort from the end of inflation until the present time is shown in 
Fig.~\ref{fig:cosmo} for a variety of different values of $\alpha$ within the
physically motivated range given in Eq.~(\ref{eq:range}).  For each of the
curves shown, we have taken $M_{\rm max}/M_{\rm min} = 10^{7}$, which yields 
the maximal duration for the stasis epoch consistent with 
Eq.~(\ref{eq:MminMmaxBounds}) for a given value of $\alpha$.  We have also 
taken $\mathcal{N}_{\rm PBH} = 10$.  The manner in which $(aH)^{-1}$ evolves 
in the standard cosmology is indicated by the dotted black line. 

The results shown in Fig.~\ref{fig:cosmo} illustrate that in cosmologies involving an 
epoch of PBH-induced stasis, $(aH)^{-1}$ can depart significantly from the 
corresponding result in the standard cosmology between the end of inflation 
and the end of the stasis epoch, and that this departure becomes more pronounced 
as $\alpha$ increases.  Indeed, this modified evolution of the comoving Hubble 
radius is ultimately responsible for many of the phenomenological 
signatures of PBH-induced stasis which we shall discuss below.

\subsection{Inflation \label{sec:Inflation}}

Cosmic inflation stands as one of the most compelling frameworks that can explain 
the extraordinary homogeneity of the cosmic microwave background (CMB) while 
generating from quantum fluctuations the metric perturbations that led to 
structure formation later in cosmic 
history~\cite{Guth:1980zm,Linde:1981mu,Albrecht:1982wi,Akrami:2018odb}. 
In its simplest realizations, a scalar field slowly rolling along its potential 
is responsible for generating the perturbation modes which give rise to the 
pattern of inhomogeneities observed in the CMB.~  In order to derive a theoretical 
prediction for this pattern of inhomogeneities in the context of a given 
inflationary model, one must determine the values of the so-called slow-roll 
parameters at the time at which the relevant perturbation modes exited 
the Hubble horizon.  However, the value of the scale factor at this time is 
entirely dependent on the subsequent expansion history of the universe. 
As discussed in Ref.~\cite{Dienes:2021woi}, the presence of a cosmic-stasis 
epoch within the cosmological timeline can affect the number of $e$-folds of 
cosmic expansion which occur between the time at which these modes exit the 
horizon and the time at which they reenter.  In this subsection, we examine
the impact that the presence of a PBH-induced stasis epoch --- along with the other 
modifications of the expansion history involved in establishing such an 
epoch --- has on predictions for inflationary observables which can be 
inferred from the properties of the CMB. 

The primary inflationary observables of this sort are the spectral index $n_s$ 
and the tensor-to-scalar ratio $r$ which characterize the spectrum of primordial 
perturbations, evaluated at some chosen pivot scale $k_\star$. 
In order to make contact with observation, we adopt the pivot scale 
$k_\star=0.002\,\mathrm{Mpc}^{-1}$ used in the analysis performed by the 
Planck Collaboration~\cite{Planck:2018jri}. 
In slow-roll inflation scenarios, these observables are related to the scalar 
potential $V(\phi)$ for the inflaton field $\phi$ through the slow-roll parameters
\begin{eqnarray}
  \epsilon &~\equiv~& \frac{M_P^2}{16\pi}
    \left[\frac{V'(\phi_\star)}{V(\phi_\star)}\right]^2 \nonumber \\
  \eta &~\equiv~& \frac{M_P^2}{8\pi}
    \left|\frac{V''(\phi_\star)}{V(\phi_\star)}\right|~,
\end{eqnarray}
where a prime denotes a derivative with respect to $\phi$ and where $\phi_\star$
denotes the value of $\phi$ at the time at which the mode with wavenumber $k_\star$
exits the horizon.  In particular, in such scenarios, one finds that 
\begin{eqnarray}
  n_s &~=~& 1 - 6\epsilon + 2\eta \nonumber \\
  r &~=~& 16\epsilon~.
\end{eqnarray}

The value of $\phi_\star$ depends sensitively on the expansion history of the 
universe after inflation.  For a given expansion history,
the value of $\phi_\star$ may be determined as follows~\cite{Liddle:2003as,
Allahverdi:2018iod,Kofman:1994rk,Ghoshal:2022ruy,Heurtier:2022rhf,Heurtier:2019eou}. 
The number of $e$-folds of expansion $\mathcal{N}_\star$ that the universe 
undergoes between the time at which the perturbation mode with comoving wavenumber 
$k_\star$ exits the horizon and the end of inflation using the relation
\beq
  k_\star ~=~ a_\star H_\star ~=~ e^{-\mathcal{N}_\star}a_{\rm end}H_\star\,,
  \label{eq:kstar}
\eeq
where $a_\star$ and $H_\star$ respectively denote the values of the scale factor and the
Hubble parameter at the moment at which this mode exits the horizon, 
and where $a_{\rm end}$ denotes the value of the scale factor at the end 
of inflation.  In the slow-roll approximation, $H_\star$ and $\phi_\star$ are 
related by 
\begin{equation}
   H^2_\star ~\approx~ \frac{8\pi V(\phi_\star)}{3M_P^2}~.
   \label{eq:Hstar}
\end{equation}
On the other hand, the slow-roll approximation also implies that
\begin{equation}
  \mathcal{N}_\star ~\approx~ 
    \log\left(\frac{a_{\rm end}}{a_\star}\right) 
    ~=~ \frac{8\pi}{M_P^2}
    \int_{\phi_{\rm end}}^{\phi_\star} \frac{V(\phi)}{V'(\phi)} d\phi~.
  \label{eq:Nstar}
\end{equation}
Taken together, Eqs.~(\ref{eq:kstar}), (\ref{eq:Hstar}), and~(\ref{eq:Nstar})
yield an integro-differential equation of the form
\begin{eqnarray}
  \frac{8\pi}{M_P^2}\int_{\phi_{\rm end}}^{\phi_\star} 
    \frac{V(\phi)}{V'(\phi)} d\phi &\,=\,&
    \frac{1}{2}\log\left(\frac{8\pi a_{\rm now}^2 V(\phi_\star)}{3M_P^2 k_\star^2}\right) 
    \nonumber \\ & & ~~ -\log\left(\frac{a_{\rm now}}{a_{\rm end}}\right)~,
  \label{eq:IntegDiffEqForPhiStar}
\end{eqnarray}
which can be solved numerically for $\phi_\star$.  It is through the second term 
on the right side of Eq.~(\ref{eq:IntegDiffEqForPhiStar}) --- which represents the total 
number of $e$-folds of expansion that occur between the end of 
inflation and the present time --- that $\phi_\star$ depends on the expansion 
history after inflation.

The results for $n_s$ and $r$ of course depend on the form of the inflaton 
potential as well as on the expansion history of the universe after inflation.
Here, for concreteness, we focus on two commonly studied classes of 
inflaton potentials.  The first are polynomial potentials which take the 
form 
\begin{equation}
  V(\phi) ~\sim~ |\phi|^p 
  \label{eq:PolyV}
\end{equation}
 for some value of $p$.  The second consists of  
 so-called $T$-model $\alpha$-attractors~\cite{Kallosh:2013hoa}, for 
 which the inflaton potential takes the form 
\begin{equation}
  V(\phi) ~\sim~ \tanh^{2n}\left(\sqrt\frac{4\pi}{3\alpha_{\rm inf}}
    \frac{\phi}{M_P}\right)~,
  \label{eq:TModelV}
\end{equation}
where $n$ and $\alpha_{\rm inf}$ are dimensionless free parameters.  We focus
here on the case in which $n=1$, for which this form of $V(\phi)$ coincides 
with the inflaton potential characteristic of Starobinsky $R^2$
models~\cite{Starobinsky:1980te,Starobinsky:1983zz}.

\begin{figure*}
  \includegraphics[width=\linewidth]{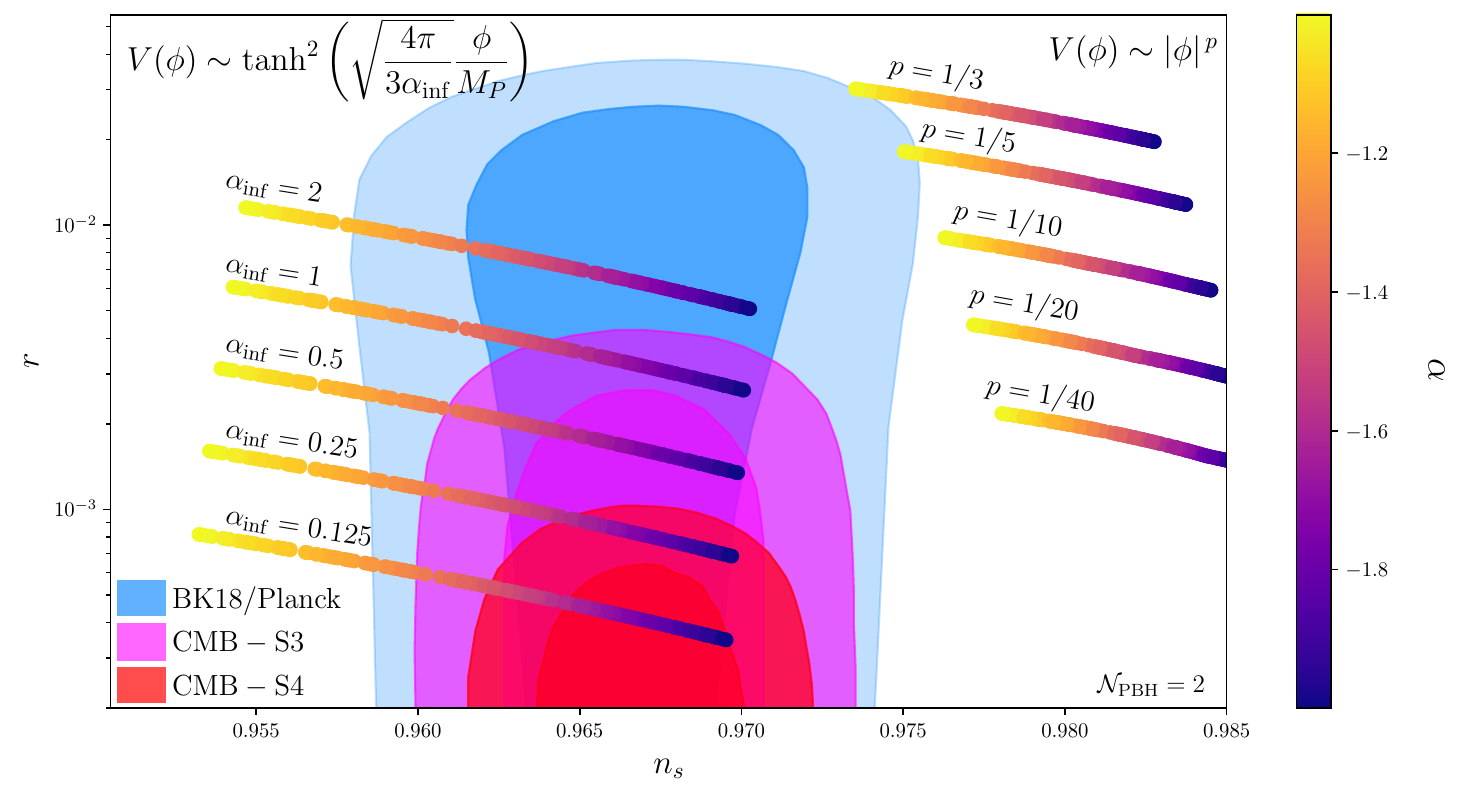}
  \caption{The spectral index $n_s$ and tensor-to-scalar 
    ratio $r$ in cosmologies involving an epoch of PBH-induced stasis, shown as 
    points in the $(n_s,r)$-plane for the two classes of inflaton potential given in 
    Eqs.~(\ref{eq:PolyV}) and~(\ref{eq:TModelV}).  Each sequence of points 
    appearing on the left side of the figure corresponds to a potential of the 
    form given in Eq.~(\ref{eq:TModelV}) with a different value $\alpha_{\rm inf}$.
    Each sequence of points appearing on the right side of the figure corresponds 
    to a different value of the parameter $p$ in Eq.~(\ref{eq:PolyV}).  The color 
    of each point within a particular such sequence indicates the value 
    of the parameter $\alpha$ in Eq.~(\ref{eq:dist}).  For all points shown in 
    the figure, we have taken $M_{\rm min} = 5\,{\rm g}$ and 
    $M_{\rm max} = 10^9\,{\rm g}$, such that the value of $\mathcal{N}_s$ is 
    maximized for each choice of $\alpha$.  We have also taken 
    $\mathcal{N}_{\rm PBH} = 2$.  The regions within which the values of $n_s$ and 
    $r$ are consistent with Planck data~\cite{Planck:2018jri} at the 68\% and 95\%~CL 
    are shaded dark and light blue, respectively.  The dark- and light-purple regions 
    represent the corresponding 68\% and 95\%~CL projections for CMB-S3 experiments
    collectively, while the dark- and light-red regions represent the corresponding 
    projections for CMB-S4~\cite{Abazajian:2019eic}.
    \label{fig:observables}}
\end{figure*}

In Fig.~\ref{fig:observables} we display the results obtained for $n_s$ and $r$
in a cosmology involving an epoch of PBH-induced stasis as points in the 
$(n_s,r)$-plane for the two classes of inflaton potential given
in Eqs.~(\ref{eq:PolyV}) and~(\ref{eq:TModelV}).  Each sequence of points 
appearing on the left side of the figure corresponds to a potential of the 
form given in Eq.~(\ref{eq:TModelV}) with a different value $\alpha_{\rm inf}$.
Likewise, each sequence of points appearing on the right side of the figure  
corresponds to a different value of the parameter $p$ in Eq.~(\ref{eq:PolyV}).
The color of each point within a particular such sequence indicates the value 
of the parameter $\alpha$ in Eq.~(\ref{eq:dist}).
In all cases, we have taken $M_{\rm min} = 5\,{\rm g}$ and 
$M_{\rm max} = 10^9\,{\rm g}$, such that the value of $\mathcal{N}_s$ is 
maximized for each choice of $\alpha$.  We have also taken 
$\mathcal{N}_{\rm PBH} = 2$.  Taken together, these choices of 
$M_{\rm min}$, $M_{\rm max}$, and $\mathcal{N}_{\rm PBH}$ specify the
duration of the PBH-formation epoch.
The regions within which the values of $n_s$ and $r$ are consistent with Planck 
data~\cite{Planck:2018jri} at the 68\% and 95\%~CL are shaded dark and light blue, 
respectively.  The dark- and light-purple regions represent the corresponding 
68\% and 95\%~CL projections for CMB-S3 experiments such as the 
Simons Observatory~\cite{SimonsObservatory:2018koc} and 
the South Pole Observatory~\cite{Moncelsi:2020ppj} collectively, 
while the dark- and light-red regions represent the corresponding projections 
for the CMB-S4 experiment~\cite{Abazajian:2019eic}.

We observe from Fig.~\ref{fig:observables} that the modifications of the 
expansion history associated with cosmologies involving a PBH-induced stasis 
epoch have a significant impact on the predictions for $r$ and $n_s$ 
obtained for a given inflaton potential.  In general, these modifications 
have the effect of decreasing $n_s$ and increasing $r$ relative to the 
corresponding values obtained in the context the standard cosmology, and 
the larger the value of $\alpha$ becomes, the more pronounced this effect 
becomes.  For the $T$-model potential in Eq.~(\ref{eq:TModelV}), these 
modifications generally increase the tension between theoretical predictions 
for $n_s$ and $r$ and CMB data.  Indeed, for $\alpha \gtrsim -1.3$, 
we find that the resulting prediction for $n_s$ and $r$ lie outside 
the Planck 95\%-CL region for essentially all values of 
$\alpha_{\rm inf}$ we consider here.
By contrast, for the polynomial potential in Eq.~(\ref{eq:PolyV}), the 
shift in $n_s$ and $r$ which results from the modification of the 
expansion history serves to partially ameliorate the tension between 
the predictions for these quantities and CMB data.  Indeed, for 
large values of $p$ the resulting shift in these quantities is 
sufficient to land them --- though just barely --- within the Planck 
95\%-CL region, thereby rendering a model which was disfavored by CMB 
(marginally) consistent with those measurements.  
We emphasize that the trends illustrated in Fig.~\ref{fig:observables} 
concerning the manner in which the values of $n_s$ and $r$ shift in 
cosmologies involving an epoch of PBH-induced stasis relative to their
values in the standard cosmology are general and should apply in
other inflationary models as well.

\subsection{Gravitational Waves \label{sec:GravitationalWaves}}

Both the presence of a PBH-induced stasis epoch within the cosmological 
timeline and the additional modifications of the expansion history involved 
in establishing such a stasis epoch also have a non-trivial impact on the 
resulting present-day spectrum of GWs.  Indeed, in cosmologies with modified 
expansion histories, this spectrum --- defined more precisely as the differential 
present-day abundance $d\Omega_{\rm GW}/d\log f$ of GWs per unit logarithmic 
present-day frequency $\log f$ --- is modified relative to 
the present-day GW spectrum that would be obtained within the context of the 
standard cosmology for the same primordial spectrum of tensor
perturbations~\cite{Opferkuch:2019zbd,Ghoshal:2022ruy}.
In particular, the manner in which $d\Omega_{\rm GW}/d\log f$ scales with $f$ 
is modified across any range of $f$ for which the GWs reenter the 
horizon during an epoch during which the universe is not radiation-dominated.
In the cosmologies we consider here, such epochs include the stasis epoch itself,
the epoch of PBH-domination which precedes it, and the epoch during which the
PBHs are initially produced.

We begin by examining the impact of the cosmological modifications involved 
in our PBH-induced stasis scenario on contributions to the GW spectrum 
generated by other sources.
For concreteness, we consider the case in which a stochastic 
background of GWs is established prior to the PBH-formation epoch.
Under standard assumptions that this stochastic GW background (SGWB) is 
homogeneous, isotropic, Gaussian, and unpolarized, the differential energy 
density $d\rho_{\rm GW}(a)/d\log k$ of GWs per unit logarithmic 
comoving-wavenumber $\log k$, expressed as a function of the 
scale factor $a$, takes the form~\cite{Caprini:2018mtu}
\beq
  \frac{d\rho_{\rm GW}(a)}{d\log k} ~=~ \frac{k^2 h_k^2(a)}{16\pi G a^2}~,
  \label{eq:GW_lnk}
\eeq
where $h_k(a)$ represents the characteristic differential amplitude of tensor
perturbations to the spacetime metric per unit $\log k$, likewise expressed as 
a function of $a$.  This differential amplitude evolves with $a$ according 
to the relation
\begin{equation}
  h_k(a)~=~\frac{a_k}{a}h_k(a_k)~,
  \label{eq:hkscal}
\end{equation}
where $a_k$ denotes the value of $a$ at the time at which a perturbation with 
comoving wavenumber $k$ reenters the horizon.  

In a flat universe, the Hubble parameter scales with $a$ 
according to the proportionality relation 
$H=\sqrt{8\pi G \rho} \propto a^{-3(1+w)/2}$
during a cosmological epoch wherein the equation-of-state 
parameter $w$ is effectively constant.  The comoving wavenumber
of the GW which enters the horizon during such an epoch therefore 
scales with $a_k$ according to the proportionality relation 
\begin{equation}
  k ~=~ a_k H_k ~\propto~ a_k^{-(1+3w)/2}~.
  \label{eq:HtoaScaling}
\end{equation}
It therefore follows from
Eqs.~(\ref{eq:GW_lnk}) and~(\ref{eq:hkscal}) that 
\beq
  \frac{d\rho_{\rm GW}(a)}{d\log k} ~\propto~ a^{-4} h_k^2(a_k) 
  k^{\xi(w)}\,,
  \label{eq:GW_slope}
\eeq
where we have defined 
\begin{equation}
  \xi(w) ~\equiv~ \frac{2(3w-1)}{(3w+1)}\,. 
  \label{eq:xi}
\end{equation}

Within the context of the standard cosmology (wherein the universe is 
radiation-dominated, with $w=1/3$, from the end of reheating until the 
time of matter-radiation equality), Eq.~(\ref{eq:GW_slope}) implies that
$d\rho_{\rm GW}(a)/d\log k \propto h_k^2(a_k)$.  
By contrast, within the context of any modified cosmology wherein the 
universe is {\it not}\/ radiation-dominated throughout this interval,
the resulting GW spectrum is distorted relative to this result.
Indeed, we see that $d\rho_{\rm GW}(a)/d\log k$ increases more rapidly 
with $k$ during an epoch wherein $w > 1/3$ than it does in the standard
cosmology, whereas it increases less rapidly than it does in the standard 
cosmology during an epoch wherein $w < 1/3$.

For concreteness, in order to illustrate how the modified expansion
history associated with a PBH-induced stasis epoch affects the GW spectrum, 
we consider a spectrum of tensor perturbations for which the initial amplitude 
$h_k(a_k)$ is independent of $k$.  Such a spectrum of tensor perturbations arises
generically in single-field inflation scenarios in which the spectrum of primordial 
curvature perturbations generated by inflation is perfectly scale 
invariant --- \ie, scenarios in which $n_s = 1$.  For simplicity, we ignore any additional 
contributions to the perturbation spectrum which might arise from other 
sources.  We shall return to discuss the effect of such contributions later in this 
subsection. 

For such an initial spectrum of tensor perturbations, the $k$-dependence of 
$d\rho_{\rm GW}/d\log k$ is determined solely by
the expansion history.  Over ranges of $k$ for which GW modes 
reenter the horizon during radiation domination, $d\rho_{\rm GW}(a)/d\log k$ 
is ``flat'' --- \ie, independent of $k$.  By contrast, over ranges of 
$k$ for which GW modes reenter the horizon during any epoch wherein 
$w \neq 1/3$, this differential energy density varies with $k$, as discussed above.
In the PBH-induced stasis scenario we are considering here, the epochs wherein 
$w\neq 1/3$ include not only the epoch of PBH-induced stasis itself, but also 
the epoch during which the PBHs are initially produced and the subsequent 
epoch of PBH domination which immediately precedes stasis.  

We now turn to assessing the potential consequences of these cosmological 
modifications for GW detection.  The present-day GW spectrum is defined 
according to the relation
\begin{equation}
  \frac{d\Omega_{\rm GW}}{d\log f}
    ~\equiv~\frac{1}{\rho_{\rm crit}(a_{\rm now})}
    \frac{d\rho_{\rm GW}(a_{\rm now})}{d\log f}~,
\end{equation}
where $\rho_{\rm crit}(a_{\rm now})$ is the present-day value of the critical 
density of the universe and where $d\rho_{\rm GW}(a_{\rm now})/d\log f$ is 
the differential present-day energy density  of GWs per unit $\log f$.
Since $f = k/(2\pi a_{\rm now})$, where $a_{\rm now}$ denotes the present-day
value of the scale factor, it follows that
\begin{equation}
  \frac{d\rho_{\rm GW}(a_{\rm now})}{d\log f} ~=~ 
    \frac{d\rho_{\rm GW}(a_{\rm now})}{d\log k}~.    
\end{equation}
In the approximation that the transitions between different cosmological 
epochs may be taken to be effectively instantaneous, we find that 
$d\Omega_{\rm GW}/d\log f$ in a cosmology with an epoch of PBH-induced 
stasis is given by the piecewise function
\begin{widetext}
\begin{equation}
  \frac{d\Omega_{\rm GW}}{d\log f}
    ~=~\frac{d\Omega_{\rm GW}^{\rm sc}}{d\log f}\times
  \begin{cases}
    1 & f\leq f_s\\
    \left(\frac{f}{f_s} \right)^{\xi(\overline{w})} 
      & f_s <f \leq f_{\rm PBH}\\
    \left(\frac{f_{\rm PBH}}{f_s} \right)^{\xi(\overline{w})}
    \left(\frac{f}{f_{\rm PBH}}\right)^{-2} & f_{\rm PBH}< f \leq f_f\\
    \left(\frac{f_{\rm PBH}}{f_s} \right)^{\xi(\overline{w})}
    \left(\frac{f}{f_{\rm PBH}}\right)^{-2} 
    \left(\frac{f}{f_f} \right)^{\xi(w_c)} 
    & f_f < f \leq f_{\rm end}\\
    0 & f_{\rm end} < f~,
  \end{cases}
  \label{eq:dOmegaGWdlnf}
  \end{equation}
\end{widetext}
where $d\Omega_{\rm GW}^{\rm sc}/d\log f$ denotes the GW spectrum 
obtained for the same spectrum of primordial tensor perturbations within the 
context of the standard cosmology and where $f_{\rm end}$, $f_f$, $f_{\rm PBH}$, and 
$f_s$ respectively indicate the values of $f$ associated with the GW modes which 
reenter the horizon at the end of inflation; at the end of the epoch immediately 
following inflation during which the PBHs are initially produced; at the end of 
the ensuing epoch of PBH domination; and at the end of the stasis epoch.  
Since $d\rho_{\rm GW}(a)/d\log f$ is independent of $f$ in the standard cosmology, 
the quantity $d\Omega_{\rm GW}^{\rm sc}/d\log f$ may be viewed simply as an overall 
normalization factor in $d\Omega_{\rm GW}/d\log f$ --- a normalization 
factor which takes the form~\cite{Caprini:2018mtu}
\begin{equation}
  \frac{d\Omega_{\rm GW}^{\rm sc}}{d\log f} ~=~ 
    \Omega_{\gamma}(a_{\rm now})
    \left(\frac{g_{\star S}(T_{\rm eq})}{g_{\star S}(T_k)}\right)^{4/3}
    \frac{g_{\star}(T_k)}{24\pi^2}\frac{H_\star^2}{M_P^2} \,,
  \label{eq:dOmegaGWdlnfsc}
\end{equation}
where $\Omega_\gamma(a_{\rm now})\simeq 5.38\times 10^{-5}$ is the present-day 
photon abundance~\cite{ParticleDataGroup:2018ovx}.

\begin{figure*}
  \includegraphics[width=0.49\linewidth]{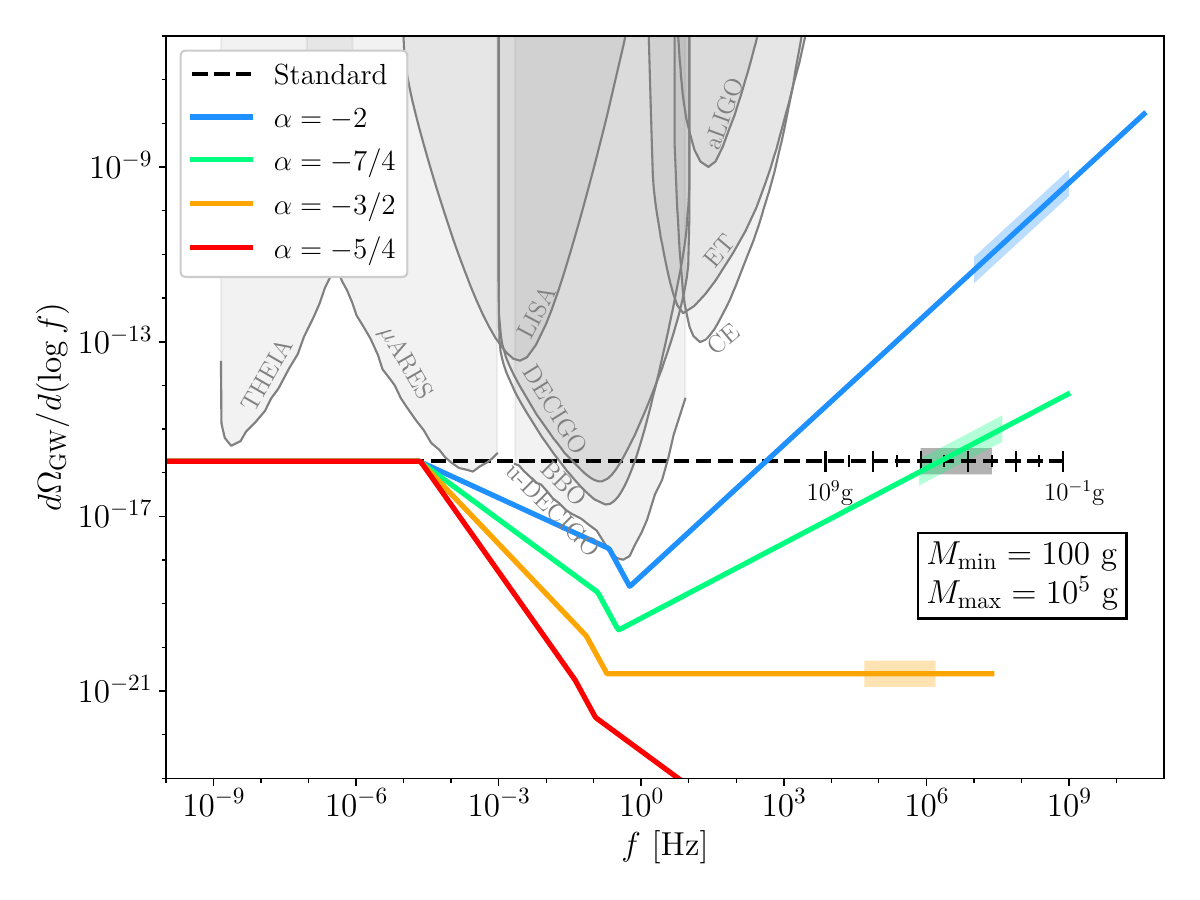}
  \includegraphics[width=0.49\linewidth]{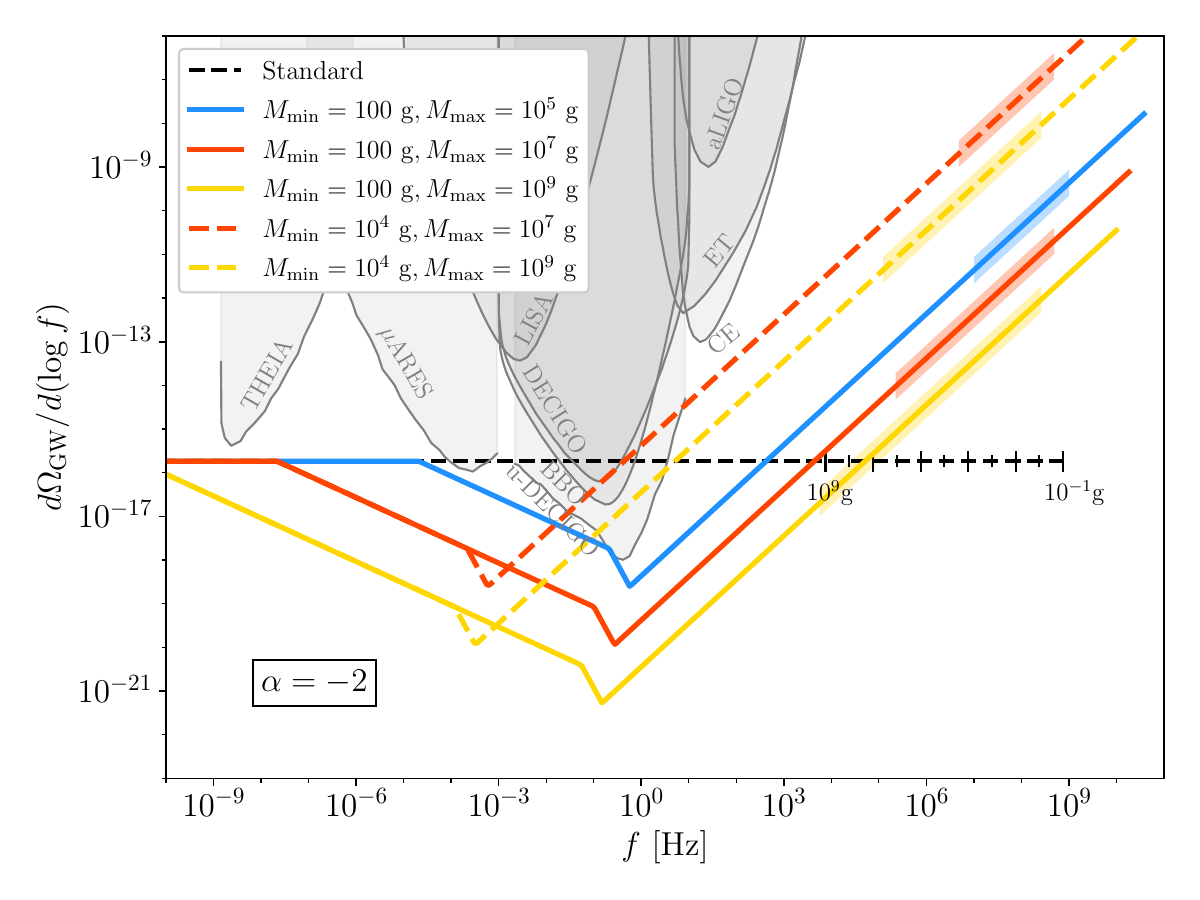}
\caption{
The GW spectra --- \ie, differential abundance of GWs per unit 
logarithmic present-day GW-frequency $\log f$ --- produced by a 
scale-invariant spectrum of primordial tensor perturbations in cosmologies 
involving an epoch of PBH-induced stasis.  The colored curves in each panel 
indicate the GW spectra obtained for different choices of the parameters 
$\alpha$, $M_{\rm min}$, and $M_{\rm max}$ which characterize the mass spectrum 
of the PBHs, while the dashed black horizontal line indicates the $k$-independent 
GW spectrum obtained within the context of the standard cosmology. 
In the left panel, we display the GW spectra obtained for several different choices 
of $\alpha$ within the physically allowed range in Eq.~(\ref{eq:AlphaRangeCombined})
for fixed $M_{\rm min} = 100$~g and $M_{\rm max} = 10^5$~g.  By contrast, in the right 
panel, we display the GW spectra obtained for several different combinations of
$M_{\rm min}$ and $M_{\rm max}$ for fixed $\alpha=-2$ (a value of $\alpha$ 
which corresponds to $\overline{w}=1/5$ and $w_c=1$).
The range of GW frequencies which is associated via Eq.~\eqref{eq:f_M_st}  
with the production of PBHs with masses in the range 
$10^{-1}~{\rm g} < M_i < 10^{9}~{\rm g}$ within the context of the standard 
cosmology is indicated by the tick-marks which appear along the right 
portion of the dashed black line.  The range of GW frequencies which is associated 
with this same range of $M_i$ within the context of a cosmology involving an epoch 
of PBH-induced stasis characterized by a particular combination of $\alpha$, 
$M_{\rm min}$, and $M_{\rm max}$ is indicated by the shaded bar which appears along 
the corresponding $d\Omega_{\rm GW}^{\rm sc}/d\log f$ curve.  Since a significant
abundance of PBHs is necessary for achieving stasis, 
$d\Omega_{\rm GW}^{\rm sc}/d\log f$ is expected to receive additional 
enhancements within this frequency range relative to the result shown here, which
pertains to the case of a perfectly scale-invariant spectrum of primordial 
tensor perturbations.
\label{fig:GW}}
\end{figure*}

In Fig.~\ref{fig:GW}, we show a variety of present-day GW spectra which arise in 
cosmologies involving a PBH-induced stasis epoch characterized by different
combinations of the parameters $\alpha$, $M_{\rm min}$, and $M_{\rm max}$.
In the left panel, we display the GW spectra obtained for several different choices 
of $\alpha$ within the physically allowed range in Eq.~(\ref{eq:AlphaRangeCombined})
for fixed $M_{\rm min} = 100$~g and $M_{\rm max} = 10^5$~g.  By contrast, in the right 
panel, we display the GW spectra obtained for several different combinations of
$M_{\rm min}$ and $M_{\rm max}$ for fixed $\alpha = -2$ (a value of $\alpha$ 
which corresponds to $\overline{w}=1/5$ and $w_c=1$).
For each of the curves shown, we have taken $H_{\rm end} = 2.5\times 10^{-5}M_P$, which 
saturates the upper bound on $H_{\rm end}$ from CMB data~\cite{Akrami:2018odb}.
For concreteness, we also take the number of $e$-folds during PBH domination to
be $\mathcal{N}_{\rm PBH}=2$.  The colored curves appearing in each panel 
indicate the GW spectra obtained for different choices of the parameters 
$\alpha$, $M_{\rm max}$, and $M_{\rm min}$ which characterize the mass spectrum 
of the PBHs, while the dashed black horizontal line indicates the $k$-independent 
GW spectrum obtained within the context of the standard cosmology.  The gray 
regions indicated in the upper left portion of each panel indicate the projected 
discovery reaches of several operating, planned, or proposed experiments capable of 
probing the GW spectrum within the range of frequencies shown, including 
LISA~\cite{LISA:2017pwj,Baker:2019nia}, 
the Big Bang Observer (BBO)~\cite{Crowder:2005nr,Corbin:2005ny,Harry:2006fi}, 
DECIGO~\cite{Seto:2001qf,Kawamura:2011zz,Yagi:2011wg}, ultimate DECIGO 
(u-DECIGO)~\cite{Seto:2001qf,Kudoh:2005as,Saikawa:2018rcs,Ringwald:2020vei}, 
THEIA~\cite{Garcia-Bellido:2021zgu}, 
$\mu$-ARES~\cite{Sesana:2019vho}, the Advanced LIGO and Advanced VIRGO detectors 
operating in concert
(aLIGO)~\cite{Harry:2010zz,LIGOScientific:2014pky,VIRGO:2014yos,LIGOScientific:2019lzm}, 
Cosmic Explorer (CE)~\cite{LIGOScientific:2016wof,Reitze:2019iox}, 
and the Einstein Telescope (ET)~\cite{Hild:2010id,Maggiore:2019uih}. 

In interpreting the results shown in Fig.~\ref{fig:GW}, we begin by noting that
each of the GW spectra appearing therein consists of four distinct segments with
different slopes --- segments which correspond to the four different frequency 
intervals in Eq.~(\ref{eq:dOmegaGWdlnf}) for which $f < f_{\rm end}$.  Each of these 
segments represents the contribution to $d\Omega_{\rm GW}/d\log f$ associated 
with GW modes which reenter the horizon during a particular cosmological epoch, 
and the slope of each segment is simply the value of the function $\xi(w)$ in 
Eq.~(\ref{eq:xi}) obtained from the equation-of-state parameter during that epoch.
By contrast, the contribution to $d\Omega_{\rm GW}/d\log f$ from modes with 
$f > f_{\rm end}$ --- modes which never exit the horizon and whose energy density is 
therefore effectively inflated away --- is negligible.  Thus, each of the 
GW spectra shown in Fig.~\ref{fig:GW} may be considered to have an ultraviolet
cutoff at $f_{\rm end}$. 

The leftmost segment of each GW spectrum, for which the slope of 
$d\Omega_{\rm GW}/d\log f$ vanishes, is 
associated with GW modes with frequencies $f < f_s$ sufficiently low that they 
reenter the horizon only after the stasis epoch has concluded.  Thus, 
for $f < f_s$, the GW spectra obtained for all values of $\alpha$, $M_{\rm min}$,
and $M_{\rm max}$ coincide with the flat spectrum obtained within the context of
the standard cosmology.

The segment immediately to the right of this first segment corresponds to 
the frequency interval $f_s < f < f_{\rm PBH}$ and is associated with GW modes 
which reenter the horizon during the stasis epoch itself.  The slope 
$\xi(\overline{w})$ of this segment depends non-trivially on $\alpha$ through
Eqs.~(\ref{eq:wBar}) and~(\ref{eq:xi}), as shown in the left panel of the figure. 
However, for any value of  $\alpha$ within the physically sensible range in Eq.~(\ref{eq:AlphaRangeCombined}), it is always the case that $\xi(w) < 0$. 
Since the slope of $d\Omega_{\rm GW}/d\log f$ is always negative for 
$f_s < f < f_{\rm PBH}$, we find that $d\Omega_{\rm GW}/d\log f$ is always 
suppressed within this frequency interval relative to 
$d\Omega_{\rm GW}^{\rm sc}/d\log f$.  
Moreover, while this slope is entirely determined by the value of $\alpha$,  
the frequency $f_s$ above which this suppression begins depends on the duration 
of the stasis epoch, and thus on the value of $M_{\rm max}$.  Taken together,
the solid blue, red, and yellow curves in the right panel of the figure 
illustrate the effect of varying $M_{\rm max}$ with $\alpha$ and $M_{\rm min}$
held fixed.  As $M_{\rm min}$ increases, $f_s$ shifts to a lower frequency.

The segment immediately to the right of this second segment corresponds 
to the frequency interval $f_{\rm PBH} < f < f_f$ and is associated with GW modes 
which reenter the horizon during the epoch of PBH-domination that immediately 
precedes stasis.  Since $\xi(0) = - 2$ and thus $\xi(0) \leq \xi(\overline{w})$ during 
this matter-dominated epoch, the downward slope of the GW spectrum steepens --- or, 
in the special case in which $\overline{w} = 0$, remains at its stasis 
value --- when $f$ increases above $f_{\rm PBH}$.  The value of this transition 
frequency depends on the point in the cosmological timeline at which 
the PBH-domination epoch begins, and thus on the value of $M_{\rm min}$.
The solid, dashed, and dotted red curves in the right panel of 
Fig.~\ref{fig:GW}, taken together, illustrate the effect on the GW spectrum of 
varying $M_{\rm min}$ with $\alpha$ and $M_{\rm max}$ held fixed.  As 
$M_{\rm min}$ decreases, $f_{\rm PBH}$ shifts to a higher frequency.
We also note, however, that the value of $f_{\rm PBH}$ depends on $\alpha$ as 
well.  This is because the rate at which perturbation modes cross the 
horizon during any particular cosmological epoch depends on $w$.  
Indeed, during any epoch wherein $w$ is effectively constant,
the scale factor scales with time according to the relation
$a\propto t^{2/(3+3w)}$.  It therefore follows from Eq.~(\ref{eq:HtoaScaling}) 
that the relationship between the wavenumber $k$ of a GW mode which enters 
the horizon during such an epoch and the time $t_k$ at which that mode enters 
the horizon is given by
\begin{equation}
  k ~\sim~ t_k^{-\frac{(1+3w)}{3(1+w)}}~.
  \label{eq:kvst}
\end{equation}
This implies that for a fixed $\log t_k$ interval, the 
corresponding $\log k$ interval --- and thus also the corresponding 
interval of $\log f$ --- increases with increasing $w$.  
As a result, the segment of the GW spectrum associated with modes which 
reenter the horizon during the stasis epoch spans a larger range of $\log f$ 
for smaller $\alpha$.  This effect, while not particularly dramatic, is evident 
in the GW spectra shown in the left panel of the Fig.~\ref{fig:GW}.

Finally, the rightmost segment of each spectrum, which  corresponds to the 
frequency interval $f_f <f < f_{\rm end}$, is associated with GW modes which reenter 
the horizon during the epoch following inflation wherein the population of 
PBHs is initially produced.  Within this frequency interval, the slope $\xi(w_c)$ 
of $d\Omega_{\rm GW}/d\log f$ can take a broad range of values, depending 
on the value of $\alpha$, as shown in the left panel of Fig.~\ref{fig:GW}.
For $\alpha > -3/2$, we find that $\xi(w_c) < 0$.  As a result, 
$d\Omega_{\rm GW}/d\log f$ remains suppressed relative to the result obtained 
within the context of the standard cosmology, even at high frequencies.
The GW spectrum also remains suppressed at high frequencies for $\alpha =-3/2$,
which corresponds to the special case in which the population of PBHs was produced 
during a radiation-dominated epoch.  In this case, $\xi(w_c) = 0$ and 
$d\Omega_{\rm GW}/d\log f$ remains flat during this epoch.
By contrast, for $\alpha < -3/2$, we find that $\xi(w_c) > 0$.  As a result
we find that $d\Omega_{\rm GW}/d\log f$ can actually become quite large as $f$ 
approaches $f_{\rm end}$.  Indeed, for many of the GW spectra shown in 
Fig.~\ref{fig:GW}, $d\Omega_{\rm GW}/d\log f$ is enhanced relative to 
$d\Omega_{\rm GW}^{\rm sc}/d\log f$ at high frequencies by several orders of 
magnitude.

Taken together, the results shown in Fig.~\ref{fig:GW} indicate that
in cosmologies involving a period of PBH-induced stasis, the prospects
for observing a stochastic background of gravitational waves can be
affected in a variety of ways, depending on the values of $\alpha$, 
$M_{\rm min}$, and $M_{\rm max}$ which characterize the population of 
PBHs.  On the one hand, a suppression of $d\Omega_{\rm GW}/d\log f$
at {\it low}\/ frequencies --- and especially within the frequency range 
$ 10^{-5}\,{\rm Hz} \lesssim f \lesssim 10^{2}\,{\rm Hz}$ for 
which discovery prospects at future GW detectors are the brightest --- 
could serve to ``hide'' the SGWB from these detectors.  Indeed, this
suppression arises due to the stasis epoch itself and to the preceding epoch 
of PBH domination.  On the other hand, the 
enhancement of $d\Omega_{\rm GW}/d\log f$ at {\it high}\/ frequencies could 
potentially enhance the prospects for SGWB discovery.  Moreover, since the  
the slope of $d\Omega_{\rm GW}/d\log f$ within the 
interval $f_s < f < f_{\rm PBH}$ and the slope of 
$d\Omega_{\rm GW}/d\log f$ within the interval $f_f < f < f_{\rm end}$ are 
both determined solely by the value of $\alpha$, the correlation
between these two slopes can be exploited as a means of testing 
whether a particular GW spectrum is potentially the result of  
an expansion history which includes an epoch of PBH-induced stasis, 
and even as a means of indirectly measuring $\alpha$.

While modifications to the expansion history of the universe can have a 
significant impact on $d\Omega_{\rm GW}/d\log f$ in cosmologies involving an epoch of 
PBH-induced stasis, other modifications to the GW spectrum are generically 
expected in such cosmologies as well.  One such 
modification is linked to the manner in which the 
population of PBHs is assumed to be produced.  The fact that a non-negligible 
abundance of PBHs arises from the collapse of primordial curvature perturbations
in such cosmologies suggests that the amplitude of such perturbations is 
enhanced on small distance scales.  However, small-scale curvature 
perturbations of this sort generically induce a contribution to the SGWB 
through their coupling to the tensor perturbations --- a coupling which 
arises at second-order in the perturbed Einstein 
equations~\cite{Matarrese:1992rp, Matarrese:1993zf, Matarrese:1997ay, 
Noh:2004bc, Carbone:2004iv, Nakamura:2004rm, Baumann:2007zm}.  This contribution 
is therefore an expected feature of the GW spectrum in cosmologies involving
an epoch of PBH-induced stasis.

As discussed 
in Sect.~\ref{sec:PBH}, the initial mass $M_i$ of a PBH produced by a perturbation 
mode with comoving wavenumber $k$ is correlated with the mass contained within the 
horizon volume at the time $t_k$ at which the mode reenters the horizon.  Within the 
context of any particular cosmological model, it then follows that each 
value of $M_i$ within the PBH mass spectrum can be associated with a particular 
value of $k$, and hence a particular frequency $f$.  
In the standard cosmology, the relationship between $M_i$ and $f$
takes the form (for details, see, \eg, Ref.~\cite{Nakama:2016gzw})
\beq
  f ~\approx~ (2.4\times 10^8\,{\rm Hz}) \times \gamma^{1/2}
    \left(\frac{g_\star}{100}\right)^{-1/12}
    \left(\frac{M_i}{1\,{\rm g}}\right)^{-1/2}\,,
    \label{eq:f_M_st}
\eeq
where $\gamma$ is the $\mathcal{O}(1)$ proportionality factor introduced
in Eq.~(\ref{eq:InitialMass}). 
Thus, for a population of PBHs with masses $0.1\,{\rm g} \lesssim M_i \lesssim 10^9\,{\rm g}$ 
within our regime of interest, the corresponding range of GW frequencies is 
$\mathcal{O}(10^3)~{\rm Hz} \lesssim f \lesssim \mathcal{O}(10^9)~{\rm Hz}$.
This frequency range lies well above the range of $f$ to which current or future 
terrestrial or space-based GW interferometers are typically sensitive.

By contrast, in scenarios in which the cosmological history departs from that of the 
standard cosmology at early times, $M_i$ and $f$ are in general not related by
Eq.~\eqref{eq:f_M_st}~\cite{Domenech:2019quo, Inomata:2020lmk, Inomata:2019ivs,
Inomata:2019zqy}.  In order to determine how these quantities are related in 
cosmologies involving a PBH-induced epoch of cosmic stasis, we begin by noting that 
$H_k$ and $M_i$ are related through Eq.~\eqref{eq:InitialMass} and that
\begin{equation}
  a_k ~=~ a_s e^{-\mathcal{N}(t_k, t_s)}~,
\end{equation}
where the subscript ``$s$'' once again indicates the value of a quantity 
at the end of the stasis epoch and where $\mathcal{N}(t_k,t_s)$
is the number of $e$-folds of expansion that occur between the time at which the 
mode with wavenumber $k$ enters the horizon and the 
end of the stasis epoch.  This latter quantity may be expressed as  
\begin{equation}
  \mathcal{N}(t_k,t_s) ~=~ 
    \mathcal{N}_c(t_k) + \mathcal{N}_{\rm PBH} +\mathcal{N}_s~,
\end{equation}
where $\mathcal{N}_c(t_k)$ denotes the number of $e$-folds of expansion that 
occur between the time at which the mode reenters the horizon and 
the end of the PBH-formation epoch.  
Furthermore, the relationship between $a_s$ and the scale 
factor $a_{\rm eq}$ at matter-radiation equality in the standard cosmology 
can be derived from the entropy-conservation relation   
\begin{equation}
  g_{\star S}(T_s)a_s^3 T_s^3 ~=~ 
    g_{\star S}(T_{\rm eq})a_{\rm eq}^3 T_{\rm eq}^3~,
\end{equation}
where $g_{\star S}(T)$ denotes the effective number of entropic degrees of 
freedom at temperature $T$.  Thus, approximating the transition from stasis to 
the radiation-domination epoch as instantaneous, we find that
\begin{equation}
  \rho_s ~=~ \frac{\pi^2}{30}g_\star(T_s)T_s^4 ~\approx~ \frac{3M_P^2H^2}{8\pi} 
    ~\approx~ \frac{M_P^2 \overline{\kappa}^2}{24\pi\tau^2(M_{\rm max})}\,,
\end{equation}
where $g_{\star}(T)$ denotes the effective number of relativistic degrees 
of freedom at temperature $T$ and where we have defined 
$\overline{\kappa}\equiv 2/(1+\overline{w})$.
Combining these individual relations, we find that the relationship 
between $f$ and $M_i$ in cosmologies involving an epoch of PBH-induced stasis 
is given by
\beqn
  f &\,=\,& \frac{\gamma}{\sqrt{\varepsilon\overline{\kappa}}} 
    \left( \frac{M_{\rm max}^3}{M_P}\right)^{1/2}
    \left[\frac{g_{\star S}(T_{\rm eq})}{g_{\star S}(T_s)}\right]^{1/3}\nn\\
  & & ~\times \left[\frac{\pi^3 g_{\star}(T_s)}{180}\right]^{1/4} 
    \frac{a_{\rm eq} T_{\rm eq}}{M_i}
    e^{-\mathcal{N}(t_k,t_s)}\nn\\
  &\,\approx\,& (1.48\times 10^{10}\, {\rm Hz}) \times \gamma 
    \left(\frac{2}{\overline{\kappa}}\right)^{1/2}\!
    \left(\frac{M_{\rm max}}{10^9\,{\rm g}}\right)^{3/2}\!
    \left(\frac{1\,{\rm g}}{M_i} \right) \nn\\
  & &~\times \left[\frac{g_\star(T_s)}{10}\right]^{-1/12} 
     e^{40-\mathcal{N}_c(t_k) -\mathcal{N}_{\rm PBH} -\mathcal{N}_s}
  \label{eq:f_M_PBHstasis}
\eeqn

The relationship between $f$ and $M_i$ given by Eq.~(\ref{eq:f_M_PBHstasis}) 
can differ significantly from the corresponding result in 
Eq.~(\ref{eq:f_M_st}) for the standard cosmology.  Indeed, the value of $f$ 
which corresponds to a given value of $M_i$ depends sensitively on the total 
number of $e$-folds that the corresponding perturbation mode experiences during 
the non-standard part of cosmological history after its horizon reentry.

The range of $f$ which corresponds to the range $M_{\rm min} < M_i < M_{\rm max}$ 
of initial PBH masses for each $d\Omega_{\rm GW}^{\rm sc}/d\log f$ curve shown 
in the figure is indicated by the shaded bar which lies along that curve.
The corresponding range of $f$ obtained for a given range of $M_i$ within the 
context of the standard cosmology is demarcated by the tick-marks
corresponding to the values of $M_{\rm max}$ (on the left) and $M_{\rm min}$ 
(on the right) which appear along the dashed black line in each panel of the 
figure.  Since a significant abundance of PBHs is necessary for achieving stasis, 
$d\Omega_{\rm GW}^{\rm sc}/d\log f$ is expected to receive additional 
enhancements within this frequency range.

The results shown in the left panel of Fig.~\ref{fig:GW} indicate not only 
that the frequency range corresponding to the range
$M_{\rm min} < M_i < M_{\rm max}$ of PBH masses broadens as 
$\alpha$ decreases for fixed $M_{\rm min}$ and $M_{\rm max}$, but also that
this frequency range shifts to larger values of $f$.  We may account for 
the latter phenomenon by noting that $f$ depends on $\alpha$ primarily 
through the quantities $\mathcal{N}_s$ and $\mathcal{N}_c(t_k)$ which appear
in the exponential factor in Eq.~(\ref{eq:f_M_PBHstasis}).  The first of
these quantities decreases with decreasing $\alpha$ according to  
Eq.~(\ref{eq:Nefold}).  The second depends on $\alpha$ through $w_c$ 
according to the relation
\begin{equation}
  \mathcal{N}_c(t_k) ~\approx~ \frac{2}{3(1+w_c)}\log\left(\frac{t_f}{t_k}\right)~,
  \label{eq:NctkScaling}
\end{equation}
where $t_f$ denotes the time at the end of the PBH-formation epoch.  Equivalently,
through use of Eq.~(\ref{eq:kvst}), we may express $\mathcal{N}_c(t_k)$ 
in terms of $k$ rather than $t_k$:
\begin{equation}
  \mathcal{N}_c(t_k) ~\approx~ \frac{2}{1+3w_c}\log\left(\frac{k_f}{k}\right)~.
  \label{eq:NctkScalingofk}
\end{equation}
Since the ratio $k_f/k$ is independent of $w_c$ for GWs which enter the horizon
during the PBH-formation epoch, Eq.~(\ref{eq:NctkScalingofk}) indicates that 
$\mathcal{N}_c(t_k)$ decreases as $w_c$ increases.  Thus, both $\mathcal{N}_s$ 
and $\mathcal{N}_c(t_k)$ decrease as $\alpha$ decreases, which accounts for 
the shift in the frequency range corresponding to $M_{\rm min} < M_i < M_{\rm max}$ 
to larger values of $f$.

In order to account for the broadening of the frequency range which corresponds to 
the range of PBH masses, we begin by noting that for any two GW modes with frequencies $f_1$ 
and $f_2$ which respectively correspond to a pair of initial PBH masses $M_1$ 
and $M_2$ such that $M_{\rm min}\leq M_1 < M_2 \leq M_{\rm max}$,
Eq.~(\ref{eq:f_M_PBHstasis}) implies that 
\begin{equation}
  \log f_2 - \log f_1 ~=~ \log\left(\frac{M_2}{M_1}\right) 
    - \Delta\mathcal{N}_c(t_1,t_2)~,
  \label{eq:Logfint}
\end{equation}
where
\begin{equation}
    \Delta \mathcal{N}_c(t_1,t_2) \equiv \mathcal{N}_c(t_2) - \mathcal{N}_c(t_1)
    ~\approx~ \frac{2}{3(1+w_c)}\log\left(\frac{t_2}{t_1}\right)
\end{equation}
represents the number of $e$-folds of cosmic expansion which occur between
the times at which these modes reenter the horizon.  Equivalently, through use of 
Eq.~(\ref{eq:InitialMass}), which implies that $M_i ~=~ 3\gamma M_P^2 (1 + w_c) t_i /4$, 
we may  express this number of $e$-folds in terms of $M_1$ and $M_2$ directly: 
\begin{equation}
  \Delta\mathcal{N}_c(t_1,t_2) ~\approx~ \frac{2}{3(1+w_c)}
    \log\left(\frac{M_2}{M_1}\right)~.
\end{equation}
We see from this relation that as $w_c$ increases, $\Delta\mathcal{N}_c(t_1,t_2)$ decreases 
and the corresponding logarithmic frequency interval in Eq.~(\ref{eq:Logfint}) increases.
Identifying $M_1$ and $M_2$ with $M_{\rm min}$ and $M_{\rm max}$, we then observe
that the range of frequencies within which we expect $d\Omega_{\rm GW}/d\log f$ to
be enhanced due to the physics underlying the formation of PBHs in cosmologies involving
an epoch of PBH-induced stasis indeed broadens as $\alpha$ decreases.

Overall, for typical parameter values, the corresponding GW frequency ranges from 
MHz to GHz, still much higher than what the space-based or terrestrial 
interferometer can reach.  However, the SGWB could potentially be probed within
this frequency range by experiments designed primarily to detect axion dark
matter~\cite{Ito:2019wcb,Ejlli:2019bqj,Berlin:2021txa,Domcke:2022rgu,Ito:2022rxn}.
We leave investigations along these lines for future work.

Finally, in addition to the contributions to the GW spectrum discussed above, 
there is an additional contribution which arises in cosmologies involving an 
epoch of PBH-induced stasis due to the presence of the PBHs themselves.
Since PBHs are discrete objects, their spatial distribution is 
granular on small scales and is therefore not strictly adiabatic.
As a result, density perturbations associated with a population of 
PBHs necessarily include an isocurvature component~\cite{Inman:2019wvr}.
These isocurvature perturbations can be converted into adiabatic 
curvature perturbations during a PBH-dominated 
epoch~\cite{Kodama:1986fg,Kodama:1986ud}, and therefore act as an additional 
source for GW production~\cite{Papanikolaou:2020qtd,Domenech:2020ssp,
Inomata:2020lmk, Kozaczuk:2021wcl,Papanikolaou:2021uhe,Bhaumik:2022pil,
Bhaumik:2022zdd,Papanikolaou:2022hkg,Papanikolaou:2022chm}.  These
processes are generically active  during any cosmological epoch wherein 
PBHs constitute a non-negligible fraction of the total energy density of the 
universe.  In cosmologies involving an epoch of PBH-induced stasis, one would 
therefore expect these processes to be active both during the stasis epoch itself 
and during the PBH-dominated epoch which immediately precedes it.  Thus, one would 
generically expect an additional contribution to the GW spectrum to arise as a result 
of this isocurvature contribution in such cosmologies.  This contribution can be
sizable even in scenarios in which the PBH mass spectrum is quite 
broad~\cite{Papanikolaou:2022chm}.  We leave the analysis 
of this contribution for future work.

\subsection{Dark Radiation} \label{sec:DarkRadiation}

In extensions of the SM involving one or more additional light 
particle species --- \eg, axions, axion-like particles, dark photons, 
sterile neutrinos, and gravitons --- particles of these species are generically 
produced as Hawking radiation.  The evaporation of a population of PBHs can 
therefore generate a potentially sizable contribution to the abundance of dark 
radiation.  This dark radiation could potentially manifest itself 
at future CMB observatories as a shift $\Delta N_{\rm eff}$ in the effective number 
of light neutrino species $N_{\rm eff}$ present in the early universe 
(for recent reviews, see, \eg, Ref.~\cite{Auffinger:2022khh} and references therein).
Particular attention has been focused on the contribution to $\Delta N_{\rm eff}$
which arises from the evaporation of a population of Kerr PBHs, since rapidly rotating 
BHs are expected to produce higher-spin particles more efficiently than non-rotating 
ones~\cite{Page:1976df,Page:1976ki,Page:1977um,Hooper:2019gtx,
Hooper:2020evu,Masina:2020xhk,Masina:2021zpu,Arbey:2021ysg, Cheek:2022dbx}. 
The contribution to $\Delta N_{\rm eff}$ which arises from a population of PBHs 
with an extended distribution of both masses and spins was recently investigated 
in Ref.~\cite{Cheek:2022mmy}.  Here, we investigate the effect that an epoch of cosmic 
stasis has on $\Delta N_{\rm eff}$.

\begin{figure}
  \includegraphics[width=\linewidth]{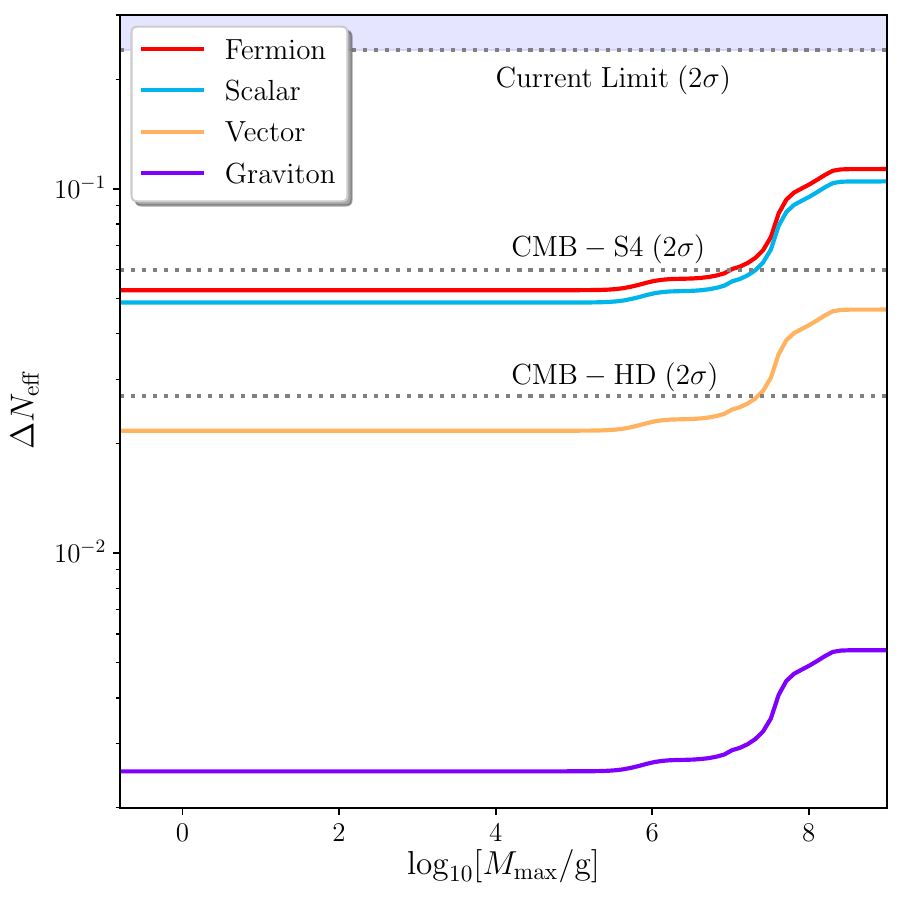}
  \caption{\label{fig:Neff} Contributions to $\Delta N_{\rm eff}$ from the 
  evaporation of a population of PBHs with a power-law distribution of masses
  into dark radiation in SM extensions involving a single additional 
  light particle species, plotted as functions of $\log_{10}(M_{\rm max}/{\rm g})$.
  The solid curves correspond to different representations of the Lorentz group
  under which particles of that species transform.
  All such curves correspond to the parameter choices 
  $M_{\rm min}=0.1\,{\rm g}$ and $\alpha =-1$; however, the 
  value of $\Delta N_{\rm eff}$ is insensitive to both $M_{\rm min}$ and $\alpha$.
  The shaded region above the uppermost dashed line is excluded by current bounds
  at the $2\sigma$ significance level.  The additional dashed lines indicate the 
  reach of future CMB experiments, as represented by the corresponding projected 
  $2\sigma$ limits.}
\end{figure}

In Fig.~\ref{fig:Neff}, we plot the contribution to $\Delta N_{\rm eff}$ that arises 
from the production of additional, light degrees of freedom via PBH evaporation 
as a function of $M_{\rm max}$ in a modified cosmology involving a stasis epoch.
The solid curves indicate the $\Delta N_{\rm eff}$ contribution associated 
with various particle species that arise in different extensions of the SM.~
The red curve represents the contribution from a gauge-singlet Majorana fermion.
The cyan curve represents the contribution from a real Lorentz scalar field.  The 
orange curve represents the contribution from a massless Lorentz vector, and
the purple curve represents the contribution from a massless graviton. 
These curves correspond to the parameter choices $M_{\rm min}=0.1\,{\rm g}$ and 
$\alpha = -1$; however, we find that the value of $\Delta N_{\rm eff}$ 
is essentially insensitive to both $M_{\rm min}$ and $\alpha$.
The shaded region above the uppermost dashed line is excluded by current bounds on
$\Delta N_{\rm eff}$ at the $2\sigma$ significance level.  The additional dashed lines 
represent the projected $2\sigma$ limits on $\Delta N_{\rm eff}$ which would be 
provided by the CMB-S4~\cite{CMB-S4:2016ple,Abazajian:2019eic} and 
CMB-HD~\cite{CMB-HD:2022bsz} experiments in the event of a non-detection --- limits 
which provide an indication of the discovery reach of these future experiments.

The results shown in Fig.~\ref{fig:Neff} indicate that current bounds on 
$\Delta N_{\rm eff}$ do not exclude the presence of a single additional
species of light, SM-gauge-singlet particle beyond the particle content 
of the SM in cosmologies involving an epoch of PBH-induced stasis.
However, depending on the properties of this additional particle species, 
future CMB experiments could be potentially be sensitive to the shift in
$N_{\rm eff}$ which results from the production of these particles via
Hawking radiation.  We also observe that $\Delta N_{\rm eff}$ is 
effectively independent of $M_{\rm max}$ when $M_{\rm max}$ is small,
but depends non-trivially on $M_{\rm max}$ when 
$M_{\rm max} \gtrsim 10^5\,{\rm g}$.  This dependence of $\Delta N_{\rm eff}$ 
on $M_{\rm max}$ is due to the lifetime $\tau(M_{\rm max})$ of the heaviest PBHs 
being sufficiently long in this regime that the electroweak phase-transition 
occurs during the stasis epoch.  Once this phase transition occurs, the effective 
number of relativistic degrees of freedom $g_\star(T)$ in the radiation bath 
begins to depends non-trivially on $T$ --- and hence on time.  Thus, for 
$M_{\rm max} \gtrsim 10^5\,{\rm g}$, the production of light particles via Hawking 
radiation remains significant after the electroweak phase transition has 
occurred --- and for $M_{\rm max} \gtrsim 10^7\,{\rm g}$, even after the 
quark-hadron phase transition has occurred --- and $g_\star(T)$ has fallen below 
its initial high-temperature value.  The contribution to 
$\Delta N_{\rm eff}$ from such a long-lived population of PBHs is therefore 
greater than the contribution from a population of short-lived PBHs which all decay 
before the electroweak phase transition occurs. 

 It is also noteworthy that these results depend neither on the value of the parameter 
 $\alpha$ nor on the value of the initial mass $M_{\rm min}$.  Indeed,  across the 
 full range of $M_i$ relevant for stasis, the corresponding initial Hawking temperature 
 $T_{\rm BH} = M_P^2/(8\pi M_i)$ is sufficiently large that all SM degrees of freedom 
 may be considered relativistic when they are produced via the evaporation of even the 
 lightest PBHs.   As a result, the ratio of the energy density of dark 
 radiation generated by PBH evaporation to the energy density of SM particles 
 generated in this way remains effectively constant across the time period during 
 which the PBHs are evaporating.  The total contribution to the energy 
 density of dark radiation at times after the end of the stasis epoch therefore 
 depends effectively only on $\tau(M_{\rm max})$.

\subsection{Dark Matter \label{sec:DarkMatter}}

In a similar vein,
it is also interesting to consider possible connections between the PBHs
which give rise to cosmic stasis in cosmologies of this sort and the 
dark matter (DM) which overwhelming evidence suggests constitutes a 
significant fraction of the present-day energy density of the universe.
The PBHs which give rise to the epoch of PBH-induced cosmic stasis
evaporate far too rapidly to contribute to the present-day abundance 
of DM themselves.  By contrast, longer-lived PBHs produced by the collapse of 
lower-frequency perturbations could of course in principle contribute to this 
abundance (see, \eg, Ref.~\cite{Escriva:2022duf} for a recent review), but such 
PBHs are not a necessary ingredient in cosmologies involving PBH-induced stasis.
Nevertheless, even though the PBHs which give rise to the stasis epoch cannot
themselves contribute to the present-day DM abundance, their evaporation can play an 
important role in producing particles that {\it do}\/ constitute the DM.~ 
Indeed, DM particles are generically expected to be produced as Hawking radiation
as PBHs evaporate, just as SM particles are~\cite{Matsas:1998zm}.  

The effect of this production mechanism on the present-day DM abundance has been 
investigated both for the case of a monochromatic PBH mass
spectrum~\cite{Bell:1998jk, Khlopov:2004tn, Fujita:2014hha, Lennon:2017tqq, 
Morrison:2018xla, Hooper:2019gtx, Baldes:2020nuv, Gondolo:2020uqv, Masina:2020xhk,  
Cheek:2021cfe, Cheek:2021odj, Masina:2021zpu} and for the case 
of an extended PBH mass spectrum, including one with the power-law profile we consider
here~\cite{Cheek:2022mmy}.  In the latter case, it was shown that for 
$\mathcal{O}(\mathrm{GeV})$ DM candidates, the broader the PBH distribution is, 
the smaller the initial abundance of PBHs must be and the shorter their lifetimes 
must be in order to guarantee that DM is not overproduced.

Considerations related to small-scale structure place additional 
constraints on the population of DM particles generated by PBH evaporation.  
If a non-negligible fraction of these particles are produced by PBHs 
with Hawking temperatures $T_{\rm PBH} \gg m_{\rm DM}$, the resulting 
contribution to the DM phase-space distribution will be highly relativistic.
If the DM is effectively decoupled from the thermal bath by the time this
contribution arises, the free-streaming effects associated with such a
population of ``hot'' DM particles can suppress the formation of structure on 
small scales.  Lyman-$\alpha$-forest 
data~\cite{SDSS:2004kjl,Becker:2006qj,Becker:2010cu,Calverley:2010tz,
Becker:2011ee,SDSS:2013qvl,Lopez:2016} imply stringent 
constraints on such a ``hot'' population of DM particles.
Indeed, scenarios in which a population of PBHs with an extended mass
spectrum comes to represent a significant fraction of the energy density 
of the universe by the time those PBHs evaporate are typically
excluded by these constraints if the DM particles are light --- in the
sense that $m_{\rm DM} \ll M_P^2/(8\pi M_{\rm max})$ --- and 
effectively decoupled~\cite{Cheek:2022mmy}.  Since PBHs collectively 
contribute an $\mathcal{O}(1)$ fraction of the total energy density
in cosmologies involving a PBH-induced stasis epoch, small-scale-structure
constraints are a serious concern.

One way of circumventing these overclosure and structure-formation constraints is
simply to posit that the DM particles interact sufficiently strongly with the
particles in the radiation bath that they remain in thermal equilibrium with
those particles until well after the last PBHs evaporate and the stasis epoch 
ends.  In this case, the DM particles produced via Hawking radiation rapidly
thermalize and an acceptable present-day DM abundance can be generated, \eg, 
through thermal freeze-out.  

However, another possible way of circumventing 
these constraints --- one in which the evaporation of the PBHs which give rise 
to stasis indeed {\it does}\/ serve as an important mechanism for DM production ---  
is to posit that the masses of the particles which constitute the DM today
significantly exceed the initial Hawking temperature of even the lightest 
PBHs --- \ie, that $m_{\rm DM} \gg M_P^2/(8\pi M_{\rm min})$.  Indeed, the production 
of such DM particles with such masses due to the evaporation of PBHs with any 
initial mass $M_{\rm min} < M_i < M_{\rm max}$ remains highly Boltzmann-suppressed 
until those PBHs lose the vast majority of their initial mass.  
Provided that $m_{\rm DM}$ is sufficiently large, the resulting population of
DM particles will not dominate the energy of the universe prematurely.  

Moreover, structure-formation constraints on the DM phase-space distribution 
are less of a concern when $m_{\rm DM} \gg M_P^2/(8\pi M_{\rm min})$.  There
are two principal reasons for this.
The first is that the majority of the DM particles in this population 
are produced by PBHs whose Hawking temperatures have only 
just crossed the threshold $T_{\rm BH} \sim m_{\rm DM}$ above which DM 
production is no longer Boltzmann-suppressed.  As a result, most of the 
DM particles produced via PBH evaporation are not highly relativistic even
at the time at which they are initially produced.
The second is that for values of $\alpha$ within the range specified in
Eq.~(\ref{eq:AlphaRangeCombined}), PBHs with smaller $M_i$ contribute a
greater portion of the total initial PBH energy density.  These lighter PBHs 
evaporate comparatively early during the stasis epoch, and the momenta of the 
DM particles that they produce are therefore redshifted to a greater extent 
by the end of the stasis epoch than are the DM particles produced at the 
end of that epoch via the evaporation of heavier PBHs.  Thus, while lighter PBHs 
lose a smaller fraction of their initial mass energy before they are capable 
of producing DM particles, this is compensated for by this redshifting effect. 
Indeed, it is known that for the case of a monochromatic initial PBH spectrum in which 
all of the PBHs have $M_i = 10^{6}\,{\rm g}$, both the overclosure bound and
the Lyman-$\alpha$ constraints on the DM phase-space distribution can be 
satisfied for $m_{\rm DM} \geq 10^{16}$~GeV~\cite{Cheek:2021odj}.

In light of these considerations,
it is therefore conceivable that Hawking radiation could serve as the
dominant mechanism for DM production in cosmologies involving an epoch
of PBH-induced stasis, provided that $m_{\rm DM}$ lies only a couple 
of orders of magnitude below the Planck scale.
That said, obtaining a reliable lower bound on $m_{\rm DM}$ for a given 
combination of $\alpha$, $M_{\rm min}$ and $M_{\rm max}$ in cosmologies 
of this sort is challenging for two reasons. 
The first of these is that the quantity $\epsilon(M)$ in Eq.~(\ref{eq:dMBHdt})  
depends non-trivially on $M$ in this case through the graybody factor.
The second is that DM particles with such large values of $m_{\rm DM}$ 
are only produced via the evaporation of a PBH with initial mass $M_i$ during 
an extremely brief time interval around $t\sim\tau(M_i)$.  An accurate assessment
of the contributions to the DM abundance and phase-space distribution from
PBH evaporation therefore requires a high degree of precision --- a degree of
precision beyond what numerical tools such as {\tt FRISBHEE} provide.  We 
leave the dedicated analysis that would be required in order to determine 
the lower bound on $m_{\rm DM}$ in cosmologies with PBH-induced stasis for
future work.

We also remark that the dark matter may not consist of particles of a single mass 
$m_{\rm DM}$.  Indeed, there exist frameworks for DM physics --- most 
notably the Dynamical Dark Matter (DDM) 
framework~\cite{Dienes:2011ja,Dienes:2011sa,Dienes:2012jb} --- wherein the 
particles which constitute the DM exhibit both an extended mass spectrum and a non-trivial 
spectrum of abundances, not unlike the PBHs we consider here.   The connection between 
PBHs and DM within these frameworks is likely to involve a rich and complicated dynamics 
which might, for example, enforce correlations between the PBH scaling exponent $\alpha$ 
and the scaling exponents and other parameters which govern the manner in which the 
abundances, lifetimes, \etc, of the DM states scale as a function of their mass.  This 
too represents an interesting direction for further study.

\subsection{Baryogenesis\label{sec:Baryogenesis}}

PBH-induced stasis can have potentially significant implications for baryogenesis as 
well.  Given the allowed range of PBH masses in Eq.~(\ref{eq:MminMmaxBounds}), it is
in principle possible for the temperature of the radiation bath during stasis 
to span a range of temperatures ranging from just above the BBN 
scale (\ie, a few MeV) up to around $10^{11}\,{\rm GeV}$. 
The modification of the expansion history of the universe across such a broad 
range of energy scales can impact baryogenesis in a number of ways.

For example, the possibility that the electroweak phase transition could take place 
during the stasis epoch has implications for electroweak baryogeneis.
Indeed, in scenarios in which the universe expands more rapidly during the 
electroweak phase transition than it does in the standard cosmology, the 
degree to which the baryon asymmetry is washed out by sphaleron 
processes is diminished~\cite{Joyce:1996cp,Joyce:1997fc,Servant:2001jh,Barenboim:2012nh}.
In baryogenesis scenarios in which the baryon asymmetry of the universe is instead 
ultimately generated via the out-of-equilibrium decays of heavy particles with 
$CP$-violating interactions --- including scenarios which achieve baryogenesis through 
leptogenesis --- it is possible that these decays could occur during the stasis 
epoch.  In leptogenesis scenarios involving right-handed neutrinos (RHN) with 
masses $M_{N} \sim \mathcal{O}(10^9 - 10^{10})\,{\rm GeV}$, for example, the thermal 
RHN population which is initially in equilibrium with the radiation bath can 
potentially decay during the stasis epoch, provided that $M_{\rm min}$ is sufficiently 
small.  The rates for inverse-decay processes and for other processes which serve 
to wash out the lepton asymmetry generated by RHN decays would therefore be modified in 
these scenarios relative to the rates obtained in the context of the standard 
cosmology.

Furthermore, in scenarios in which the baryon asymmetry is generated by the 
out-of-equilibrium decays of heavy particles, there is an additional effect 
which can significantly impact this asymmetry in cosmologies involving 
an epoch of PBH-induced stasis.  This is that these heavy particles can be 
produced via the evaporation of the PBHs themselves.  Indeed, even if 
these particles are sufficiently massive that they are not produced by the 
thermal bath during the stasis epoch, they could nevertheless be produced as
Hawking radiation. 

The contribution to the baryon asymmetry of the universe which results 
from the production of heavy, $CP$-violating particles as Hawking radiation 
was considered in Refs.~\cite{Toussaint:1978br,Hawking:1982ga,Turner:1979bt,
Grillo:1980rt,Fujita:2014hha,Hook:2014mla,Hamada:2016jnq,
Morrison:2018xla,Perez-Gonzalez:2020vnz,Datta:2020bht,Hooper:2020otu,
Granelli:2020pim,Chaudhuri:2020wjo,JyotiDas:2021shi,Bernal:2022pue} for the 
case of a monochromatic PBH mass spectrum.  On the one hand, the effect of 
entropy dilution from PBH decay renders it challenging to obtain a sufficient
baryon asymmetry, \eg, in Majorana leptogenesis scenarios in which 
$10^6\,{\rm GeV} \lesssim M_N \lesssim 10^8\,{\rm GeV}$~\cite{Perez-Gonzalez:2020vnz,
JyotiDas:2021shi}.  On the other hand, since the interaction rates associated with 
lepton-number-violating washout processes involving particles in the thermal bath are 
suppressed at low temperatures, the contribution to the baryon asymmetry from RHN 
produced at late 
times from PBH evaporation is impacted far less by such processes than the 
contribution from the decay of a thermal population of RHN~\cite{Bernal:2022pue}.  
As a result, leptogenesis scenarios in which $M_N \gtrsim 10^{15}\,{\rm GeV}$, 
which would not yield a sufficient baryon asymmetry within the context for the 
standard cosmology, can become viable if RHN production from PBH evaporation is 
significant.   

Both the dynamics of particle production from PBH evaporation and 
entropy generation in cosmologies involving an extended mass spectrum and a 
protracted period of stasis can differ drastically from the corresponding 
dynamics in the cosmologies considered in these studies.  Thus, assessing 
the effect of a PBH-induced stasis epoch on the baryon asymmetry in baryogenesis
scenarios of this sort would require a dedicated analysis, which we leave for 
future work.

\subsection{Effects of Accretion and Mergers}\label{sec:mergers}

In our analysis of PBH dynamics in the early universe, as discussed in 
Sect.~\ref{sec:PBH}, we focused on the regime in
which the only processes which have a significant impact on the evolution of 
$f(M,t)$ are cosmic expansion and Hawking evaporation. 
In general, however, $f(M,t)$ can be affected by other processes as well.  
In particular, black holes can acquire 
additional mass through the accretion of radiation or other forms of ambient 
energy density. Likewise, black holes can also merge with other black holes that 
they encounter, thereby forming more massive black holes.  

A detailed study of the constraints that these considerations impose on the 
parameter space of our PBH-induced stasis scenario can be found in 
Ref.~\cite{Dienes:2025qdw}.  In particular, it is shown in
Ref.~\cite{Dienes:2025qdw} that the 
impact of mergers on 
$f_{\rm BH}(M,t)$ is negligible across the entirety of the parameter space
of this scenario that is not already excluded by other constraints.  

By contrast, the requirement that accretion not appreciably distort the 
PBH mass spectrum prior to the end of the stasis epoch does place 
additional constraints on the parameter space of our scenario.
In particular, within the regime in which $ M_{\rm min} \ll M_{\rm max}$,
$\mathcal{N}_{\rm PBH} \gtrsim 1$, and in which $\alpha$ is not too close to  
the upper end of the range in Eq.~(\ref{eq:AlphaRangeCombined}), 
one finds~\cite{Dienes:2025qdw} that 
accretion has a negligible 
effect across the entire PBH mass spectrum so long as  
\begin{equation}
  \frac{M_{\rm min}^3}{M_{\rm max}} ~\gg~ 
  \left[\frac{48\sqrt{3}\overline{w}}{(1+\overline{w})^2} 
   + \frac{(1+w_c)A(w_c)}{1+2w_c}e^{3\mathcal{N}_{\rm PBH}/2}\right]
    \epsilon M_P^2~,
  \label{eq:AccretionCondit}
\end{equation}
where we have defined
\begin{equation}
  A(w_c) ~\equiv~ \frac{(1+3w_c)^{(1+3w_c)/(2w_c)}}{4w_c^{3/2}}~.
\end{equation}

The condition in Eq.~(\ref{eq:AccretionCondit}) turns out to be satisfied throughout 
the majority of our parameter-space region of interest for stasis.  Indeed, it is 
violated only in situations in which $\mathcal{N}_{\rm PBH}$ is quite large and in 
which the range of masses within the PBH mass spectrum is almost maximally 
broad~\cite{Dienes:2025qdw}.  Indeed, all of the results shown in 
Figs.~\ref{fig:observables}~and~\ref{fig:GW} correspond to choices of $\alpha$, 
$M_{\rm max}$, $M_{\rm min}$, and $\mathcal{N}_{\rm PBH}$ for which the effect of 
accretion on $f_{\rm BH}(M,t)$ is negligible.  The only exception is that 
in Fig.~\ref{fig:observables} we have included results for values of $\alpha$ very 
close to $-1$ --- values of $\alpha$ for which the effect of accretion is likely 
significant but difficult to estimate precisely due to theoretical uncertainties, 
in order to illustrate overall trends.


\section{Conclusions\label{sec:conclusions}}


A population of PBHs with a mass spectrum of the sort that arises naturally in a 
variety of cosmological scenarios can give rise to {\it cosmic stasis}\/ --- \ie, 
an extended period during which multiple cosmological components with different equations 
of state (in this case, black holes and radiation) have abundances that remain 
effectively constant even as the universe expands.
For example, the spectrum of PBHs generated by the collapse of density fluctuations 
with a scale-invariant power spectrum generically gives rise to stasis, 
provided that this collapse occurs during an epoch wherein the effective 
equation-of-state parameter for the universe lies within the range 
$0 \leq w_c \leq 1$.  This is in large part due to the fact that stasis is a 
global attractor for the dynamical system that governs the evolution of the 
cosmological abundances of matter --- \ie, the PBHs --- and radiation.
Moreover, even though stasis is a global attractor, it is not eternal --- indeed, 
it terminates when the heaviest PBHs evaporate.   Thus, since a period of PBH-induced stasis 
has not only a beginning but also an end, it constitutes a true cosmological epoch.
Such a stasis epoch is therefore expected to be a common feature in the 
cosmological timeline in scenarios involving PBHs with such a spectrum.
It is therefore crucial to understand the consequences of such an epoch.

In this paper, we have investigated the phenomenological implications of a 
PBH-induced stasis epoch.  We have shown that this epoch can represent a significant 
number of $e$-folds of cosmic expansion and span a range of energy scales 
ranging from $T \sim 10^{11}\,{\rm GeV}$ down to just above the BBN scale.  
As a result, the modification of the expansion history associated with cosmologies
of this sort can have a significant impact on a number of astrophysical observables.
We have found, for example, that these modifications can have a significant impact on 
the values of the spectral tilt $n_s$ and tensor-to-scalar ratio $r$ of the 
primordial perturbation spectrum obtained from CMB data.   
Moreover, we have also found that these modifications can have a highly non-trivial 
impact on the stochastic background of GWs which arises from primordial 
perturbations after inflation.  In particular, we have shown that the resulting GW spectrum
can be modified in a variety of ways within the frequency range most relevant for 
detection at future GW-interferometry experiments, depending on the values of the 
parameters $\alpha$, $M_{\rm min}$, and $M_{\rm max}$ which characterize the mass spectrum 
of the PBHs.  However, we have also shown that the scaling behaviors which the GW spectrum 
exhibits within different frequency intervals are correlated in cosmologies of this sort.
These correlations therefore not only provide an observational handle that can be used
to verify whether or not a particular SGWB spectrum is consistent with a cosmology involving 
an epoch of PBH stasis, but also provide information about the values of $\alpha$, 
$M_{\rm min}$, and $M_{\rm max}$ themselves. 

We have also considered several additional phenomenological consequences of a 
PBH-induced stasis epoch in the context of new-physics scenarios which involve 
additional particle species beyond the particle content of the SM.~  In particular, 
we have investigated the shift $\Delta N_{\rm eff}$ in the effective number 
of light neutrino species which arises due to the production of particles which 
behave as dark radiation via PBH evaporation in cosmologies of this sort.  We have 
shown that the contribution to $\Delta N_{\rm eff}$ from a single species of 
dark-radiation particle is typically consistent with current observational bounds, 
but also that future CMB experiments could potentially be sensitive to this contribution, 
especially if $M_{\rm max} \gtrsim 10^7$\,g.  Moreover, we have also discussed the 
potential consequences of particle-production via Hawking radiation during a 
PBH-induced stasis epoch for baryogenesis and for the production of DM.~  
In doing so, we have highlighted the particular ways in which the presence of the 
stasis epoch can play a potentially important role in determining the values 
of quantities such as the baryon asymmetry of the universe and the present-day DM
abundance obtained for any particular new-physics scenario.

In this paper, we have frequently approximated the transitions 
between successive epochs within our PBH-induced stasis scenarios as instantaneous.  In 
reality, the equation-of-state parameter $w$ for the universe evolves smoothly during each 
such transition from the approximately constant value associated with the earlier epoch to 
the approximately constant value associated with the later one.  For example, as 
shown in Fig.~\ref{fig:results}, there is a brief period during which the universe evolves 
under the influence of the stasis attractor from a state in which $\Omega_{\rm BH} = 1$ and 
$w = 0$ to a state in which $\Omega_{\rm BH} = \barOmega_{\rm BH}$ and 
$w = \overline{w}$.  While accounting for the full time-dependence of these transitions 
would not have a significant impact on our results, it would allow for a more precise 
determination of the discovery reach for potential signals of PBH-induced stasis future 
experiments.  We leave such an analysis for future work.

In closing, we note that another potentially observable consequence of cosmic 
stasis stems from the fact that density perturbations evolve differently in cosmologies 
involving a stasis epoch than they do in the standard cosmology.  This is due
not only to the modification of the expansion history, but also  
to the fact that when energy density is transferred from matter to 
radiation --- \eg, via the decays of PBHs --- perturbations in the energy 
density of matter serve as a source  for perturbations in the energy density 
of radiation.  Both of these effects are present in cosmologies involving an 
epoch of early matter domination as well, where they can modify the 
spectrum of perturbations in the radiation density.  In scenarios 
in which the particle species which will eventually come to constitute the dark
matter are still relativistic and in equilibrium with the radiation bath
throughout the stasis epoch, this modification can in turn lead to 
a modification of the shape of the matter power
spectrum~\cite{Erickcek:2011us,Fan:2014zua,Georg:2016yxa}.
One would therefore expect similar modifications of the matter power spectrum
to arise in cosmologies involving a stasis epoch.


\bigskip
\bigskip
\section*{Acknowledgements}


The research activities of KRD are supported in part 
by the U.S.\ Department of Energy under Grant DE-FG02-13ER41976 / DE-SC0009913 
and by the U.S.\ National Science Foundation through its employee IR/D program.
The work of LH is funded by the UK Science and Technology Facilities Council 
(STFC) under Grant ST/P001246/1. LH also acknowledges the support of the Institut 
Pascal at Universit\'{e} Paris-Saclay during the Paris-Saclay Astroparticle 
Symposium 2022, with the support of the IN2P3 master projet UCMN, the P2IO 
Laboratory of Excellence (program ``Investissements d'avenir'' 
ANR-11-IDEX-0003-01 Paris-Saclay and ANR-10-LABX-0038), the P2I axis of the 
Graduate School Physics of Universit\'{e} Paris-Saclay, as well as IJCLab, CEA, 
IPhT, APPEC,  and ANR-11-IDEX-0003-01 Paris-Saclay and ANR-10-LABX-0038.   
The work of FH and TMPT is supported in part by the U.S.\ National Science 
Foundation under Grant PHY-1915005, while the work of FH is also supported 
in part by the International Postdoctoral Exchange Fellowship Program,
and the work of TMPT is also supported in part by the U.S.\ National Science 
Foundation under Grant PHY-2210283.
The work of DK is supported by the U.S.\ Department of Energy Grant DE-SC0010813.
The research activities of BT are supported in 
part by the U.S.\ National Science Foundation under Grant PHY-2014104.
The opinions and conclusions expressed herein are those of the authors and do 
not represent any funding agencies. 

\vfill\eject

\bibliography{ref}

\begin{thebibliography}{195}%
\makeatletter
\providecommand \@ifxundefined [1]{%
 \@ifx{#1\undefined}
}%
\providecommand \@ifnum [1]{%
 \ifnum #1\expandafter \@firstoftwo
 \else \expandafter \@secondoftwo
 \fi
}%
\providecommand \@ifx [1]{%
 \ifx #1\expandafter \@firstoftwo
 \else \expandafter \@secondoftwo
 \fi
}%
\providecommand \natexlab [1]{#1}%
\providecommand \enquote  [1]{``#1''}%
\providecommand \bibnamefont  [1]{#1}%
\providecommand \bibfnamefont [1]{#1}%
\providecommand \citenamefont [1]{#1}%
\providecommand \href@noop [0]{\@secondoftwo}%
\providecommand \href [0]{\begingroup \@sanitize@url \@href}%
\providecommand \@href[1]{\@@startlink{#1}\@@href}%
\providecommand \@@href[1]{\endgroup#1\@@endlink}%
\providecommand \@sanitize@url [0]{\catcode `\\12\catcode `\$12\catcode
  `\&12\catcode `\#12\catcode `\^12\catcode `\_12\catcode `\%12\relax}%
\providecommand \@@startlink[1]{}%
\providecommand \@@endlink[0]{}%
\providecommand \url  [0]{\begingroup\@sanitize@url \@url }%
\providecommand \@url [1]{\endgroup\@href {#1}{\urlprefix }}%
\providecommand \urlprefix  [0]{URL }%
\providecommand \Eprint [0]{\href }%
\providecommand \doibase [0]{http://dx.doi.org/}%
\providecommand \selectlanguage [0]{\@gobble}%
\providecommand \bibinfo  [0]{\@secondoftwo}%
\providecommand \bibfield  [0]{\@secondoftwo}%
\providecommand \translation [1]{[#1]}%
\providecommand \BibitemOpen [0]{}%
\providecommand \bibitemStop [0]{}%
\providecommand \bibitemNoStop [0]{.\EOS\space}%
\providecommand \EOS [0]{\spacefactor3000\relax}%
\providecommand \BibitemShut  [1]{\csname bibitem#1\endcsname}%
\let\auto@bib@innerbib\@empty
\bibitem [{\citenamefont {Hawking}(1974)}]{Hawking:1974rv}%
  \BibitemOpen
  \bibfield  {author} {\bibinfo {author} {\bibfnamefont {S.~W.}\ \bibnamefont
  {Hawking}},\ }\href {\doibase 10.1038/248030a0} {\bibfield  {journal}
  {\bibinfo  {journal} {Nature}\ }\textbf {\bibinfo {volume} {248}},\ \bibinfo
  {pages} {30} (\bibinfo {year} {1974})}\BibitemShut {NoStop}%
\bibitem [{\citenamefont {Hawking}(1975)}]{Hawking:1975vcx}%
  \BibitemOpen
  \bibfield  {author} {\bibinfo {author} {\bibfnamefont {S.~W.}\ \bibnamefont
  {Hawking}},\ }\href {\doibase 10.1007/BF02345020} {\bibfield  {journal}
  {\bibinfo  {journal} {Commun. Math. Phys.}\ }\textbf {\bibinfo {volume}
  {43}},\ \bibinfo {pages} {199} (\bibinfo {year} {1975})},\ \bibinfo {note}
  {[Erratum: Commun.Math.Phys. 46, 206 (1976)]}\BibitemShut {NoStop}%
\bibitem [{\citenamefont {Escriv\`a}\ \emph {et~al.}(2022)\citenamefont
  {Escriv\`a}, \citenamefont {Kuhnel},\ and\ \citenamefont
  {Tada}}]{Escriva:2022duf}%
  \BibitemOpen
  \bibfield  {author} {\bibinfo {author} {\bibfnamefont {A.}~\bibnamefont
  {Escriv\`a}}, \bibinfo {author} {\bibfnamefont {F.}~\bibnamefont {Kuhnel}}, \
  and\ \bibinfo {author} {\bibfnamefont {Y.}~\bibnamefont {Tada}},\ }\href@noop
  {} {\  (\bibinfo {year} {2022})},\ \Eprint {http://arxiv.org/abs/2211.05767}
  {arXiv:2211.05767 [astro-ph.CO]} \BibitemShut {NoStop}%
\bibitem [{\citenamefont {Carr}\ \emph {et~al.}(2021)\citenamefont {Carr},
  \citenamefont {Kohri}, \citenamefont {Sendouda},\ and\ \citenamefont
  {Yokoyama}}]{Carr:2020gox}%
  \BibitemOpen
  \bibfield  {author} {\bibinfo {author} {\bibfnamefont {B.}~\bibnamefont
  {Carr}}, \bibinfo {author} {\bibfnamefont {K.}~\bibnamefont {Kohri}},
  \bibinfo {author} {\bibfnamefont {Y.}~\bibnamefont {Sendouda}}, \ and\
  \bibinfo {author} {\bibfnamefont {J.}~\bibnamefont {Yokoyama}},\ }\href
  {\doibase 10.1088/1361-6633/ac1e31} {\bibfield  {journal} {\bibinfo
  {journal} {Rept. Prog. Phys.}\ }\textbf {\bibinfo {volume} {84}},\ \bibinfo
  {pages} {116902} (\bibinfo {year} {2021})},\ \Eprint
  {http://arxiv.org/abs/2002.12778} {arXiv:2002.12778 [astro-ph.CO]}
  \BibitemShut {NoStop}%
\bibitem [{\citenamefont {Green}\ and\ \citenamefont
  {Kavanagh}(2021)}]{Green:2020jor}%
  \BibitemOpen
  \bibfield  {author} {\bibinfo {author} {\bibfnamefont {A.~M.}\ \bibnamefont
  {Green}}\ and\ \bibinfo {author} {\bibfnamefont {B.~J.}\ \bibnamefont
  {Kavanagh}},\ }\href {\doibase 10.1088/1361-6471/abc534} {\bibfield
  {journal} {\bibinfo  {journal} {J. Phys. G}\ }\textbf {\bibinfo {volume}
  {48}},\ \bibinfo {pages} {4} (\bibinfo {year} {2021})},\ \Eprint
  {http://arxiv.org/abs/2007.10722} {arXiv:2007.10722 [astro-ph.CO]}
  \BibitemShut {NoStop}%
\bibitem [{\citenamefont {Villanueva-Domingo}\ \emph
  {et~al.}(2021)\citenamefont {Villanueva-Domingo}, \citenamefont {Mena},\ and\
  \citenamefont {Palomares-Ruiz}}]{Villanueva-Domingo:2021spv}%
  \BibitemOpen
  \bibfield  {author} {\bibinfo {author} {\bibfnamefont {P.}~\bibnamefont
  {Villanueva-Domingo}}, \bibinfo {author} {\bibfnamefont {O.}~\bibnamefont
  {Mena}}, \ and\ \bibinfo {author} {\bibfnamefont {S.}~\bibnamefont
  {Palomares-Ruiz}},\ }\href {\doibase 10.3389/fspas.2021.681084} {\bibfield
  {journal} {\bibinfo  {journal} {Front. Astron. Space Sci.}\ }\textbf
  {\bibinfo {volume} {8}},\ \bibinfo {pages} {87} (\bibinfo {year} {2021})},\
  \Eprint {http://arxiv.org/abs/2103.12087} {arXiv:2103.12087 [astro-ph.CO]}
  \BibitemShut {NoStop}%
\bibitem [{\citenamefont {Carr}\ \emph {et~al.}(2010)\citenamefont {Carr},
  \citenamefont {Kohri}, \citenamefont {Sendouda},\ and\ \citenamefont
  {Yokoyama}}]{Carr:2009jm}%
  \BibitemOpen
  \bibfield  {author} {\bibinfo {author} {\bibfnamefont {B.~J.}\ \bibnamefont
  {Carr}}, \bibinfo {author} {\bibfnamefont {K.}~\bibnamefont {Kohri}},
  \bibinfo {author} {\bibfnamefont {Y.}~\bibnamefont {Sendouda}}, \ and\
  \bibinfo {author} {\bibfnamefont {J.}~\bibnamefont {Yokoyama}},\ }\href
  {\doibase 10.1103/PhysRevD.81.104019} {\bibfield  {journal} {\bibinfo
  {journal} {Phys. Rev. D}\ }\textbf {\bibinfo {volume} {81}},\ \bibinfo
  {pages} {104019} (\bibinfo {year} {2010})},\ \Eprint
  {http://arxiv.org/abs/0912.5297} {arXiv:0912.5297 [astro-ph.CO]} \BibitemShut
  {NoStop}%
\bibitem [{\citenamefont {Keith}\ \emph {et~al.}(2020)\citenamefont {Keith},
  \citenamefont {Hooper}, \citenamefont {Blinov},\ and\ \citenamefont
  {McDermott}}]{Keith:2020jww}%
  \BibitemOpen
  \bibfield  {author} {\bibinfo {author} {\bibfnamefont {C.}~\bibnamefont
  {Keith}}, \bibinfo {author} {\bibfnamefont {D.}~\bibnamefont {Hooper}},
  \bibinfo {author} {\bibfnamefont {N.}~\bibnamefont {Blinov}}, \ and\ \bibinfo
  {author} {\bibfnamefont {S.~D.}\ \bibnamefont {McDermott}},\ }\href {\doibase
  10.1103/PhysRevD.102.103512} {\bibfield  {journal} {\bibinfo  {journal}
  {Phys. Rev. D}\ }\textbf {\bibinfo {volume} {102}},\ \bibinfo {pages}
  {103512} (\bibinfo {year} {2020})},\ \Eprint
  {http://arxiv.org/abs/2006.03608} {arXiv:2006.03608 [astro-ph.CO]}
  \BibitemShut {NoStop}%
\bibitem [{\citenamefont {Hooper}\ \emph {et~al.}(2019)\citenamefont {Hooper},
  \citenamefont {Krnjaic},\ and\ \citenamefont {McDermott}}]{Hooper:2019gtx}%
  \BibitemOpen
  \bibfield  {author} {\bibinfo {author} {\bibfnamefont {D.}~\bibnamefont
  {Hooper}}, \bibinfo {author} {\bibfnamefont {G.}~\bibnamefont {Krnjaic}}, \
  and\ \bibinfo {author} {\bibfnamefont {S.~D.}\ \bibnamefont {McDermott}},\
  }\href {\doibase 10.1007/JHEP08(2019)001} {\bibfield  {journal} {\bibinfo
  {journal} {JHEP}\ }\textbf {\bibinfo {volume} {08}},\ \bibinfo {pages} {001}
  (\bibinfo {year} {2019})},\ \Eprint {http://arxiv.org/abs/1905.01301}
  {arXiv:1905.01301 [hep-ph]} \BibitemShut {NoStop}%
\bibitem [{\citenamefont {Masina}(2020)}]{Masina:2020xhk}%
  \BibitemOpen
  \bibfield  {author} {\bibinfo {author} {\bibfnamefont {I.}~\bibnamefont
  {Masina}},\ }\href {\doibase 10.1140/epjp/s13360-020-00564-9} {\bibfield
  {journal} {\bibinfo  {journal} {Eur. Phys. J. Plus}\ }\textbf {\bibinfo
  {volume} {135}},\ \bibinfo {pages} {552} (\bibinfo {year} {2020})},\ \Eprint
  {http://arxiv.org/abs/2004.04740} {arXiv:2004.04740 [hep-ph]} \BibitemShut
  {NoStop}%
\bibitem [{\citenamefont {Baldes}\ \emph {et~al.}(2020)\citenamefont {Baldes},
  \citenamefont {Decant}, \citenamefont {Hooper},\ and\ \citenamefont
  {Lopez-Honorez}}]{Baldes:2020nuv}%
  \BibitemOpen
  \bibfield  {author} {\bibinfo {author} {\bibfnamefont {I.}~\bibnamefont
  {Baldes}}, \bibinfo {author} {\bibfnamefont {Q.}~\bibnamefont {Decant}},
  \bibinfo {author} {\bibfnamefont {D.~C.}\ \bibnamefont {Hooper}}, \ and\
  \bibinfo {author} {\bibfnamefont {L.}~\bibnamefont {Lopez-Honorez}},\ }\href
  {\doibase 10.1088/1475-7516/2020/08/045} {\bibfield  {journal} {\bibinfo
  {journal} {JCAP}\ }\textbf {\bibinfo {volume} {08}},\ \bibinfo {pages} {045}
  (\bibinfo {year} {2020})},\ \Eprint {http://arxiv.org/abs/2004.14773}
  {arXiv:2004.14773 [astro-ph.CO]} \BibitemShut {NoStop}%
\bibitem [{\citenamefont {Gondolo}\ \emph {et~al.}(2020)\citenamefont
  {Gondolo}, \citenamefont {Sandick},\ and\ \citenamefont {Shams
  Es~Haghi}}]{Gondolo:2020uqv}%
  \BibitemOpen
  \bibfield  {author} {\bibinfo {author} {\bibfnamefont {P.}~\bibnamefont
  {Gondolo}}, \bibinfo {author} {\bibfnamefont {P.}~\bibnamefont {Sandick}}, \
  and\ \bibinfo {author} {\bibfnamefont {B.}~\bibnamefont {Shams Es~Haghi}},\
  }\href {\doibase 10.1103/PhysRevD.102.095018} {\bibfield  {journal} {\bibinfo
   {journal} {Phys. Rev. D}\ }\textbf {\bibinfo {volume} {102}},\ \bibinfo
  {pages} {095018} (\bibinfo {year} {2020})},\ \Eprint
  {http://arxiv.org/abs/2009.02424} {arXiv:2009.02424 [hep-ph]} \BibitemShut
  {NoStop}%
\bibitem [{\citenamefont {Bernal}\ and\ \citenamefont
  {Zapata}(2021{\natexlab{a}})}]{Bernal:2020kse}%
  \BibitemOpen
  \bibfield  {author} {\bibinfo {author} {\bibfnamefont {N.}~\bibnamefont
  {Bernal}}\ and\ \bibinfo {author} {\bibfnamefont {O.}~\bibnamefont
  {Zapata}},\ }\href {\doibase 10.1088/1475-7516/2021/03/007} {\bibfield
  {journal} {\bibinfo  {journal} {JCAP}\ }\textbf {\bibinfo {volume} {03}},\
  \bibinfo {pages} {007} (\bibinfo {year} {2021}{\natexlab{a}})},\ \Eprint
  {http://arxiv.org/abs/2010.09725} {arXiv:2010.09725 [hep-ph]} \BibitemShut
  {NoStop}%
\bibitem [{\citenamefont {Bernal}\ and\ \citenamefont
  {Zapata}(2021{\natexlab{b}})}]{Bernal:2020ili}%
  \BibitemOpen
  \bibfield  {author} {\bibinfo {author} {\bibfnamefont {N.}~\bibnamefont
  {Bernal}}\ and\ \bibinfo {author} {\bibfnamefont {O.}~\bibnamefont
  {Zapata}},\ }\href {\doibase 10.1016/j.physletb.2021.136129} {\bibfield
  {journal} {\bibinfo  {journal} {Phys. Lett. B}\ }\textbf {\bibinfo {volume}
  {815}},\ \bibinfo {pages} {136129} (\bibinfo {year} {2021}{\natexlab{b}})},\
  \Eprint {http://arxiv.org/abs/2011.02510} {arXiv:2011.02510 [hep-ph]}
  \BibitemShut {NoStop}%
\bibitem [{\citenamefont {Bernal}\ and\ \citenamefont
  {Zapata}(2021{\natexlab{c}})}]{Bernal:2020bjf}%
  \BibitemOpen
  \bibfield  {author} {\bibinfo {author} {\bibfnamefont {N.}~\bibnamefont
  {Bernal}}\ and\ \bibinfo {author} {\bibfnamefont {O.}~\bibnamefont
  {Zapata}},\ }\href {\doibase 10.1088/1475-7516/2021/03/015} {\bibfield
  {journal} {\bibinfo  {journal} {JCAP}\ }\textbf {\bibinfo {volume} {03}},\
  \bibinfo {pages} {015} (\bibinfo {year} {2021}{\natexlab{c}})},\ \Eprint
  {http://arxiv.org/abs/2011.12306} {arXiv:2011.12306 [astro-ph.CO]}
  \BibitemShut {NoStop}%
\bibitem [{\citenamefont {Masina}(2021)}]{Masina:2021zpu}%
  \BibitemOpen
  \bibfield  {author} {\bibinfo {author} {\bibfnamefont {I.}~\bibnamefont
  {Masina}},\ }\href {\doibase 10.1134/S0202289321040101} {\bibfield  {journal}
  {\bibinfo  {journal} {Grav. Cosmol.}\ }\textbf {\bibinfo {volume} {27}},\
  \bibinfo {pages} {315} (\bibinfo {year} {2021})},\ \Eprint
  {http://arxiv.org/abs/2103.13825} {arXiv:2103.13825 [gr-qc]} \BibitemShut
  {NoStop}%
\bibitem [{\citenamefont {Arbey}\ \emph {et~al.}(2021)\citenamefont {Arbey},
  \citenamefont {Auffinger}, \citenamefont {Sandick}, \citenamefont {Shams
  Es~Haghi},\ and\ \citenamefont {Sinha}}]{Arbey:2021ysg}%
  \BibitemOpen
  \bibfield  {author} {\bibinfo {author} {\bibfnamefont {A.}~\bibnamefont
  {Arbey}}, \bibinfo {author} {\bibfnamefont {J.}~\bibnamefont {Auffinger}},
  \bibinfo {author} {\bibfnamefont {P.}~\bibnamefont {Sandick}}, \bibinfo
  {author} {\bibfnamefont {B.}~\bibnamefont {Shams Es~Haghi}}, \ and\ \bibinfo
  {author} {\bibfnamefont {K.}~\bibnamefont {Sinha}},\ }\href {\doibase
  10.1103/PhysRevD.103.123549} {\bibfield  {journal} {\bibinfo  {journal}
  {Phys. Rev. D}\ }\textbf {\bibinfo {volume} {103}},\ \bibinfo {pages}
  {123549} (\bibinfo {year} {2021})},\ \Eprint
  {http://arxiv.org/abs/2104.04051} {arXiv:2104.04051 [astro-ph.CO]}
  \BibitemShut {NoStop}%
\bibitem [{\citenamefont {Jyoti~Das}\ \emph {et~al.}(2021)\citenamefont
  {Jyoti~Das}, \citenamefont {Mahanta},\ and\ \citenamefont
  {Borah}}]{JyotiDas:2021shi}%
  \BibitemOpen
  \bibfield  {author} {\bibinfo {author} {\bibfnamefont {S.}~\bibnamefont
  {Jyoti~Das}}, \bibinfo {author} {\bibfnamefont {D.}~\bibnamefont {Mahanta}},
  \ and\ \bibinfo {author} {\bibfnamefont {D.}~\bibnamefont {Borah}},\ }\href
  {\doibase 10.1088/1475-7516/2021/11/019} {\bibfield  {journal} {\bibinfo
  {journal} {JCAP}\ }\textbf {\bibinfo {volume} {11}},\ \bibinfo {pages} {019}
  (\bibinfo {year} {2021})},\ \Eprint {http://arxiv.org/abs/2104.14496}
  {arXiv:2104.14496 [hep-ph]} \BibitemShut {NoStop}%
\bibitem [{\citenamefont {Cheek}\ \emph
  {et~al.}(2022{\natexlab{a}})\citenamefont {Cheek}, \citenamefont {Heurtier},
  \citenamefont {Perez-Gonzalez},\ and\ \citenamefont
  {Turner}}]{Cheek:2021odj}%
  \BibitemOpen
  \bibfield  {author} {\bibinfo {author} {\bibfnamefont {A.}~\bibnamefont
  {Cheek}}, \bibinfo {author} {\bibfnamefont {L.}~\bibnamefont {Heurtier}},
  \bibinfo {author} {\bibfnamefont {Y.~F.}\ \bibnamefont {Perez-Gonzalez}}, \
  and\ \bibinfo {author} {\bibfnamefont {J.}~\bibnamefont {Turner}},\ }\href
  {\doibase 10.1103/PhysRevD.105.015022} {\bibfield  {journal} {\bibinfo
  {journal} {Phys. Rev. D}\ }\textbf {\bibinfo {volume} {105}},\ \bibinfo
  {pages} {015022} (\bibinfo {year} {2022}{\natexlab{a}})},\ \Eprint
  {http://arxiv.org/abs/2107.00013} {arXiv:2107.00013 [hep-ph]} \BibitemShut
  {NoStop}%
\bibitem [{\citenamefont {Cheek}\ \emph
  {et~al.}(2022{\natexlab{b}})\citenamefont {Cheek}, \citenamefont {Heurtier},
  \citenamefont {Perez-Gonzalez},\ and\ \citenamefont
  {Turner}}]{Cheek:2021cfe}%
  \BibitemOpen
  \bibfield  {author} {\bibinfo {author} {\bibfnamefont {A.}~\bibnamefont
  {Cheek}}, \bibinfo {author} {\bibfnamefont {L.}~\bibnamefont {Heurtier}},
  \bibinfo {author} {\bibfnamefont {Y.~F.}\ \bibnamefont {Perez-Gonzalez}}, \
  and\ \bibinfo {author} {\bibfnamefont {J.}~\bibnamefont {Turner}},\ }\href
  {\doibase 10.1103/PhysRevD.105.015023} {\bibfield  {journal} {\bibinfo
  {journal} {Phys. Rev. D}\ }\textbf {\bibinfo {volume} {105}},\ \bibinfo
  {pages} {015023} (\bibinfo {year} {2022}{\natexlab{b}})},\ \Eprint
  {http://arxiv.org/abs/2107.00016} {arXiv:2107.00016 [hep-ph]} \BibitemShut
  {NoStop}%
\bibitem [{\citenamefont {Sandick}\ \emph {et~al.}(2021)\citenamefont
  {Sandick}, \citenamefont {Es~Haghi},\ and\ \citenamefont
  {Sinha}}]{Sandick:2021gew}%
  \BibitemOpen
  \bibfield  {author} {\bibinfo {author} {\bibfnamefont {P.}~\bibnamefont
  {Sandick}}, \bibinfo {author} {\bibfnamefont {B.~S.}\ \bibnamefont
  {Es~Haghi}}, \ and\ \bibinfo {author} {\bibfnamefont {K.}~\bibnamefont
  {Sinha}},\ }\href {\doibase 10.1103/PhysRevD.104.083523} {\bibfield
  {journal} {\bibinfo  {journal} {Phys. Rev. D}\ }\textbf {\bibinfo {volume}
  {104}},\ \bibinfo {pages} {083523} (\bibinfo {year} {2021})},\ \Eprint
  {http://arxiv.org/abs/2108.08329} {arXiv:2108.08329 [astro-ph.CO]}
  \BibitemShut {NoStop}%
\bibitem [{\citenamefont {Schiavone}\ \emph {et~al.}(2021)\citenamefont
  {Schiavone}, \citenamefont {Montanino}, \citenamefont {Mirizzi},\ and\
  \citenamefont {Capozzi}}]{Schiavone:2021imu}%
  \BibitemOpen
  \bibfield  {author} {\bibinfo {author} {\bibfnamefont {F.}~\bibnamefont
  {Schiavone}}, \bibinfo {author} {\bibfnamefont {D.}~\bibnamefont
  {Montanino}}, \bibinfo {author} {\bibfnamefont {A.}~\bibnamefont {Mirizzi}},
  \ and\ \bibinfo {author} {\bibfnamefont {F.}~\bibnamefont {Capozzi}},\ }\href
  {\doibase 10.1088/1475-7516/2021/08/063} {\bibfield  {journal} {\bibinfo
  {journal} {JCAP}\ }\textbf {\bibinfo {volume} {08}},\ \bibinfo {pages} {063}
  (\bibinfo {year} {2021})},\ \Eprint {http://arxiv.org/abs/2107.03420}
  {arXiv:2107.03420 [hep-ph]} \BibitemShut {NoStop}%
\bibitem [{\citenamefont {Bernal}\ \emph
  {et~al.}(2021{\natexlab{a}})\citenamefont {Bernal}, \citenamefont
  {Hajkarim},\ and\ \citenamefont {Xu}}]{Bernal:2021yyb}%
  \BibitemOpen
  \bibfield  {author} {\bibinfo {author} {\bibfnamefont {N.}~\bibnamefont
  {Bernal}}, \bibinfo {author} {\bibfnamefont {F.}~\bibnamefont {Hajkarim}}, \
  and\ \bibinfo {author} {\bibfnamefont {Y.}~\bibnamefont {Xu}},\ }\href
  {\doibase 10.1103/PhysRevD.104.075007} {\bibfield  {journal} {\bibinfo
  {journal} {Phys. Rev. D}\ }\textbf {\bibinfo {volume} {104}},\ \bibinfo
  {pages} {075007} (\bibinfo {year} {2021}{\natexlab{a}})},\ \Eprint
  {http://arxiv.org/abs/2107.13575} {arXiv:2107.13575 [hep-ph]} \BibitemShut
  {NoStop}%
\bibitem [{\citenamefont {Cheek}\ \emph
  {et~al.}(2022{\natexlab{c}})\citenamefont {Cheek}, \citenamefont {Heurtier},
  \citenamefont {Perez-Gonzalez},\ and\ \citenamefont
  {Turner}}]{Cheek:2022dbx}%
  \BibitemOpen
  \bibfield  {author} {\bibinfo {author} {\bibfnamefont {A.}~\bibnamefont
  {Cheek}}, \bibinfo {author} {\bibfnamefont {L.}~\bibnamefont {Heurtier}},
  \bibinfo {author} {\bibfnamefont {Y.~F.}\ \bibnamefont {Perez-Gonzalez}}, \
  and\ \bibinfo {author} {\bibfnamefont {J.}~\bibnamefont {Turner}},\ }\href
  {\doibase 10.1103/PhysRevD.106.103012} {\bibfield  {journal} {\bibinfo
  {journal} {Phys. Rev. D}\ }\textbf {\bibinfo {volume} {106}},\ \bibinfo
  {pages} {103012} (\bibinfo {year} {2022}{\natexlab{c}})},\ \Eprint
  {http://arxiv.org/abs/2207.09462} {arXiv:2207.09462 [astro-ph.CO]}
  \BibitemShut {NoStop}%
\bibitem [{\citenamefont {Bernal}\ \emph
  {et~al.}(2022{\natexlab{a}})\citenamefont {Bernal}, \citenamefont
  {Perez-Gonzalez},\ and\ \citenamefont {Xu}}]{Bernal:2022oha}%
  \BibitemOpen
  \bibfield  {author} {\bibinfo {author} {\bibfnamefont {N.}~\bibnamefont
  {Bernal}}, \bibinfo {author} {\bibfnamefont {Y.~F.}\ \bibnamefont
  {Perez-Gonzalez}}, \ and\ \bibinfo {author} {\bibfnamefont {Y.}~\bibnamefont
  {Xu}},\ }\href {\doibase 10.1103/PhysRevD.106.015020} {\bibfield  {journal}
  {\bibinfo  {journal} {Phys. Rev. D}\ }\textbf {\bibinfo {volume} {106}},\
  \bibinfo {pages} {015020} (\bibinfo {year} {2022}{\natexlab{a}})},\ \Eprint
  {http://arxiv.org/abs/2205.11522} {arXiv:2205.11522 [hep-ph]} \BibitemShut
  {NoStop}%
\bibitem [{\citenamefont {Morrison}\ \emph {et~al.}(2019)\citenamefont
  {Morrison}, \citenamefont {Profumo},\ and\ \citenamefont
  {Yu}}]{Morrison:2018xla}%
  \BibitemOpen
  \bibfield  {author} {\bibinfo {author} {\bibfnamefont {L.}~\bibnamefont
  {Morrison}}, \bibinfo {author} {\bibfnamefont {S.}~\bibnamefont {Profumo}}, \
  and\ \bibinfo {author} {\bibfnamefont {Y.}~\bibnamefont {Yu}},\ }\href
  {\doibase 10.1088/1475-7516/2019/05/005} {\bibfield  {journal} {\bibinfo
  {journal} {JCAP}\ }\textbf {\bibinfo {volume} {1905}},\ \bibinfo {pages}
  {005} (\bibinfo {year} {2019})},\ \Eprint {http://arxiv.org/abs/1812.10606}
  {arXiv:1812.10606 [astro-ph.CO]} \BibitemShut {NoStop}%
\bibitem [{\citenamefont {Auffinger}\ \emph {et~al.}(2021)\citenamefont
  {Auffinger}, \citenamefont {Masina},\ and\ \citenamefont
  {Orlando}}]{Auffinger:2020afu}%
  \BibitemOpen
  \bibfield  {author} {\bibinfo {author} {\bibfnamefont {J.}~\bibnamefont
  {Auffinger}}, \bibinfo {author} {\bibfnamefont {I.}~\bibnamefont {Masina}}, \
  and\ \bibinfo {author} {\bibfnamefont {G.}~\bibnamefont {Orlando}},\ }\href
  {\doibase 10.1140/epjp/s13360-021-01247-9} {\bibfield  {journal} {\bibinfo
  {journal} {Eur. Phys. J. Plus}\ }\textbf {\bibinfo {volume} {136}},\ \bibinfo
  {pages} {261} (\bibinfo {year} {2021})},\ \Eprint
  {http://arxiv.org/abs/2012.09867} {arXiv:2012.09867 [hep-ph]} \BibitemShut
  {NoStop}%
\bibitem [{\citenamefont {Khlopov}\ \emph {et~al.}(2006)\citenamefont
  {Khlopov}, \citenamefont {Barrau},\ and\ \citenamefont
  {Grain}}]{Khlopov:2004tn}%
  \BibitemOpen
  \bibfield  {author} {\bibinfo {author} {\bibfnamefont {M.~Y.}\ \bibnamefont
  {Khlopov}}, \bibinfo {author} {\bibfnamefont {A.}~\bibnamefont {Barrau}}, \
  and\ \bibinfo {author} {\bibfnamefont {J.}~\bibnamefont {Grain}},\ }\href
  {\doibase 10.1088/0264-9381/23/6/004} {\bibfield  {journal} {\bibinfo
  {journal} {Class. Quant. Grav.}\ }\textbf {\bibinfo {volume} {23}},\ \bibinfo
  {pages} {1875} (\bibinfo {year} {2006})},\ \Eprint
  {http://arxiv.org/abs/astro-ph/0406621} {arXiv:astro-ph/0406621} \BibitemShut
  {NoStop}%
\bibitem [{\citenamefont {Allahverdi}\ \emph
  {et~al.}(2018{\natexlab{a}})\citenamefont {Allahverdi}, \citenamefont
  {Dent},\ and\ \citenamefont {Osinski}}]{Allahverdi:2017sks}%
  \BibitemOpen
  \bibfield  {author} {\bibinfo {author} {\bibfnamefont {R.}~\bibnamefont
  {Allahverdi}}, \bibinfo {author} {\bibfnamefont {J.}~\bibnamefont {Dent}}, \
  and\ \bibinfo {author} {\bibfnamefont {J.}~\bibnamefont {Osinski}},\ }\href
  {\doibase 10.1103/PhysRevD.97.055013} {\bibfield  {journal} {\bibinfo
  {journal} {Phys. Rev.}\ }\textbf {\bibinfo {volume} {D97}},\ \bibinfo {pages}
  {055013} (\bibinfo {year} {2018}{\natexlab{a}})},\ \Eprint
  {http://arxiv.org/abs/1711.10511} {arXiv:1711.10511 [astro-ph.CO]}
  \BibitemShut {NoStop}%
\bibitem [{\citenamefont {Lennon}\ \emph {et~al.}(2018)\citenamefont {Lennon},
  \citenamefont {March-Russell}, \citenamefont {Petrossian-Byrne},\ and\
  \citenamefont {Tillim}}]{Lennon:2017tqq}%
  \BibitemOpen
  \bibfield  {author} {\bibinfo {author} {\bibfnamefont {O.}~\bibnamefont
  {Lennon}}, \bibinfo {author} {\bibfnamefont {J.}~\bibnamefont
  {March-Russell}}, \bibinfo {author} {\bibfnamefont {R.}~\bibnamefont
  {Petrossian-Byrne}}, \ and\ \bibinfo {author} {\bibfnamefont
  {H.}~\bibnamefont {Tillim}},\ }\href {\doibase 10.1088/1475-7516/2018/04/009}
  {\bibfield  {journal} {\bibinfo  {journal} {JCAP}\ }\textbf {\bibinfo
  {volume} {1804}},\ \bibinfo {pages} {009} (\bibinfo {year} {2018})},\ \Eprint
  {http://arxiv.org/abs/1712.07664} {arXiv:1712.07664 [hep-ph]} \BibitemShut
  {NoStop}%
\bibitem [{\citenamefont {Kitabayashi}(2021)}]{Kitabayashi:2021hox}%
  \BibitemOpen
  \bibfield  {author} {\bibinfo {author} {\bibfnamefont {T.}~\bibnamefont
  {Kitabayashi}},\ }\href {\doibase 10.1142/S0217751X21501396} {\bibfield
  {journal} {\bibinfo  {journal} {Int. J. Mod. Phys. A}\ }\textbf {\bibinfo
  {volume} {36}},\ \bibinfo {pages} {2150139} (\bibinfo {year} {2021})},\
  \Eprint {http://arxiv.org/abs/2101.01921} {arXiv:2101.01921 [hep-ph]}
  \BibitemShut {NoStop}%
\bibitem [{\citenamefont {Barrow}\ \emph
  {et~al.}(1991{\natexlab{a}})\citenamefont {Barrow}, \citenamefont {Copeland},
  \citenamefont {Kolb},\ and\ \citenamefont {Liddle}}]{Barrow:1990he}%
  \BibitemOpen
  \bibfield  {author} {\bibinfo {author} {\bibfnamefont {J.~D.}\ \bibnamefont
  {Barrow}}, \bibinfo {author} {\bibfnamefont {E.~J.}\ \bibnamefont
  {Copeland}}, \bibinfo {author} {\bibfnamefont {E.~W.}\ \bibnamefont {Kolb}},
  \ and\ \bibinfo {author} {\bibfnamefont {A.~R.}\ \bibnamefont {Liddle}},\
  }\href {\doibase 10.1103/PhysRevD.43.984} {\bibfield  {journal} {\bibinfo
  {journal} {Phys. Rev. D}\ }\textbf {\bibinfo {volume} {43}},\ \bibinfo
  {pages} {984} (\bibinfo {year} {1991}{\natexlab{a}})}\BibitemShut {NoStop}%
\bibitem [{\citenamefont {Hamada}\ and\ \citenamefont
  {Iso}(2017)}]{Hamada:2016jnq}%
  \BibitemOpen
  \bibfield  {author} {\bibinfo {author} {\bibfnamefont {Y.}~\bibnamefont
  {Hamada}}\ and\ \bibinfo {author} {\bibfnamefont {S.}~\bibnamefont {Iso}},\
  }\href {\doibase 10.1093/ptep/ptx011} {\bibfield  {journal} {\bibinfo
  {journal} {PTEP}\ }\textbf {\bibinfo {volume} {2017}},\ \bibinfo {pages}
  {033B02} (\bibinfo {year} {2017})},\ \Eprint
  {http://arxiv.org/abs/1610.02586} {arXiv:1610.02586 [hep-ph]} \BibitemShut
  {NoStop}%
\bibitem [{\citenamefont {Hooper}\ and\ \citenamefont
  {Krnjaic}(2021)}]{Hooper:2020otu}%
  \BibitemOpen
  \bibfield  {author} {\bibinfo {author} {\bibfnamefont {D.}~\bibnamefont
  {Hooper}}\ and\ \bibinfo {author} {\bibfnamefont {G.}~\bibnamefont
  {Krnjaic}},\ }\href {\doibase 10.1103/PhysRevD.103.043504} {\bibfield
  {journal} {\bibinfo  {journal} {Phys. Rev. D}\ }\textbf {\bibinfo {volume}
  {103}},\ \bibinfo {pages} {043504} (\bibinfo {year} {2021})},\ \Eprint
  {http://arxiv.org/abs/2010.01134} {arXiv:2010.01134 [hep-ph]} \BibitemShut
  {NoStop}%
\bibitem [{\citenamefont {Perez-Gonzalez}\ and\ \citenamefont
  {Turner}(2021)}]{Perez-Gonzalez:2020vnz}%
  \BibitemOpen
  \bibfield  {author} {\bibinfo {author} {\bibfnamefont {Y.~F.}\ \bibnamefont
  {Perez-Gonzalez}}\ and\ \bibinfo {author} {\bibfnamefont {J.}~\bibnamefont
  {Turner}},\ }\href {\doibase 10.1103/PhysRevD.104.103021} {\bibfield
  {journal} {\bibinfo  {journal} {Phys. Rev. D}\ }\textbf {\bibinfo {volume}
  {104}},\ \bibinfo {pages} {103021} (\bibinfo {year} {2021})},\ \Eprint
  {http://arxiv.org/abs/2010.03565} {arXiv:2010.03565 [hep-ph]} \BibitemShut
  {NoStop}%
\bibitem [{\citenamefont {Datta}\ \emph {et~al.}(2021)\citenamefont {Datta},
  \citenamefont {Ghosal},\ and\ \citenamefont {Samanta}}]{Datta:2020bht}%
  \BibitemOpen
  \bibfield  {author} {\bibinfo {author} {\bibfnamefont {S.}~\bibnamefont
  {Datta}}, \bibinfo {author} {\bibfnamefont {A.}~\bibnamefont {Ghosal}}, \
  and\ \bibinfo {author} {\bibfnamefont {R.}~\bibnamefont {Samanta}},\ }\href
  {\doibase 10.1088/1475-7516/2021/08/021} {\bibfield  {journal} {\bibinfo
  {journal} {JCAP}\ }\textbf {\bibinfo {volume} {08}},\ \bibinfo {pages} {021}
  (\bibinfo {year} {2021})},\ \Eprint {http://arxiv.org/abs/2012.14981}
  {arXiv:2012.14981 [hep-ph]} \BibitemShut {NoStop}%
\bibitem [{\citenamefont {Gehrman}\ \emph {et~al.}(2023)\citenamefont
  {Gehrman}, \citenamefont {Shams Es~Haghi}, \citenamefont {Sinha},\ and\
  \citenamefont {Xu}}]{Gehrman:2022imk}%
  \BibitemOpen
  \bibfield  {author} {\bibinfo {author} {\bibfnamefont {T.~C.}\ \bibnamefont
  {Gehrman}}, \bibinfo {author} {\bibfnamefont {B.}~\bibnamefont {Shams
  Es~Haghi}}, \bibinfo {author} {\bibfnamefont {K.}~\bibnamefont {Sinha}}, \
  and\ \bibinfo {author} {\bibfnamefont {T.}~\bibnamefont {Xu}},\ }\href
  {\doibase 10.1088/1475-7516/2023/02/062} {\bibfield  {journal} {\bibinfo
  {journal} {JCAP}\ }\textbf {\bibinfo {volume} {02}},\ \bibinfo {pages} {062}
  (\bibinfo {year} {2023})},\ \Eprint {http://arxiv.org/abs/2211.08431}
  {arXiv:2211.08431 [hep-ph]} \BibitemShut {NoStop}%
\bibitem [{\citenamefont {Bernal}\ \emph
  {et~al.}(2021{\natexlab{b}})\citenamefont {Bernal}, \citenamefont
  {Perez-Gonzalez}, \citenamefont {Xu},\ and\ \citenamefont
  {Zapata}}]{Bernal:2021bbv}%
  \BibitemOpen
  \bibfield  {author} {\bibinfo {author} {\bibfnamefont {N.}~\bibnamefont
  {Bernal}}, \bibinfo {author} {\bibfnamefont {Y.~F.}\ \bibnamefont
  {Perez-Gonzalez}}, \bibinfo {author} {\bibfnamefont {Y.}~\bibnamefont {Xu}},
  \ and\ \bibinfo {author} {\bibfnamefont {O.}~\bibnamefont {Zapata}},\ }\href
  {\doibase 10.1103/PhysRevD.104.123536} {\bibfield  {journal} {\bibinfo
  {journal} {Phys. Rev. D}\ }\textbf {\bibinfo {volume} {104}},\ \bibinfo
  {pages} {123536} (\bibinfo {year} {2021}{\natexlab{b}})},\ \Eprint
  {http://arxiv.org/abs/2110.04312} {arXiv:2110.04312 [hep-ph]} \BibitemShut
  {NoStop}%
\bibitem [{\citenamefont {Barrow}\ \emph
  {et~al.}(1991{\natexlab{b}})\citenamefont {Barrow}, \citenamefont
  {Copeland},\ and\ \citenamefont {Liddle}}]{Barrow:1991dn}%
  \BibitemOpen
  \bibfield  {author} {\bibinfo {author} {\bibfnamefont {J.~D.}\ \bibnamefont
  {Barrow}}, \bibinfo {author} {\bibfnamefont {E.~J.}\ \bibnamefont
  {Copeland}}, \ and\ \bibinfo {author} {\bibfnamefont {A.~R.}\ \bibnamefont
  {Liddle}},\ }\href@noop {} {\bibfield  {journal} {\bibinfo  {journal} {Mon.
  Not. Roy. Astron. Soc.}\ }\textbf {\bibinfo {volume} {253}},\ \bibinfo
  {pages} {675} (\bibinfo {year} {1991}{\natexlab{b}})}\BibitemShut {NoStop}%
\bibitem [{\citenamefont {Dienes}\ \emph {et~al.}(2022)\citenamefont {Dienes},
  \citenamefont {Heurtier}, \citenamefont {Huang}, \citenamefont {Kim},
  \citenamefont {Tait},\ and\ \citenamefont {Thomas}}]{Dienes:2021woi}%
  \BibitemOpen
  \bibfield  {author} {\bibinfo {author} {\bibfnamefont {K.~R.}\ \bibnamefont
  {Dienes}}, \bibinfo {author} {\bibfnamefont {L.}~\bibnamefont {Heurtier}},
  \bibinfo {author} {\bibfnamefont {F.}~\bibnamefont {Huang}}, \bibinfo
  {author} {\bibfnamefont {D.}~\bibnamefont {Kim}}, \bibinfo {author}
  {\bibfnamefont {T.~M.~P.}\ \bibnamefont {Tait}}, \ and\ \bibinfo {author}
  {\bibfnamefont {B.}~\bibnamefont {Thomas}},\ }\href {\doibase
  10.1103/PhysRevD.105.023530} {\bibfield  {journal} {\bibinfo  {journal}
  {Phys. Rev. D}\ }\textbf {\bibinfo {volume} {105}},\ \bibinfo {pages}
  {023530} (\bibinfo {year} {2022})},\ \Eprint
  {http://arxiv.org/abs/2111.04753} {arXiv:2111.04753 [astro-ph.CO]}
  \BibitemShut {NoStop}%
\bibitem [{\citenamefont {Byrnes}\ and\ \citenamefont
  {Cole}(2021)}]{Byrnes:2021jka}%
  \BibitemOpen
  \bibfield  {author} {\bibinfo {author} {\bibfnamefont {C.~T.}\ \bibnamefont
  {Byrnes}}\ and\ \bibinfo {author} {\bibfnamefont {P.~S.}\ \bibnamefont
  {Cole}}\ }(\bibinfo {year} {2021})\ \Eprint {http://arxiv.org/abs/2112.05716}
  {arXiv:2112.05716 [astro-ph.CO]} \BibitemShut {NoStop}%
\bibitem [{\citenamefont {Carr}\ and\ \citenamefont
  {Kuhnel}(2020)}]{Carr:2020xqk}%
  \BibitemOpen
  \bibfield  {author} {\bibinfo {author} {\bibfnamefont {B.}~\bibnamefont
  {Carr}}\ and\ \bibinfo {author} {\bibfnamefont {F.}~\bibnamefont {Kuhnel}},\
  }\href {\doibase 10.1146/annurev-nucl-050520-125911} {\bibfield  {journal}
  {\bibinfo  {journal} {Ann. Rev. Nucl. Part. Sci.}\ }\textbf {\bibinfo
  {volume} {70}},\ \bibinfo {pages} {355} (\bibinfo {year} {2020})},\ \Eprint
  {http://arxiv.org/abs/2006.02838} {arXiv:2006.02838 [astro-ph.CO]}
  \BibitemShut {NoStop}%
\bibitem [{\citenamefont {Carr}(1975)}]{Carr:1975qj}%
  \BibitemOpen
  \bibfield  {author} {\bibinfo {author} {\bibfnamefont {B.~J.}\ \bibnamefont
  {Carr}},\ }\href {\doibase 10.1086/153853} {\bibfield  {journal} {\bibinfo
  {journal} {Astrophys. J.}\ }\textbf {\bibinfo {volume} {201}},\ \bibinfo
  {pages} {1} (\bibinfo {year} {1975})}\BibitemShut {NoStop}%
\bibitem [{\citenamefont {Clesse}\ and\ \citenamefont
  {Garc\'\i{}a-Bellido}(2015)}]{Clesse:2015wea}%
  \BibitemOpen
  \bibfield  {author} {\bibinfo {author} {\bibfnamefont {S.}~\bibnamefont
  {Clesse}}\ and\ \bibinfo {author} {\bibfnamefont {J.}~\bibnamefont
  {Garc\'\i{}a-Bellido}},\ }\href {\doibase 10.1103/PhysRevD.92.023524}
  {\bibfield  {journal} {\bibinfo  {journal} {Phys. Rev. D}\ }\textbf {\bibinfo
  {volume} {92}},\ \bibinfo {pages} {023524} (\bibinfo {year} {2015})},\
  \Eprint {http://arxiv.org/abs/1501.07565} {arXiv:1501.07565 [astro-ph.CO]}
  \BibitemShut {NoStop}%
\bibitem [{\citenamefont {Dolgov}\ and\ \citenamefont
  {Silk}(1993)}]{Dolgov:1992pu}%
  \BibitemOpen
  \bibfield  {author} {\bibinfo {author} {\bibfnamefont {A.}~\bibnamefont
  {Dolgov}}\ and\ \bibinfo {author} {\bibfnamefont {J.}~\bibnamefont {Silk}},\
  }\href {\doibase 10.1103/PhysRevD.47.4244} {\bibfield  {journal} {\bibinfo
  {journal} {Phys. Rev. D}\ }\textbf {\bibinfo {volume} {47}},\ \bibinfo
  {pages} {4244} (\bibinfo {year} {1993})}\BibitemShut {NoStop}%
\bibitem [{\citenamefont {Green}(2016)}]{Green:2016xgy}%
  \BibitemOpen
  \bibfield  {author} {\bibinfo {author} {\bibfnamefont {A.~M.}\ \bibnamefont
  {Green}},\ }\href {\doibase 10.1103/PhysRevD.94.063530} {\bibfield  {journal}
  {\bibinfo  {journal} {Phys. Rev. D}\ }\textbf {\bibinfo {volume} {94}},\
  \bibinfo {pages} {063530} (\bibinfo {year} {2016})},\ \Eprint
  {http://arxiv.org/abs/1609.01143} {arXiv:1609.01143 [astro-ph.CO]}
  \BibitemShut {NoStop}%
\bibitem [{\citenamefont {Carr}\ and\ \citenamefont
  {Lidsey}(1993)}]{Carr:1993aq}%
  \BibitemOpen
  \bibfield  {author} {\bibinfo {author} {\bibfnamefont {B.~J.}\ \bibnamefont
  {Carr}}\ and\ \bibinfo {author} {\bibfnamefont {J.~E.}\ \bibnamefont
  {Lidsey}},\ }\href {\doibase 10.1103/PhysRevD.48.543} {\bibfield  {journal}
  {\bibinfo  {journal} {Phys. Rev. D}\ }\textbf {\bibinfo {volume} {48}},\
  \bibinfo {pages} {543} (\bibinfo {year} {1993})}\BibitemShut {NoStop}%
\bibitem [{\citenamefont {Ivanov}\ \emph {et~al.}(1994)\citenamefont {Ivanov},
  \citenamefont {Naselsky},\ and\ \citenamefont {Novikov}}]{Ivanov:1994pa}%
  \BibitemOpen
  \bibfield  {author} {\bibinfo {author} {\bibfnamefont {P.}~\bibnamefont
  {Ivanov}}, \bibinfo {author} {\bibfnamefont {P.}~\bibnamefont {Naselsky}}, \
  and\ \bibinfo {author} {\bibfnamefont {I.}~\bibnamefont {Novikov}},\ }\href
  {\doibase 10.1103/PhysRevD.50.7173} {\bibfield  {journal} {\bibinfo
  {journal} {Phys. Rev. D}\ }\textbf {\bibinfo {volume} {50}},\ \bibinfo
  {pages} {7173} (\bibinfo {year} {1994})}\BibitemShut {NoStop}%
\bibitem [{\citenamefont {Garcia-Bellido}\ \emph {et~al.}(1996)\citenamefont
  {Garcia-Bellido}, \citenamefont {Linde},\ and\ \citenamefont
  {Wands}}]{Garcia-Bellido:1996mdl}%
  \BibitemOpen
  \bibfield  {author} {\bibinfo {author} {\bibfnamefont {J.}~\bibnamefont
  {Garcia-Bellido}}, \bibinfo {author} {\bibfnamefont {A.~D.}\ \bibnamefont
  {Linde}}, \ and\ \bibinfo {author} {\bibfnamefont {D.}~\bibnamefont
  {Wands}},\ }\href {\doibase 10.1103/PhysRevD.54.6040} {\bibfield  {journal}
  {\bibinfo  {journal} {Phys. Rev. D}\ }\textbf {\bibinfo {volume} {54}},\
  \bibinfo {pages} {6040} (\bibinfo {year} {1996})},\ \Eprint
  {http://arxiv.org/abs/astro-ph/9605094} {arXiv:astro-ph/9605094} \BibitemShut
  {NoStop}%
\bibitem [{\citenamefont {Randall}\ \emph {et~al.}(1996)\citenamefont
  {Randall}, \citenamefont {Soljacic},\ and\ \citenamefont
  {Guth}}]{Randall:1995dj}%
  \BibitemOpen
  \bibfield  {author} {\bibinfo {author} {\bibfnamefont {L.}~\bibnamefont
  {Randall}}, \bibinfo {author} {\bibfnamefont {M.}~\bibnamefont {Soljacic}}, \
  and\ \bibinfo {author} {\bibfnamefont {A.~H.}\ \bibnamefont {Guth}},\ }\href
  {\doibase 10.1016/0550-3213(96)00174-5} {\bibfield  {journal} {\bibinfo
  {journal} {Nucl. Phys. B}\ }\textbf {\bibinfo {volume} {472}},\ \bibinfo
  {pages} {377} (\bibinfo {year} {1996})},\ \Eprint
  {http://arxiv.org/abs/hep-ph/9512439} {arXiv:hep-ph/9512439} \BibitemShut
  {NoStop}%
\bibitem [{\citenamefont {Heurtier}\ \emph {et~al.}(2023)\citenamefont
  {Heurtier}, \citenamefont {Moursy},\ and\ \citenamefont
  {Wacquez}}]{Heurtier:2022rhf}%
  \BibitemOpen
  \bibfield  {author} {\bibinfo {author} {\bibfnamefont {L.}~\bibnamefont
  {Heurtier}}, \bibinfo {author} {\bibfnamefont {A.}~\bibnamefont {Moursy}}, \
  and\ \bibinfo {author} {\bibfnamefont {L.}~\bibnamefont {Wacquez}},\ }\href
  {\doibase 10.1088/1475-7516/2023/03/020} {\bibfield  {journal} {\bibinfo
  {journal} {JCAP}\ }\textbf {\bibinfo {volume} {03}},\ \bibinfo {pages} {020}
  (\bibinfo {year} {2023})},\ \Eprint {http://arxiv.org/abs/2207.11502}
  {arXiv:2207.11502 [hep-th]} \BibitemShut {NoStop}%
\bibitem [{\citenamefont {Dimopoulos}(2017)}]{Dimopoulos:2017ged}%
  \BibitemOpen
  \bibfield  {author} {\bibinfo {author} {\bibfnamefont {K.}~\bibnamefont
  {Dimopoulos}},\ }\href {\doibase 10.1016/j.physletb.2017.10.066} {\bibfield
  {journal} {\bibinfo  {journal} {Phys. Lett. B}\ }\textbf {\bibinfo {volume}
  {775}},\ \bibinfo {pages} {262} (\bibinfo {year} {2017})},\ \Eprint
  {http://arxiv.org/abs/1707.05644} {arXiv:1707.05644 [hep-ph]} \BibitemShut
  {NoStop}%
\bibitem [{\citenamefont {Ballesteros}\ and\ \citenamefont
  {Taoso}(2018)}]{Ballesteros:2017fsr}%
  \BibitemOpen
  \bibfield  {author} {\bibinfo {author} {\bibfnamefont {G.}~\bibnamefont
  {Ballesteros}}\ and\ \bibinfo {author} {\bibfnamefont {M.}~\bibnamefont
  {Taoso}},\ }\href {\doibase 10.1103/PhysRevD.97.023501} {\bibfield  {journal}
  {\bibinfo  {journal} {Phys. Rev. D}\ }\textbf {\bibinfo {volume} {97}},\
  \bibinfo {pages} {023501} (\bibinfo {year} {2018})},\ \Eprint
  {http://arxiv.org/abs/1709.05565} {arXiv:1709.05565 [hep-ph]} \BibitemShut
  {NoStop}%
\bibitem [{\citenamefont {Karam}\ \emph {et~al.}(2023)\citenamefont {Karam},
  \citenamefont {Koivunen}, \citenamefont {Tomberg}, \citenamefont {Vaskonen},\
  and\ \citenamefont {Veerm{\"a}e}}]{Karam:2022nym}%
  \BibitemOpen
  \bibfield  {author} {\bibinfo {author} {\bibfnamefont {A.}~\bibnamefont
  {Karam}}, \bibinfo {author} {\bibfnamefont {N.}~\bibnamefont {Koivunen}},
  \bibinfo {author} {\bibfnamefont {E.}~\bibnamefont {Tomberg}}, \bibinfo
  {author} {\bibfnamefont {V.}~\bibnamefont {Vaskonen}}, \ and\ \bibinfo
  {author} {\bibfnamefont {H.}~\bibnamefont {Veerm{\"a}e}},\ }\href {\doibase
  10.1088/1475-7516/2023/03/013} {\bibfield  {journal} {\bibinfo  {journal}
  {JCAP}\ }\textbf {\bibinfo {volume} {03}},\ \bibinfo {pages} {013} (\bibinfo
  {year} {2023})},\ \Eprint {http://arxiv.org/abs/2205.13540} {arXiv:2205.13540
  [astro-ph.CO]} \BibitemShut {NoStop}%
\bibitem [{\citenamefont {Dalianis}\ \emph {et~al.}(2019)\citenamefont
  {Dalianis}, \citenamefont {Kehagias},\ and\ \citenamefont
  {Tringas}}]{Dalianis:2018frf}%
  \BibitemOpen
  \bibfield  {author} {\bibinfo {author} {\bibfnamefont {I.}~\bibnamefont
  {Dalianis}}, \bibinfo {author} {\bibfnamefont {A.}~\bibnamefont {Kehagias}},
  \ and\ \bibinfo {author} {\bibfnamefont {G.}~\bibnamefont {Tringas}},\ }\href
  {\doibase 10.1088/1475-7516/2019/01/037} {\bibfield  {journal} {\bibinfo
  {journal} {JCAP}\ }\textbf {\bibinfo {volume} {01}},\ \bibinfo {pages} {037}
  (\bibinfo {year} {2019})},\ \Eprint {http://arxiv.org/abs/1805.09483}
  {arXiv:1805.09483 [astro-ph.CO]} \BibitemShut {NoStop}%
\bibitem [{\citenamefont {Kannike}\ \emph {et~al.}(2017)\citenamefont
  {Kannike}, \citenamefont {Marzola}, \citenamefont {Raidal},\ and\
  \citenamefont {Veerm\"ae}}]{Kannike:2017bxn}%
  \BibitemOpen
  \bibfield  {author} {\bibinfo {author} {\bibfnamefont {K.}~\bibnamefont
  {Kannike}}, \bibinfo {author} {\bibfnamefont {L.}~\bibnamefont {Marzola}},
  \bibinfo {author} {\bibfnamefont {M.}~\bibnamefont {Raidal}}, \ and\ \bibinfo
  {author} {\bibfnamefont {H.}~\bibnamefont {Veerm\"ae}},\ }\href {\doibase
  10.1088/1475-7516/2017/09/020} {\bibfield  {journal} {\bibinfo  {journal}
  {JCAP}\ }\textbf {\bibinfo {volume} {09}},\ \bibinfo {pages} {020} (\bibinfo
  {year} {2017})},\ \Eprint {http://arxiv.org/abs/1705.06225} {arXiv:1705.06225
  [astro-ph.CO]} \BibitemShut {NoStop}%
\bibitem [{\citenamefont {Crawford}\ and\ \citenamefont
  {Schramm}(1982)}]{Crawford:1982yz}%
  \BibitemOpen
  \bibfield  {author} {\bibinfo {author} {\bibfnamefont {M.}~\bibnamefont
  {Crawford}}\ and\ \bibinfo {author} {\bibfnamefont {D.~N.}\ \bibnamefont
  {Schramm}},\ }\href {\doibase 10.1038/298538a0} {\bibfield  {journal}
  {\bibinfo  {journal} {Nature}\ }\textbf {\bibinfo {volume} {298}},\ \bibinfo
  {pages} {538} (\bibinfo {year} {1982})}\BibitemShut {NoStop}%
\bibitem [{\citenamefont {Jedamzik}(1997)}]{Jedamzik:1996mr}%
  \BibitemOpen
  \bibfield  {author} {\bibinfo {author} {\bibfnamefont {K.}~\bibnamefont
  {Jedamzik}},\ }\href {\doibase 10.1103/PhysRevD.55.R5871} {\bibfield
  {journal} {\bibinfo  {journal} {Phys. Rev. D}\ }\textbf {\bibinfo {volume}
  {55}},\ \bibinfo {pages} {5871} (\bibinfo {year} {1997})},\ \Eprint
  {http://arxiv.org/abs/astro-ph/9605152} {arXiv:astro-ph/9605152} \BibitemShut
  {NoStop}%
\bibitem [{\citenamefont {Schmid}\ \emph {et~al.}(1999)\citenamefont {Schmid},
  \citenamefont {Schwarz},\ and\ \citenamefont {Widerin}}]{Schmid:1998mx}%
  \BibitemOpen
  \bibfield  {author} {\bibinfo {author} {\bibfnamefont {C.}~\bibnamefont
  {Schmid}}, \bibinfo {author} {\bibfnamefont {D.~J.}\ \bibnamefont {Schwarz}},
  \ and\ \bibinfo {author} {\bibfnamefont {P.}~\bibnamefont {Widerin}},\ }\href
  {\doibase 10.1103/PhysRevD.59.043517} {\bibfield  {journal} {\bibinfo
  {journal} {Phys. Rev. D}\ }\textbf {\bibinfo {volume} {59}},\ \bibinfo
  {pages} {043517} (\bibinfo {year} {1999})},\ \Eprint
  {http://arxiv.org/abs/astro-ph/9807257} {arXiv:astro-ph/9807257} \BibitemShut
  {NoStop}%
\bibitem [{\citenamefont {Widerin}\ and\ \citenamefont
  {Schmid}(1998)}]{Widerin:1998my}%
  \BibitemOpen
  \bibfield  {author} {\bibinfo {author} {\bibfnamefont {P.}~\bibnamefont
  {Widerin}}\ and\ \bibinfo {author} {\bibfnamefont {C.}~\bibnamefont
  {Schmid}},\ }\href@noop {} {\  (\bibinfo {year} {1998})},\ \Eprint
  {http://arxiv.org/abs/astro-ph/9808142} {arXiv:astro-ph/9808142} \BibitemShut
  {NoStop}%
\bibitem [{\citenamefont {Hawking}\ \emph {et~al.}(1982)\citenamefont
  {Hawking}, \citenamefont {Moss},\ and\ \citenamefont
  {Stewart}}]{Hawking:1982ga}%
  \BibitemOpen
  \bibfield  {author} {\bibinfo {author} {\bibfnamefont {S.~W.}\ \bibnamefont
  {Hawking}}, \bibinfo {author} {\bibfnamefont {I.~G.}\ \bibnamefont {Moss}}, \
  and\ \bibinfo {author} {\bibfnamefont {J.~M.}\ \bibnamefont {Stewart}},\
  }\href {\doibase 10.1103/PhysRevD.26.2681} {\bibfield  {journal} {\bibinfo
  {journal} {Phys. Rev. D}\ }\textbf {\bibinfo {volume} {26}},\ \bibinfo
  {pages} {2681} (\bibinfo {year} {1982})}\BibitemShut {NoStop}%
\bibitem [{\citenamefont {Kodama}\ \emph {et~al.}(1982)\citenamefont {Kodama},
  \citenamefont {Sasaki},\ and\ \citenamefont {Sato}}]{Kodama:1982sf}%
  \BibitemOpen
  \bibfield  {author} {\bibinfo {author} {\bibfnamefont {H.}~\bibnamefont
  {Kodama}}, \bibinfo {author} {\bibfnamefont {M.}~\bibnamefont {Sasaki}}, \
  and\ \bibinfo {author} {\bibfnamefont {K.}~\bibnamefont {Sato}},\ }\href
  {\doibase 10.1143/PTP.68.1979} {\bibfield  {journal} {\bibinfo  {journal}
  {Prog. Theor. Phys.}\ }\textbf {\bibinfo {volume} {68}},\ \bibinfo {pages}
  {1979} (\bibinfo {year} {1982})}\BibitemShut {NoStop}%
\bibitem [{\citenamefont {Leach}\ \emph {et~al.}(2000)\citenamefont {Leach},
  \citenamefont {Grivell},\ and\ \citenamefont {Liddle}}]{Leach:2000ea}%
  \BibitemOpen
  \bibfield  {author} {\bibinfo {author} {\bibfnamefont {S.~M.}\ \bibnamefont
  {Leach}}, \bibinfo {author} {\bibfnamefont {I.~J.}\ \bibnamefont {Grivell}},
  \ and\ \bibinfo {author} {\bibfnamefont {A.~R.}\ \bibnamefont {Liddle}},\
  }\href {\doibase 10.1103/PhysRevD.62.043516} {\bibfield  {journal} {\bibinfo
  {journal} {Phys. Rev. D}\ }\textbf {\bibinfo {volume} {62}},\ \bibinfo
  {pages} {043516} (\bibinfo {year} {2000})},\ \Eprint
  {http://arxiv.org/abs/astro-ph/0004296} {arXiv:astro-ph/0004296} \BibitemShut
  {NoStop}%
\bibitem [{\citenamefont {Moss}(1994)}]{Moss:1994iq}%
  \BibitemOpen
  \bibfield  {author} {\bibinfo {author} {\bibfnamefont {I.~G.}\ \bibnamefont
  {Moss}},\ }\href {\doibase 10.1103/PhysRevD.50.676} {\bibfield  {journal}
  {\bibinfo  {journal} {Phys. Rev. D}\ }\textbf {\bibinfo {volume} {50}},\
  \bibinfo {pages} {676} (\bibinfo {year} {1994})}\BibitemShut {NoStop}%
\bibitem [{\citenamefont {Kitajima}\ and\ \citenamefont
  {Takahashi}(2020)}]{Kitajima:2020kig}%
  \BibitemOpen
  \bibfield  {author} {\bibinfo {author} {\bibfnamefont {N.}~\bibnamefont
  {Kitajima}}\ and\ \bibinfo {author} {\bibfnamefont {F.}~\bibnamefont
  {Takahashi}},\ }\href {\doibase 10.1088/1475-7516/2020/11/060} {\bibfield
  {journal} {\bibinfo  {journal} {JCAP}\ }\textbf {\bibinfo {volume} {11}},\
  \bibinfo {pages} {060} (\bibinfo {year} {2020})},\ \Eprint
  {http://arxiv.org/abs/2006.13137} {arXiv:2006.13137 [hep-ph]} \BibitemShut
  {NoStop}%
\bibitem [{\citenamefont {Kodama}\ \emph {et~al.}(1981)\citenamefont {Kodama},
  \citenamefont {Sasaki}, \citenamefont {Sato},\ and\ \citenamefont
  {Maeda}}]{Kodama:1981gu}%
  \BibitemOpen
  \bibfield  {author} {\bibinfo {author} {\bibfnamefont {H.}~\bibnamefont
  {Kodama}}, \bibinfo {author} {\bibfnamefont {M.}~\bibnamefont {Sasaki}},
  \bibinfo {author} {\bibfnamefont {K.}~\bibnamefont {Sato}}, \ and\ \bibinfo
  {author} {\bibfnamefont {K.-i.}\ \bibnamefont {Maeda}},\ }\href {\doibase
  10.1143/PTP.66.2052} {\bibfield  {journal} {\bibinfo  {journal} {Prog. Theor.
  Phys.}\ }\textbf {\bibinfo {volume} {66}},\ \bibinfo {pages} {2052} (\bibinfo
  {year} {1981})}\BibitemShut {NoStop}%
\bibitem [{\citenamefont {Maeda}(1986)}]{Maeda:1985gz}%
  \BibitemOpen
  \bibfield  {author} {\bibinfo {author} {\bibfnamefont {K.-i.}\ \bibnamefont
  {Maeda}},\ }\href {\doibase 10.1088/0264-9381/3/4/019} {\bibfield  {journal}
  {\bibinfo  {journal} {Class. Quant. Grav.}\ }\textbf {\bibinfo {volume}
  {3}},\ \bibinfo {pages} {651} (\bibinfo {year} {1986})}\BibitemShut {NoStop}%
\bibitem [{\citenamefont {Khlopov}\ \emph {et~al.}(1998)\citenamefont
  {Khlopov}, \citenamefont {Konoplich}, \citenamefont {Rubin},\ and\
  \citenamefont {Sakharov}}]{Khlopov:1998nm}%
  \BibitemOpen
  \bibfield  {author} {\bibinfo {author} {\bibfnamefont {M.~Y.}\ \bibnamefont
  {Khlopov}}, \bibinfo {author} {\bibfnamefont {R.~V.}\ \bibnamefont
  {Konoplich}}, \bibinfo {author} {\bibfnamefont {S.~G.}\ \bibnamefont
  {Rubin}}, \ and\ \bibinfo {author} {\bibfnamefont {A.~S.}\ \bibnamefont
  {Sakharov}},\ }\href@noop {} {\  (\bibinfo {year} {1998})},\ \Eprint
  {http://arxiv.org/abs/hep-ph/9807343} {arXiv:hep-ph/9807343} \BibitemShut
  {NoStop}%
\bibitem [{\citenamefont {Konoplich}\ \emph {et~al.}(1998)\citenamefont
  {Konoplich}, \citenamefont {Rubin}, \citenamefont {Sakharov},\ and\
  \citenamefont {Khlopov}}]{Konoplich:1998ugi}%
  \BibitemOpen
  \bibfield  {author} {\bibinfo {author} {\bibfnamefont {R.~V.}\ \bibnamefont
  {Konoplich}}, \bibinfo {author} {\bibfnamefont {S.~G.}\ \bibnamefont
  {Rubin}}, \bibinfo {author} {\bibfnamefont {A.~S.}\ \bibnamefont {Sakharov}},
  \ and\ \bibinfo {author} {\bibfnamefont {M.~Y.}\ \bibnamefont {Khlopov}},\
  }\href@noop {} {\bibfield  {journal} {\bibinfo  {journal} {Astron. Lett.}\
  }\textbf {\bibinfo {volume} {24}},\ \bibinfo {pages} {413} (\bibinfo {year}
  {1998})}\BibitemShut {NoStop}%
\bibitem [{\citenamefont {Pavsic}(1996)}]{Pavsic:1995jt}%
  \BibitemOpen
  \bibfield  {author} {\bibinfo {author} {\bibfnamefont {M.}~\bibnamefont
  {Pavsic}},\ }\href@noop {} {\bibfield  {journal} {\bibinfo  {journal} {Grav.
  Cosmol.}\ }\textbf {\bibinfo {volume} {2}},\ \bibinfo {pages} {1} (\bibinfo
  {year} {1996})},\ \Eprint {http://arxiv.org/abs/gr-qc/9511020}
  {arXiv:gr-qc/9511020} \BibitemShut {NoStop}%
\bibitem [{\citenamefont {Khlopov}\ \emph {et~al.}(2000)\citenamefont
  {Khlopov}, \citenamefont {Konoplich}, \citenamefont {Rubin},\ and\
  \citenamefont {Sakharov}}]{Khlopov:2000js}%
  \BibitemOpen
  \bibfield  {author} {\bibinfo {author} {\bibfnamefont {M.~Y.}\ \bibnamefont
  {Khlopov}}, \bibinfo {author} {\bibfnamefont {R.~V.}\ \bibnamefont
  {Konoplich}}, \bibinfo {author} {\bibfnamefont {S.~G.}\ \bibnamefont
  {Rubin}}, \ and\ \bibinfo {author} {\bibfnamefont {A.~S.}\ \bibnamefont
  {Sakharov}},\ }\href@noop {} {\bibfield  {journal} {\bibinfo  {journal}
  {Grav. Cosmol.}\ }\textbf {\bibinfo {volume} {6}},\ \bibinfo {pages} {153}
  (\bibinfo {year} {2000})}\BibitemShut {NoStop}%
\bibitem [{\citenamefont {Dokuchaev}\ \emph {et~al.}(2005)\citenamefont
  {Dokuchaev}, \citenamefont {Eroshenko},\ and\ \citenamefont
  {Rubin}}]{Dokuchaev:2004kr}%
  \BibitemOpen
  \bibfield  {author} {\bibinfo {author} {\bibfnamefont {V.}~\bibnamefont
  {Dokuchaev}}, \bibinfo {author} {\bibfnamefont {Y.}~\bibnamefont
  {Eroshenko}}, \ and\ \bibinfo {author} {\bibfnamefont {S.}~\bibnamefont
  {Rubin}},\ }\href@noop {} {\bibfield  {journal} {\bibinfo  {journal} {Grav.
  Cosmol.}\ }\textbf {\bibinfo {volume} {11}},\ \bibinfo {pages} {99} (\bibinfo
  {year} {2005})},\ \Eprint {http://arxiv.org/abs/astro-ph/0412418}
  {arXiv:astro-ph/0412418} \BibitemShut {NoStop}%
\bibitem [{\citenamefont {Rubin}\ \emph {et~al.}(2001)\citenamefont {Rubin},
  \citenamefont {Sakharov},\ and\ \citenamefont {Khlopov}}]{Rubin:2001yw}%
  \BibitemOpen
  \bibfield  {author} {\bibinfo {author} {\bibfnamefont {S.~G.}\ \bibnamefont
  {Rubin}}, \bibinfo {author} {\bibfnamefont {A.~S.}\ \bibnamefont {Sakharov}},
  \ and\ \bibinfo {author} {\bibfnamefont {M.~Y.}\ \bibnamefont {Khlopov}},\
  }\href {\doibase 10.1134/1.1385631} {\bibfield  {journal} {\bibinfo
  {journal} {J. Exp. Theor. Phys.}\ }\textbf {\bibinfo {volume} {91}},\
  \bibinfo {pages} {921} (\bibinfo {year} {2001})},\ \Eprint
  {http://arxiv.org/abs/hep-ph/0106187} {arXiv:hep-ph/0106187} \BibitemShut
  {NoStop}%
\bibitem [{\citenamefont {Garriga}\ \emph {et~al.}(2016)\citenamefont
  {Garriga}, \citenamefont {Vilenkin},\ and\ \citenamefont
  {Zhang}}]{Garriga:2015fdk}%
  \BibitemOpen
  \bibfield  {author} {\bibinfo {author} {\bibfnamefont {J.}~\bibnamefont
  {Garriga}}, \bibinfo {author} {\bibfnamefont {A.}~\bibnamefont {Vilenkin}}, \
  and\ \bibinfo {author} {\bibfnamefont {J.}~\bibnamefont {Zhang}},\ }\href
  {\doibase 10.1088/1475-7516/2016/02/064} {\bibfield  {journal} {\bibinfo
  {journal} {JCAP}\ }\textbf {\bibinfo {volume} {02}},\ \bibinfo {pages} {064}
  (\bibinfo {year} {2016})},\ \Eprint {http://arxiv.org/abs/1512.01819}
  {arXiv:1512.01819 [hep-th]} \BibitemShut {NoStop}%
\bibitem [{\citenamefont {Hawking}(1989)}]{Hawking:1987bn}%
  \BibitemOpen
  \bibfield  {author} {\bibinfo {author} {\bibfnamefont {S.~W.}\ \bibnamefont
  {Hawking}},\ }\href {\doibase 10.1016/0370-2693(89)90206-2} {\bibfield
  {journal} {\bibinfo  {journal} {Phys. Lett. B}\ }\textbf {\bibinfo {volume}
  {231}},\ \bibinfo {pages} {237} (\bibinfo {year} {1989})}\BibitemShut
  {NoStop}%
\bibitem [{\citenamefont {Polnarev}\ and\ \citenamefont
  {Zembowicz}(1991)}]{Polnarev:1988dh}%
  \BibitemOpen
  \bibfield  {author} {\bibinfo {author} {\bibfnamefont {A.}~\bibnamefont
  {Polnarev}}\ and\ \bibinfo {author} {\bibfnamefont {R.}~\bibnamefont
  {Zembowicz}},\ }\href {\doibase 10.1103/PhysRevD.43.1106} {\bibfield
  {journal} {\bibinfo  {journal} {Phys. Rev. D}\ }\textbf {\bibinfo {volume}
  {43}},\ \bibinfo {pages} {1106} (\bibinfo {year} {1991})}\BibitemShut
  {NoStop}%
\bibitem [{\citenamefont {Garriga}\ and\ \citenamefont
  {Sakellariadou}(1993)}]{Garriga:1993gj}%
  \BibitemOpen
  \bibfield  {author} {\bibinfo {author} {\bibfnamefont {J.}~\bibnamefont
  {Garriga}}\ and\ \bibinfo {author} {\bibfnamefont {M.}~\bibnamefont
  {Sakellariadou}},\ }\href {\doibase 10.1103/PhysRevD.48.2502} {\bibfield
  {journal} {\bibinfo  {journal} {Phys. Rev. D}\ }\textbf {\bibinfo {volume}
  {48}},\ \bibinfo {pages} {2502} (\bibinfo {year} {1993})},\ \Eprint
  {http://arxiv.org/abs/hep-th/9303024} {arXiv:hep-th/9303024} \BibitemShut
  {NoStop}%
\bibitem [{\citenamefont {Caldwell}\ and\ \citenamefont
  {Casper}(1996)}]{Caldwell:1995fu}%
  \BibitemOpen
  \bibfield  {author} {\bibinfo {author} {\bibfnamefont {R.~R.}\ \bibnamefont
  {Caldwell}}\ and\ \bibinfo {author} {\bibfnamefont {P.}~\bibnamefont
  {Casper}},\ }\href {\doibase 10.1103/PhysRevD.53.3002} {\bibfield  {journal}
  {\bibinfo  {journal} {Phys. Rev. D}\ }\textbf {\bibinfo {volume} {53}},\
  \bibinfo {pages} {3002} (\bibinfo {year} {1996})},\ \Eprint
  {http://arxiv.org/abs/gr-qc/9509012} {arXiv:gr-qc/9509012} \BibitemShut
  {NoStop}%
\bibitem [{\citenamefont {MacGibbon}\ \emph {et~al.}(1998)\citenamefont
  {MacGibbon}, \citenamefont {Brandenberger},\ and\ \citenamefont
  {Wichoski}}]{MacGibbon:1997pu}%
  \BibitemOpen
  \bibfield  {author} {\bibinfo {author} {\bibfnamefont {J.~H.}\ \bibnamefont
  {MacGibbon}}, \bibinfo {author} {\bibfnamefont {R.~H.}\ \bibnamefont
  {Brandenberger}}, \ and\ \bibinfo {author} {\bibfnamefont {U.~F.}\
  \bibnamefont {Wichoski}},\ }\href {\doibase 10.1103/PhysRevD.57.2158}
  {\bibfield  {journal} {\bibinfo  {journal} {Phys. Rev. D}\ }\textbf {\bibinfo
  {volume} {57}},\ \bibinfo {pages} {2158} (\bibinfo {year} {1998})},\ \Eprint
  {http://arxiv.org/abs/astro-ph/9707146} {arXiv:astro-ph/9707146} \BibitemShut
  {NoStop}%
\bibitem [{\citenamefont {Jenkins}\ and\ \citenamefont
  {Sakellariadou}(2020)}]{Jenkins:2020ctp}%
  \BibitemOpen
  \bibfield  {author} {\bibinfo {author} {\bibfnamefont {A.~C.}\ \bibnamefont
  {Jenkins}}\ and\ \bibinfo {author} {\bibfnamefont {M.}~\bibnamefont
  {Sakellariadou}},\ }\href@noop {} {\  (\bibinfo {year} {2020})},\ \Eprint
  {http://arxiv.org/abs/2006.16249} {arXiv:2006.16249 [astro-ph.CO]}
  \BibitemShut {NoStop}%
\bibitem [{\citenamefont {Martin}\ \emph
  {et~al.}(2020{\natexlab{a}})\citenamefont {Martin}, \citenamefont
  {Papanikolaou},\ and\ \citenamefont {Vennin}}]{Martin:2019nuw}%
  \BibitemOpen
  \bibfield  {author} {\bibinfo {author} {\bibfnamefont {J.}~\bibnamefont
  {Martin}}, \bibinfo {author} {\bibfnamefont {T.}~\bibnamefont
  {Papanikolaou}}, \ and\ \bibinfo {author} {\bibfnamefont {V.}~\bibnamefont
  {Vennin}},\ }\href {\doibase 10.1088/1475-7516/2020/01/024} {\bibfield
  {journal} {\bibinfo  {journal} {JCAP}\ }\textbf {\bibinfo {volume} {01}},\
  \bibinfo {pages} {024} (\bibinfo {year} {2020}{\natexlab{a}})},\ \Eprint
  {http://arxiv.org/abs/1907.04236} {arXiv:1907.04236 [astro-ph.CO]}
  \BibitemShut {NoStop}%
\bibitem [{\citenamefont {Martin}\ \emph
  {et~al.}(2020{\natexlab{b}})\citenamefont {Martin}, \citenamefont
  {Papanikolaou}, \citenamefont {Pinol},\ and\ \citenamefont
  {Vennin}}]{Martin:2020fgl}%
  \BibitemOpen
  \bibfield  {author} {\bibinfo {author} {\bibfnamefont {J.}~\bibnamefont
  {Martin}}, \bibinfo {author} {\bibfnamefont {T.}~\bibnamefont
  {Papanikolaou}}, \bibinfo {author} {\bibfnamefont {L.}~\bibnamefont {Pinol}},
  \ and\ \bibinfo {author} {\bibfnamefont {V.}~\bibnamefont {Vennin}},\ }\href
  {\doibase 10.1088/1475-7516/2020/05/003} {\bibfield  {journal} {\bibinfo
  {journal} {JCAP}\ }\textbf {\bibinfo {volume} {05}},\ \bibinfo {pages} {003}
  (\bibinfo {year} {2020}{\natexlab{b}})},\ \Eprint
  {http://arxiv.org/abs/2002.01820} {arXiv:2002.01820 [astro-ph.CO]}
  \BibitemShut {NoStop}%
\bibitem [{\citenamefont {Green}\ and\ \citenamefont
  {Liddle}(1997)}]{Green:1997sz}%
  \BibitemOpen
  \bibfield  {author} {\bibinfo {author} {\bibfnamefont {A.~M.}\ \bibnamefont
  {Green}}\ and\ \bibinfo {author} {\bibfnamefont {A.~R.}\ \bibnamefont
  {Liddle}},\ }\href {\doibase 10.1103/PhysRevD.56.6166} {\bibfield  {journal}
  {\bibinfo  {journal} {Phys. Rev. D}\ }\textbf {\bibinfo {volume} {56}},\
  \bibinfo {pages} {6166} (\bibinfo {year} {1997})},\ \Eprint
  {http://arxiv.org/abs/astro-ph/9704251} {arXiv:astro-ph/9704251} \BibitemShut
  {NoStop}%
\bibitem [{\citenamefont {Bringmann}\ \emph {et~al.}(2002)\citenamefont
  {Bringmann}, \citenamefont {Kiefer},\ and\ \citenamefont
  {Polarski}}]{Bringmann:2001yp}%
  \BibitemOpen
  \bibfield  {author} {\bibinfo {author} {\bibfnamefont {T.}~\bibnamefont
  {Bringmann}}, \bibinfo {author} {\bibfnamefont {C.}~\bibnamefont {Kiefer}}, \
  and\ \bibinfo {author} {\bibfnamefont {D.}~\bibnamefont {Polarski}},\ }\href
  {\doibase 10.1103/PhysRevD.65.024008} {\bibfield  {journal} {\bibinfo
  {journal} {Phys. Rev. D}\ }\textbf {\bibinfo {volume} {65}},\ \bibinfo
  {pages} {024008} (\bibinfo {year} {2002})},\ \Eprint
  {http://arxiv.org/abs/astro-ph/0109404} {arXiv:astro-ph/0109404} \BibitemShut
  {NoStop}%
\bibitem [{\citenamefont {Kim}\ \emph {et~al.}(1999)\citenamefont {Kim},
  \citenamefont {Lee},\ and\ \citenamefont {MacGibbon}}]{Kim:1999iv}%
  \BibitemOpen
  \bibfield  {author} {\bibinfo {author} {\bibfnamefont {H.~I.}\ \bibnamefont
  {Kim}}, \bibinfo {author} {\bibfnamefont {C.~H.}\ \bibnamefont {Lee}}, \ and\
  \bibinfo {author} {\bibfnamefont {J.~H.}\ \bibnamefont {MacGibbon}},\ }\href
  {\doibase 10.1103/PhysRevD.59.063004} {\bibfield  {journal} {\bibinfo
  {journal} {Phys. Rev. D}\ }\textbf {\bibinfo {volume} {59}},\ \bibinfo
  {pages} {063004} (\bibinfo {year} {1999})},\ \Eprint
  {http://arxiv.org/abs/astro-ph/9901030} {arXiv:astro-ph/9901030} \BibitemShut
  {NoStop}%
\bibitem [{\citenamefont {Carr}\ \emph {et~al.}(2017)\citenamefont {Carr},
  \citenamefont {Raidal}, \citenamefont {Tenkanen}, \citenamefont {Vaskonen},\
  and\ \citenamefont {Veerm\"ae}}]{Carr:2017jsz}%
  \BibitemOpen
  \bibfield  {author} {\bibinfo {author} {\bibfnamefont {B.}~\bibnamefont
  {Carr}}, \bibinfo {author} {\bibfnamefont {M.}~\bibnamefont {Raidal}},
  \bibinfo {author} {\bibfnamefont {T.}~\bibnamefont {Tenkanen}}, \bibinfo
  {author} {\bibfnamefont {V.}~\bibnamefont {Vaskonen}}, \ and\ \bibinfo
  {author} {\bibfnamefont {H.}~\bibnamefont {Veerm\"ae}},\ }\href {\doibase
  10.1103/PhysRevD.96.023514} {\bibfield  {journal} {\bibinfo  {journal} {Phys.
  Rev. D}\ }\textbf {\bibinfo {volume} {96}},\ \bibinfo {pages} {023514}
  (\bibinfo {year} {2017})},\ \Eprint {http://arxiv.org/abs/1705.05567}
  {arXiv:1705.05567 [astro-ph.CO]} \BibitemShut {NoStop}%
\bibitem [{\citenamefont {Dienes}\ \emph {et~al.}(2019)\citenamefont {Dienes},
  \citenamefont {Kumar}, \citenamefont {Stengel},\ and\ \citenamefont
  {Thomas}}]{Dienes:2018yoq}%
  \BibitemOpen
  \bibfield  {author} {\bibinfo {author} {\bibfnamefont {K.~R.}\ \bibnamefont
  {Dienes}}, \bibinfo {author} {\bibfnamefont {J.}~\bibnamefont {Kumar}},
  \bibinfo {author} {\bibfnamefont {P.}~\bibnamefont {Stengel}}, \ and\
  \bibinfo {author} {\bibfnamefont {B.}~\bibnamefont {Thomas}},\ }\href
  {\doibase 10.1103/PhysRevD.99.043513} {\bibfield  {journal} {\bibinfo
  {journal} {Phys. Rev. D}\ }\textbf {\bibinfo {volume} {99}},\ \bibinfo
  {pages} {043513} (\bibinfo {year} {2019})},\ \Eprint
  {http://arxiv.org/abs/1810.10587} {arXiv:1810.10587 [hep-ph]} \BibitemShut
  {NoStop}%
\bibitem [{\citenamefont {Akrami}\ \emph
  {et~al.}(2020{\natexlab{a}})\citenamefont {Akrami} \emph
  {et~al.}}]{Akrami:2018odb}%
  \BibitemOpen
  \bibfield  {author} {\bibinfo {author} {\bibfnamefont {Y.}~\bibnamefont
  {Akrami}} \emph {et~al.} (\bibinfo {collaboration} {Planck}),\ }\href
  {\doibase 10.1051/0004-6361/201833887} {\bibfield  {journal} {\bibinfo
  {journal} {Astron. Astrophys.}\ }\textbf {\bibinfo {volume} {641}},\ \bibinfo
  {pages} {A10} (\bibinfo {year} {2020}{\natexlab{a}})},\ \Eprint
  {http://arxiv.org/abs/1807.06211} {arXiv:1807.06211 [astro-ph.CO]}
  \BibitemShut {NoStop}%
\bibitem [{\citenamefont {Cole}\ \emph {et~al.}(2024)\citenamefont {Cole},
  \citenamefont {Gow}, \citenamefont {Byrnes},\ and\ \citenamefont
  {Patil}}]{Cole:2022xqc}%
  \BibitemOpen
  \bibfield  {author} {\bibinfo {author} {\bibfnamefont {P.~S.}\ \bibnamefont
  {Cole}}, \bibinfo {author} {\bibfnamefont {A.~D.}\ \bibnamefont {Gow}},
  \bibinfo {author} {\bibfnamefont {C.~T.}\ \bibnamefont {Byrnes}}, \ and\
  \bibinfo {author} {\bibfnamefont {S.~P.}\ \bibnamefont {Patil}},\ }\href
  {\doibase 10.1088/1475-7516/2024/05/022} {\bibfield  {journal} {\bibinfo
  {journal} {JCAP}\ }\textbf {\bibinfo {volume} {05}},\ \bibinfo {pages} {022}
  (\bibinfo {year} {2024})},\ \Eprint {http://arxiv.org/abs/2204.07573}
  {arXiv:2204.07573 [astro-ph.CO]} \BibitemShut {NoStop}%
\bibitem [{\citenamefont {MacGibbon}\ and\ \citenamefont
  {Webber}(1990)}]{PhysRevD.41.3052}%
  \BibitemOpen
  \bibfield  {author} {\bibinfo {author} {\bibfnamefont {J.~H.}\ \bibnamefont
  {MacGibbon}}\ and\ \bibinfo {author} {\bibfnamefont {B.~R.}\ \bibnamefont
  {Webber}},\ }\href {\doibase 10.1103/PhysRevD.41.3052} {\bibfield  {journal}
  {\bibinfo  {journal} {Phys. Rev. D}\ }\textbf {\bibinfo {volume} {41}},\
  \bibinfo {pages} {3052} (\bibinfo {year} {1990})}\BibitemShut {NoStop}%
\bibitem [{\citenamefont {MacGibbon}(1991)}]{PhysRevD.44.376}%
  \BibitemOpen
  \bibfield  {author} {\bibinfo {author} {\bibfnamefont {J.~H.}\ \bibnamefont
  {MacGibbon}},\ }\href {\doibase 10.1103/PhysRevD.44.376} {\bibfield
  {journal} {\bibinfo  {journal} {Phys. Rev. D}\ }\textbf {\bibinfo {volume}
  {44}},\ \bibinfo {pages} {376} (\bibinfo {year} {1991})}\BibitemShut
  {NoStop}%
\bibitem [{\citenamefont {Cheek}\ \emph {et~al.}(2023)\citenamefont {Cheek},
  \citenamefont {Heurtier}, \citenamefont {Perez-Gonzalez},\ and\ \citenamefont
  {Turner}}]{Cheek:2022mmy}%
  \BibitemOpen
  \bibfield  {author} {\bibinfo {author} {\bibfnamefont {A.}~\bibnamefont
  {Cheek}}, \bibinfo {author} {\bibfnamefont {L.}~\bibnamefont {Heurtier}},
  \bibinfo {author} {\bibfnamefont {Y.~F.}\ \bibnamefont {Perez-Gonzalez}}, \
  and\ \bibinfo {author} {\bibfnamefont {J.}~\bibnamefont {Turner}},\ }\href
  {\doibase 10.1103/PhysRevD.108.015005} {\bibfield  {journal} {\bibinfo
  {journal} {Phys. Rev. D}\ }\textbf {\bibinfo {volume} {108}},\ \bibinfo
  {pages} {015005} (\bibinfo {year} {2023})},\ \Eprint
  {http://arxiv.org/abs/2212.03878} {arXiv:2212.03878 [hep-ph]} \BibitemShut
  {NoStop}%
\bibitem [{\citenamefont {Cheek}\ \emph {et~al.}()\citenamefont {Cheek},
  \citenamefont {Heurtier}, \citenamefont {Perez-Gonzalez},\ and\ \citenamefont
  {Turner}}]{frisbhee}%
  \BibitemOpen
  \bibfield  {author} {\bibinfo {author} {\bibfnamefont {A.}~\bibnamefont
  {Cheek}}, \bibinfo {author} {\bibfnamefont {L.}~\bibnamefont {Heurtier}},
  \bibinfo {author} {\bibfnamefont {Y.~F.}\ \bibnamefont {Perez-Gonzalez}}, \
  and\ \bibinfo {author} {\bibfnamefont {J.}~\bibnamefont {Turner}},\
  }\href@noop {} {\enquote {\bibinfo {title} {Friedmann solver for black hole
  evaporation in the early universe},}\ }\bibinfo {howpublished}
  {\url{https://github.com/yfperezg/frisbhee}}\BibitemShut {NoStop}%
\bibitem [{\citenamefont {Guth}(1981)}]{Guth:1980zm}%
  \BibitemOpen
  \bibfield  {author} {\bibinfo {author} {\bibfnamefont {A.~H.}\ \bibnamefont
  {Guth}},\ }\href {\doibase 10.1103/PhysRevD.23.347} {\bibfield  {journal}
  {\bibinfo  {journal} {Phys. Rev. D}\ }\textbf {\bibinfo {volume} {23}},\
  \bibinfo {pages} {347} (\bibinfo {year} {1981})}\BibitemShut {NoStop}%
\bibitem [{\citenamefont {Linde}(1982)}]{Linde:1981mu}%
  \BibitemOpen
  \bibfield  {author} {\bibinfo {author} {\bibfnamefont {A.~D.}\ \bibnamefont
  {Linde}},\ }\href {\doibase 10.1016/0370-2693(82)91219-9} {\bibfield
  {journal} {\bibinfo  {journal} {Phys. Lett. B}\ }\textbf {\bibinfo {volume}
  {108}},\ \bibinfo {pages} {389} (\bibinfo {year} {1982})}\BibitemShut
  {NoStop}%
\bibitem [{\citenamefont {Albrecht}\ and\ \citenamefont
  {Steinhardt}(1982)}]{Albrecht:1982wi}%
  \BibitemOpen
  \bibfield  {author} {\bibinfo {author} {\bibfnamefont {A.}~\bibnamefont
  {Albrecht}}\ and\ \bibinfo {author} {\bibfnamefont {P.~J.}\ \bibnamefont
  {Steinhardt}},\ }\href {\doibase 10.1103/PhysRevLett.48.1220} {\bibfield
  {journal} {\bibinfo  {journal} {Phys. Rev. Lett.}\ }\textbf {\bibinfo
  {volume} {48}},\ \bibinfo {pages} {1220} (\bibinfo {year}
  {1982})}\BibitemShut {NoStop}%
\bibitem [{\citenamefont {Akrami}\ \emph
  {et~al.}(2020{\natexlab{b}})\citenamefont {Akrami} \emph
  {et~al.}}]{Planck:2018jri}%
  \BibitemOpen
  \bibfield  {author} {\bibinfo {author} {\bibfnamefont {Y.}~\bibnamefont
  {Akrami}} \emph {et~al.} (\bibinfo {collaboration} {Planck}),\ }\href
  {\doibase 10.1051/0004-6361/201833887} {\bibfield  {journal} {\bibinfo
  {journal} {Astron. Astrophys.}\ }\textbf {\bibinfo {volume} {641}},\ \bibinfo
  {pages} {A10} (\bibinfo {year} {2020}{\natexlab{b}})},\ \Eprint
  {http://arxiv.org/abs/1807.06211} {arXiv:1807.06211 [astro-ph.CO]}
  \BibitemShut {NoStop}%
\bibitem [{\citenamefont {Liddle}\ and\ \citenamefont
  {Leach}(2003)}]{Liddle:2003as}%
  \BibitemOpen
  \bibfield  {author} {\bibinfo {author} {\bibfnamefont {A.~R.}\ \bibnamefont
  {Liddle}}\ and\ \bibinfo {author} {\bibfnamefont {S.~M.}\ \bibnamefont
  {Leach}},\ }\href {\doibase 10.1103/PhysRevD.68.103503} {\bibfield  {journal}
  {\bibinfo  {journal} {Phys. Rev. D}\ }\textbf {\bibinfo {volume} {68}},\
  \bibinfo {pages} {103503} (\bibinfo {year} {2003})},\ \Eprint
  {http://arxiv.org/abs/astro-ph/0305263} {arXiv:astro-ph/0305263} \BibitemShut
  {NoStop}%
\bibitem [{\citenamefont {Allahverdi}\ \emph
  {et~al.}(2018{\natexlab{b}})\citenamefont {Allahverdi}, \citenamefont
  {Dutta},\ and\ \citenamefont {Maharana}}]{Allahverdi:2018iod}%
  \BibitemOpen
  \bibfield  {author} {\bibinfo {author} {\bibfnamefont {R.}~\bibnamefont
  {Allahverdi}}, \bibinfo {author} {\bibfnamefont {K.}~\bibnamefont {Dutta}}, \
  and\ \bibinfo {author} {\bibfnamefont {A.}~\bibnamefont {Maharana}},\ }\href
  {\doibase 10.1088/1475-7516/2018/10/038} {\bibfield  {journal} {\bibinfo
  {journal} {JCAP}\ }\textbf {\bibinfo {volume} {10}},\ \bibinfo {pages} {038}
  (\bibinfo {year} {2018}{\natexlab{b}})},\ \Eprint
  {http://arxiv.org/abs/1808.02659} {arXiv:1808.02659 [astro-ph.CO]}
  \BibitemShut {NoStop}%
\bibitem [{\citenamefont {Kofman}\ \emph {et~al.}(1994)\citenamefont {Kofman},
  \citenamefont {Linde},\ and\ \citenamefont {Starobinsky}}]{Kofman:1994rk}%
  \BibitemOpen
  \bibfield  {author} {\bibinfo {author} {\bibfnamefont {L.}~\bibnamefont
  {Kofman}}, \bibinfo {author} {\bibfnamefont {A.~D.}\ \bibnamefont {Linde}}, \
  and\ \bibinfo {author} {\bibfnamefont {A.~A.}\ \bibnamefont {Starobinsky}},\
  }\href {\doibase 10.1103/PhysRevLett.73.3195} {\bibfield  {journal} {\bibinfo
   {journal} {Phys. Rev. Lett.}\ }\textbf {\bibinfo {volume} {73}},\ \bibinfo
  {pages} {3195} (\bibinfo {year} {1994})},\ \Eprint
  {http://arxiv.org/abs/hep-th/9405187} {arXiv:hep-th/9405187} \BibitemShut
  {NoStop}%
\bibitem [{\citenamefont {Ghoshal}\ \emph {et~al.}(2022)\citenamefont
  {Ghoshal}, \citenamefont {Heurtier},\ and\ \citenamefont
  {Paul}}]{Ghoshal:2022ruy}%
  \BibitemOpen
  \bibfield  {author} {\bibinfo {author} {\bibfnamefont {A.}~\bibnamefont
  {Ghoshal}}, \bibinfo {author} {\bibfnamefont {L.}~\bibnamefont {Heurtier}}, \
  and\ \bibinfo {author} {\bibfnamefont {A.}~\bibnamefont {Paul}},\ }\href
  {\doibase 10.1007/JHEP12(2022)105} {\bibfield  {journal} {\bibinfo  {journal}
  {JHEP}\ }\textbf {\bibinfo {volume} {12}},\ \bibinfo {pages} {105} (\bibinfo
  {year} {2022})},\ \Eprint {http://arxiv.org/abs/2208.01670} {arXiv:2208.01670
  [hep-ph]} \BibitemShut {NoStop}%
\bibitem [{\citenamefont {Heurtier}\ and\ \citenamefont
  {Huang}(2019)}]{Heurtier:2019eou}%
  \BibitemOpen
  \bibfield  {author} {\bibinfo {author} {\bibfnamefont {L.}~\bibnamefont
  {Heurtier}}\ and\ \bibinfo {author} {\bibfnamefont {F.}~\bibnamefont
  {Huang}},\ }\href {\doibase 10.1103/PhysRevD.100.043507} {\bibfield
  {journal} {\bibinfo  {journal} {Phys. Rev. D}\ }\textbf {\bibinfo {volume}
  {100}},\ \bibinfo {pages} {043507} (\bibinfo {year} {2019})},\ \Eprint
  {http://arxiv.org/abs/1905.05191} {arXiv:1905.05191 [hep-ph]} \BibitemShut
  {NoStop}%
\bibitem [{\citenamefont {Kallosh}\ and\ \citenamefont
  {Linde}(2013)}]{Kallosh:2013hoa}%
  \BibitemOpen
  \bibfield  {author} {\bibinfo {author} {\bibfnamefont {R.}~\bibnamefont
  {Kallosh}}\ and\ \bibinfo {author} {\bibfnamefont {A.}~\bibnamefont
  {Linde}},\ }\href {\doibase 10.1088/1475-7516/2013/07/002} {\bibfield
  {journal} {\bibinfo  {journal} {JCAP}\ }\textbf {\bibinfo {volume} {07}},\
  \bibinfo {pages} {002} (\bibinfo {year} {2013})},\ \Eprint
  {http://arxiv.org/abs/1306.5220} {arXiv:1306.5220 [hep-th]} \BibitemShut
  {NoStop}%
\bibitem [{\citenamefont {Starobinsky}(1980)}]{Starobinsky:1980te}%
  \BibitemOpen
  \bibfield  {author} {\bibinfo {author} {\bibfnamefont {A.~A.}\ \bibnamefont
  {Starobinsky}},\ }\href {\doibase 10.1016/0370-2693(80)90670-X} {\bibfield
  {journal} {\bibinfo  {journal} {Phys. Lett. B}\ }\textbf {\bibinfo {volume}
  {91}},\ \bibinfo {pages} {99} (\bibinfo {year} {1980})}\BibitemShut {NoStop}%
\bibitem [{\citenamefont {Starobinsky}(1983)}]{Starobinsky:1983zz}%
  \BibitemOpen
  \bibfield  {author} {\bibinfo {author} {\bibfnamefont {A.~A.}\ \bibnamefont
  {Starobinsky}},\ }\href@noop {} {\bibfield  {journal} {\bibinfo  {journal}
  {Sov. Astron. Lett.}\ }\textbf {\bibinfo {volume} {9}},\ \bibinfo {pages}
  {302} (\bibinfo {year} {1983})}\BibitemShut {NoStop}%
\bibitem [{\citenamefont {Ade}\ \emph {et~al.}(2019)\citenamefont {Ade} \emph
  {et~al.}}]{SimonsObservatory:2018koc}%
  \BibitemOpen
  \bibfield  {author} {\bibinfo {author} {\bibfnamefont {P.}~\bibnamefont
  {Ade}} \emph {et~al.} (\bibinfo {collaboration} {Simons Observatory}),\
  }\href {\doibase 10.1088/1475-7516/2019/02/056} {\bibfield  {journal}
  {\bibinfo  {journal} {JCAP}\ }\textbf {\bibinfo {volume} {02}},\ \bibinfo
  {pages} {056} (\bibinfo {year} {2019})},\ \Eprint
  {http://arxiv.org/abs/1808.07445} {arXiv:1808.07445 [astro-ph.CO]}
  \BibitemShut {NoStop}%
\bibitem [{\citenamefont {Moncelsi}\ \emph {et~al.}(2020)\citenamefont
  {Moncelsi} \emph {et~al.}}]{Moncelsi:2020ppj}%
  \BibitemOpen
  \bibfield  {author} {\bibinfo {author} {\bibfnamefont {L.}~\bibnamefont
  {Moncelsi}} \emph {et~al.},\ }\href {\doibase 10.1117/12.2561995} {\bibfield
  {journal} {\bibinfo  {journal} {Proc. SPIE Int. Soc. Opt. Eng.}\ }\textbf
  {\bibinfo {volume} {11453}},\ \bibinfo {pages} {1145314} (\bibinfo {year}
  {2020})},\ \Eprint {http://arxiv.org/abs/2012.04047} {arXiv:2012.04047
  [astro-ph.IM]} \BibitemShut {NoStop}%
\bibitem [{\citenamefont {Abazajian}\ \emph {et~al.}(2019)\citenamefont
  {Abazajian} \emph {et~al.}}]{Abazajian:2019eic}%
  \BibitemOpen
  \bibfield  {author} {\bibinfo {author} {\bibfnamefont {K.}~\bibnamefont
  {Abazajian}} \emph {et~al.},\ }\href@noop {} {\  (\bibinfo {year} {2019})},\
  \Eprint {http://arxiv.org/abs/1907.04473} {arXiv:1907.04473 [astro-ph.IM]}
  \BibitemShut {NoStop}%
\bibitem [{\citenamefont {Opferkuch}\ \emph {et~al.}(2019)\citenamefont
  {Opferkuch}, \citenamefont {Schwaller},\ and\ \citenamefont
  {Stefanek}}]{Opferkuch:2019zbd}%
  \BibitemOpen
  \bibfield  {author} {\bibinfo {author} {\bibfnamefont {T.}~\bibnamefont
  {Opferkuch}}, \bibinfo {author} {\bibfnamefont {P.}~\bibnamefont
  {Schwaller}}, \ and\ \bibinfo {author} {\bibfnamefont {B.~A.}\ \bibnamefont
  {Stefanek}},\ }\href {\doibase 10.1088/1475-7516/2019/07/016} {\bibfield
  {journal} {\bibinfo  {journal} {JCAP}\ }\textbf {\bibinfo {volume} {07}},\
  \bibinfo {pages} {016} (\bibinfo {year} {2019})},\ \Eprint
  {http://arxiv.org/abs/1905.06823} {arXiv:1905.06823 [gr-qc]} \BibitemShut
  {NoStop}%
\bibitem [{\citenamefont {Caprini}\ and\ \citenamefont
  {Figueroa}(2018)}]{Caprini:2018mtu}%
  \BibitemOpen
  \bibfield  {author} {\bibinfo {author} {\bibfnamefont {C.}~\bibnamefont
  {Caprini}}\ and\ \bibinfo {author} {\bibfnamefont {D.~G.}\ \bibnamefont
  {Figueroa}},\ }\href {\doibase 10.1088/1361-6382/aac608} {\bibfield
  {journal} {\bibinfo  {journal} {Class. Quant. Grav.}\ }\textbf {\bibinfo
  {volume} {35}},\ \bibinfo {pages} {163001} (\bibinfo {year} {2018})},\
  \Eprint {http://arxiv.org/abs/1801.04268} {arXiv:1801.04268 [astro-ph.CO]}
  \BibitemShut {NoStop}%
\bibitem [{\citenamefont {Tanabashi}\ \emph {et~al.}(2018)\citenamefont
  {Tanabashi} \emph {et~al.}}]{ParticleDataGroup:2018ovx}%
  \BibitemOpen
  \bibfield  {author} {\bibinfo {author} {\bibfnamefont {M.}~\bibnamefont
  {Tanabashi}} \emph {et~al.} (\bibinfo {collaboration} {Particle Data
  Group}),\ }\href {\doibase 10.1103/PhysRevD.98.030001} {\bibfield  {journal}
  {\bibinfo  {journal} {Phys. Rev. D}\ }\textbf {\bibinfo {volume} {98}},\
  \bibinfo {pages} {030001} (\bibinfo {year} {2018})}\BibitemShut {NoStop}%
\bibitem [{\citenamefont {Amaro-Seoane}\ \emph {et~al.}(2017)\citenamefont
  {Amaro-Seoane} \emph {et~al.}}]{LISA:2017pwj}%
  \BibitemOpen
  \bibfield  {author} {\bibinfo {author} {\bibfnamefont {P.}~\bibnamefont
  {Amaro-Seoane}} \emph {et~al.} (\bibinfo {collaboration} {LISA}),\
  }\href@noop {} {\  (\bibinfo {year} {2017})},\ \Eprint
  {http://arxiv.org/abs/1702.00786} {arXiv:1702.00786 [astro-ph.IM]}
  \BibitemShut {NoStop}%
\bibitem [{\citenamefont {Baker}\ \emph {et~al.}(2019)\citenamefont {Baker}
  \emph {et~al.}}]{Baker:2019nia}%
  \BibitemOpen
  \bibfield  {author} {\bibinfo {author} {\bibfnamefont {J.}~\bibnamefont
  {Baker}} \emph {et~al.},\ }\href@noop {} {\  (\bibinfo {year} {2019})},\
  \Eprint {http://arxiv.org/abs/1907.06482} {arXiv:1907.06482 [astro-ph.IM]}
  \BibitemShut {NoStop}%
\bibitem [{\citenamefont {Crowder}\ and\ \citenamefont
  {Cornish}(2005)}]{Crowder:2005nr}%
  \BibitemOpen
  \bibfield  {author} {\bibinfo {author} {\bibfnamefont {J.}~\bibnamefont
  {Crowder}}\ and\ \bibinfo {author} {\bibfnamefont {N.~J.}\ \bibnamefont
  {Cornish}},\ }\href {\doibase 10.1103/PhysRevD.72.083005} {\bibfield
  {journal} {\bibinfo  {journal} {Phys. Rev. D}\ }\textbf {\bibinfo {volume}
  {72}},\ \bibinfo {pages} {083005} (\bibinfo {year} {2005})},\ \Eprint
  {http://arxiv.org/abs/gr-qc/0506015} {arXiv:gr-qc/0506015} \BibitemShut
  {NoStop}%
\bibitem [{\citenamefont {Corbin}\ and\ \citenamefont
  {Cornish}(2006)}]{Corbin:2005ny}%
  \BibitemOpen
  \bibfield  {author} {\bibinfo {author} {\bibfnamefont {V.}~\bibnamefont
  {Corbin}}\ and\ \bibinfo {author} {\bibfnamefont {N.~J.}\ \bibnamefont
  {Cornish}},\ }\href {\doibase 10.1088/0264-9381/23/7/014} {\bibfield
  {journal} {\bibinfo  {journal} {Class. Quant. Grav.}\ }\textbf {\bibinfo
  {volume} {23}},\ \bibinfo {pages} {2435} (\bibinfo {year} {2006})},\ \Eprint
  {http://arxiv.org/abs/gr-qc/0512039} {arXiv:gr-qc/0512039} \BibitemShut
  {NoStop}%
\bibitem [{\citenamefont {Harry}\ \emph {et~al.}(2006)\citenamefont {Harry},
  \citenamefont {Fritschel}, \citenamefont {Shaddock}, \citenamefont
  {Folkner},\ and\ \citenamefont {Phinney}}]{Harry:2006fi}%
  \BibitemOpen
  \bibfield  {author} {\bibinfo {author} {\bibfnamefont {G.~M.}\ \bibnamefont
  {Harry}}, \bibinfo {author} {\bibfnamefont {P.}~\bibnamefont {Fritschel}},
  \bibinfo {author} {\bibfnamefont {D.~A.}\ \bibnamefont {Shaddock}}, \bibinfo
  {author} {\bibfnamefont {W.}~\bibnamefont {Folkner}}, \ and\ \bibinfo
  {author} {\bibfnamefont {E.~S.}\ \bibnamefont {Phinney}},\ }\href {\doibase
  10.1088/0264-9381/23/15/008} {\bibfield  {journal} {\bibinfo  {journal}
  {Class. Quant. Grav.}\ }\textbf {\bibinfo {volume} {23}},\ \bibinfo {pages}
  {4887} (\bibinfo {year} {2006})},\ \bibinfo {note} {[Erratum:
  Class.Quant.Grav. 23, 7361 (2006)]}\BibitemShut {NoStop}%
\bibitem [{\citenamefont {Seto}\ \emph {et~al.}(2001)\citenamefont {Seto},
  \citenamefont {Kawamura},\ and\ \citenamefont {Nakamura}}]{Seto:2001qf}%
  \BibitemOpen
  \bibfield  {author} {\bibinfo {author} {\bibfnamefont {N.}~\bibnamefont
  {Seto}}, \bibinfo {author} {\bibfnamefont {S.}~\bibnamefont {Kawamura}}, \
  and\ \bibinfo {author} {\bibfnamefont {T.}~\bibnamefont {Nakamura}},\ }\href
  {\doibase 10.1103/PhysRevLett.87.221103} {\bibfield  {journal} {\bibinfo
  {journal} {Phys. Rev. Lett.}\ }\textbf {\bibinfo {volume} {87}},\ \bibinfo
  {pages} {221103} (\bibinfo {year} {2001})},\ \Eprint
  {http://arxiv.org/abs/astro-ph/0108011} {arXiv:astro-ph/0108011} \BibitemShut
  {NoStop}%
\bibitem [{\citenamefont {Kawamura}\ \emph {et~al.}(2011)\citenamefont
  {Kawamura} \emph {et~al.}}]{Kawamura:2011zz}%
  \BibitemOpen
  \bibfield  {author} {\bibinfo {author} {\bibfnamefont {S.}~\bibnamefont
  {Kawamura}} \emph {et~al.},\ }\href {\doibase 10.1088/0264-9381/28/9/094011}
  {\bibfield  {journal} {\bibinfo  {journal} {Class. Quant. Grav.}\ }\textbf
  {\bibinfo {volume} {28}},\ \bibinfo {pages} {094011} (\bibinfo {year}
  {2011})}\BibitemShut {NoStop}%
\bibitem [{\citenamefont {Yagi}\ and\ \citenamefont
  {Seto}(2011)}]{Yagi:2011wg}%
  \BibitemOpen
  \bibfield  {author} {\bibinfo {author} {\bibfnamefont {K.}~\bibnamefont
  {Yagi}}\ and\ \bibinfo {author} {\bibfnamefont {N.}~\bibnamefont {Seto}},\
  }\href {\doibase 10.1103/PhysRevD.83.044011} {\bibfield  {journal} {\bibinfo
  {journal} {Phys. Rev. D}\ }\textbf {\bibinfo {volume} {83}},\ \bibinfo
  {pages} {044011} (\bibinfo {year} {2011})},\ \bibinfo {note} {[Erratum:
  Phys.Rev.D 95, 109901 (2017)]},\ \Eprint {http://arxiv.org/abs/1101.3940}
  {arXiv:1101.3940 [astro-ph.CO]} \BibitemShut {NoStop}%
\bibitem [{\citenamefont {Kudoh}\ \emph {et~al.}(2006)\citenamefont {Kudoh},
  \citenamefont {Taruya}, \citenamefont {Hiramatsu},\ and\ \citenamefont
  {Himemoto}}]{Kudoh:2005as}%
  \BibitemOpen
  \bibfield  {author} {\bibinfo {author} {\bibfnamefont {H.}~\bibnamefont
  {Kudoh}}, \bibinfo {author} {\bibfnamefont {A.}~\bibnamefont {Taruya}},
  \bibinfo {author} {\bibfnamefont {T.}~\bibnamefont {Hiramatsu}}, \ and\
  \bibinfo {author} {\bibfnamefont {Y.}~\bibnamefont {Himemoto}},\ }\href
  {\doibase 10.1103/PhysRevD.73.064006} {\bibfield  {journal} {\bibinfo
  {journal} {Phys. Rev. D}\ }\textbf {\bibinfo {volume} {73}},\ \bibinfo
  {pages} {064006} (\bibinfo {year} {2006})},\ \Eprint
  {http://arxiv.org/abs/gr-qc/0511145} {arXiv:gr-qc/0511145} \BibitemShut
  {NoStop}%
\bibitem [{\citenamefont {Saikawa}\ and\ \citenamefont
  {Shirai}(2018)}]{Saikawa:2018rcs}%
  \BibitemOpen
  \bibfield  {author} {\bibinfo {author} {\bibfnamefont {K.}~\bibnamefont
  {Saikawa}}\ and\ \bibinfo {author} {\bibfnamefont {S.}~\bibnamefont
  {Shirai}},\ }\href {\doibase 10.1088/1475-7516/2018/05/035} {\bibfield
  {journal} {\bibinfo  {journal} {JCAP}\ }\textbf {\bibinfo {volume} {05}},\
  \bibinfo {pages} {035} (\bibinfo {year} {2018})},\ \Eprint
  {http://arxiv.org/abs/1803.01038} {arXiv:1803.01038 [hep-ph]} \BibitemShut
  {NoStop}%
\bibitem [{\citenamefont {Ringwald}\ \emph {et~al.}(2021)\citenamefont
  {Ringwald}, \citenamefont {Saikawa},\ and\ \citenamefont
  {Tamarit}}]{Ringwald:2020vei}%
  \BibitemOpen
  \bibfield  {author} {\bibinfo {author} {\bibfnamefont {A.}~\bibnamefont
  {Ringwald}}, \bibinfo {author} {\bibfnamefont {K.}~\bibnamefont {Saikawa}}, \
  and\ \bibinfo {author} {\bibfnamefont {C.}~\bibnamefont {Tamarit}},\ }\href
  {\doibase 10.1088/1475-7516/2021/02/046} {\bibfield  {journal} {\bibinfo
  {journal} {JCAP}\ }\textbf {\bibinfo {volume} {02}},\ \bibinfo {pages} {046}
  (\bibinfo {year} {2021})},\ \Eprint {http://arxiv.org/abs/2009.02050}
  {arXiv:2009.02050 [hep-ph]} \BibitemShut {NoStop}%
\bibitem [{\citenamefont {Garcia-Bellido}\ \emph {et~al.}(2021)\citenamefont
  {Garcia-Bellido}, \citenamefont {Murayama},\ and\ \citenamefont
  {White}}]{Garcia-Bellido:2021zgu}%
  \BibitemOpen
  \bibfield  {author} {\bibinfo {author} {\bibfnamefont {J.}~\bibnamefont
  {Garcia-Bellido}}, \bibinfo {author} {\bibfnamefont {H.}~\bibnamefont
  {Murayama}}, \ and\ \bibinfo {author} {\bibfnamefont {G.}~\bibnamefont
  {White}},\ }\href {\doibase 10.1088/1475-7516/2021/12/023} {\bibfield
  {journal} {\bibinfo  {journal} {JCAP}\ }\textbf {\bibinfo {volume} {12}},\
  \bibinfo {pages} {023} (\bibinfo {year} {2021})},\ \Eprint
  {http://arxiv.org/abs/2104.04778} {arXiv:2104.04778 [hep-ph]} \BibitemShut
  {NoStop}%
\bibitem [{\citenamefont {Sesana}\ \emph {et~al.}(2021)\citenamefont {Sesana}
  \emph {et~al.}}]{Sesana:2019vho}%
  \BibitemOpen
  \bibfield  {author} {\bibinfo {author} {\bibfnamefont {A.}~\bibnamefont
  {Sesana}} \emph {et~al.},\ }\href {\doibase 10.1007/s10686-021-09709-9}
  {\bibfield  {journal} {\bibinfo  {journal} {Exper. Astron.}\ }\textbf
  {\bibinfo {volume} {51}},\ \bibinfo {pages} {1333} (\bibinfo {year}
  {2021})},\ \Eprint {http://arxiv.org/abs/1908.11391} {arXiv:1908.11391
  [astro-ph.IM]} \BibitemShut {NoStop}%
\bibitem [{\citenamefont {Harry}(2010)}]{Harry:2010zz}%
  \BibitemOpen
  \bibfield  {author} {\bibinfo {author} {\bibfnamefont {G.~M.}\ \bibnamefont
  {Harry}} (\bibinfo {collaboration} {LIGO Scientific}),\ }\href {\doibase
  10.1088/0264-9381/27/8/084006} {\bibfield  {journal} {\bibinfo  {journal}
  {Class. Quant. Grav.}\ }\textbf {\bibinfo {volume} {27}},\ \bibinfo {pages}
  {084006} (\bibinfo {year} {2010})}\BibitemShut {NoStop}%
\bibitem [{\citenamefont {Aasi}\ \emph {et~al.}(2015)\citenamefont {Aasi} \emph
  {et~al.}}]{LIGOScientific:2014pky}%
  \BibitemOpen
  \bibfield  {author} {\bibinfo {author} {\bibfnamefont {J.}~\bibnamefont
  {Aasi}} \emph {et~al.} (\bibinfo {collaboration} {LIGO Scientific}),\ }\href
  {\doibase 10.1088/0264-9381/32/7/074001} {\bibfield  {journal} {\bibinfo
  {journal} {Class. Quant. Grav.}\ }\textbf {\bibinfo {volume} {32}},\ \bibinfo
  {pages} {074001} (\bibinfo {year} {2015})},\ \Eprint
  {http://arxiv.org/abs/1411.4547} {arXiv:1411.4547 [gr-qc]} \BibitemShut
  {NoStop}%
\bibitem [{\citenamefont {Acernese}\ \emph {et~al.}(2015)\citenamefont
  {Acernese} \emph {et~al.}}]{VIRGO:2014yos}%
  \BibitemOpen
  \bibfield  {author} {\bibinfo {author} {\bibfnamefont {F.}~\bibnamefont
  {Acernese}} \emph {et~al.} (\bibinfo {collaboration} {VIRGO}),\ }\href
  {\doibase 10.1088/0264-9381/32/2/024001} {\bibfield  {journal} {\bibinfo
  {journal} {Class. Quant. Grav.}\ }\textbf {\bibinfo {volume} {32}},\ \bibinfo
  {pages} {024001} (\bibinfo {year} {2015})},\ \Eprint
  {http://arxiv.org/abs/1408.3978} {arXiv:1408.3978 [gr-qc]} \BibitemShut
  {NoStop}%
\bibitem [{\citenamefont {Abbott}\ \emph {et~al.}(2021)\citenamefont {Abbott}
  \emph {et~al.}}]{LIGOScientific:2019lzm}%
  \BibitemOpen
  \bibfield  {author} {\bibinfo {author} {\bibfnamefont {R.}~\bibnamefont
  {Abbott}} \emph {et~al.} (\bibinfo {collaboration} {LIGO Scientific,
  Virgo}),\ }\href {\doibase 10.1016/j.softx.2021.100658} {\bibfield  {journal}
  {\bibinfo  {journal} {SoftwareX}\ }\textbf {\bibinfo {volume} {13}},\
  \bibinfo {pages} {100658} (\bibinfo {year} {2021})},\ \Eprint
  {http://arxiv.org/abs/1912.11716} {arXiv:1912.11716 [gr-qc]} \BibitemShut
  {NoStop}%
\bibitem [{\citenamefont {Abbott}\ \emph {et~al.}(2017)\citenamefont {Abbott}
  \emph {et~al.}}]{LIGOScientific:2016wof}%
  \BibitemOpen
  \bibfield  {author} {\bibinfo {author} {\bibfnamefont {B.~P.}\ \bibnamefont
  {Abbott}} \emph {et~al.} (\bibinfo {collaboration} {LIGO Scientific}),\
  }\href {\doibase 10.1088/1361-6382/aa51f4} {\bibfield  {journal} {\bibinfo
  {journal} {Class. Quant. Grav.}\ }\textbf {\bibinfo {volume} {34}},\ \bibinfo
  {pages} {044001} (\bibinfo {year} {2017})},\ \Eprint
  {http://arxiv.org/abs/1607.08697} {arXiv:1607.08697 [astro-ph.IM]}
  \BibitemShut {NoStop}%
\bibitem [{\citenamefont {Reitze}\ \emph {et~al.}(2019)\citenamefont {Reitze}
  \emph {et~al.}}]{Reitze:2019iox}%
  \BibitemOpen
  \bibfield  {author} {\bibinfo {author} {\bibfnamefont {D.}~\bibnamefont
  {Reitze}} \emph {et~al.},\ }\href@noop {} {\bibfield  {journal} {\bibinfo
  {journal} {Bull. Am. Astron. Soc.}\ }\textbf {\bibinfo {volume} {51}},\
  \bibinfo {pages} {035} (\bibinfo {year} {2019})},\ \Eprint
  {http://arxiv.org/abs/1907.04833} {arXiv:1907.04833 [astro-ph.IM]}
  \BibitemShut {NoStop}%
\bibitem [{\citenamefont {Hild}\ \emph {et~al.}(2011)\citenamefont {Hild} \emph
  {et~al.}}]{Hild:2010id}%
  \BibitemOpen
  \bibfield  {author} {\bibinfo {author} {\bibfnamefont {S.}~\bibnamefont
  {Hild}} \emph {et~al.},\ }\href {\doibase 10.1088/0264-9381/28/9/094013}
  {\bibfield  {journal} {\bibinfo  {journal} {Class. Quant. Grav.}\ }\textbf
  {\bibinfo {volume} {28}},\ \bibinfo {pages} {094013} (\bibinfo {year}
  {2011})},\ \Eprint {http://arxiv.org/abs/1012.0908} {arXiv:1012.0908 [gr-qc]}
  \BibitemShut {NoStop}%
\bibitem [{\citenamefont {Maggiore}\ \emph {et~al.}(2020)\citenamefont
  {Maggiore} \emph {et~al.}}]{Maggiore:2019uih}%
  \BibitemOpen
  \bibfield  {author} {\bibinfo {author} {\bibfnamefont {M.}~\bibnamefont
  {Maggiore}} \emph {et~al.},\ }\href {\doibase 10.1088/1475-7516/2020/03/050}
  {\bibfield  {journal} {\bibinfo  {journal} {JCAP}\ }\textbf {\bibinfo
  {volume} {03}},\ \bibinfo {pages} {050} (\bibinfo {year} {2020})},\ \Eprint
  {http://arxiv.org/abs/1912.02622} {arXiv:1912.02622 [astro-ph.CO]}
  \BibitemShut {NoStop}%
\bibitem [{\citenamefont {Matarrese}\ \emph {et~al.}(1993)\citenamefont
  {Matarrese}, \citenamefont {Pantano},\ and\ \citenamefont
  {Saez}}]{Matarrese:1992rp}%
  \BibitemOpen
  \bibfield  {author} {\bibinfo {author} {\bibfnamefont {S.}~\bibnamefont
  {Matarrese}}, \bibinfo {author} {\bibfnamefont {O.}~\bibnamefont {Pantano}},
  \ and\ \bibinfo {author} {\bibfnamefont {D.}~\bibnamefont {Saez}},\ }\href
  {\doibase 10.1103/PhysRevD.47.1311} {\bibfield  {journal} {\bibinfo
  {journal} {Phys. Rev. D}\ }\textbf {\bibinfo {volume} {47}},\ \bibinfo
  {pages} {1311} (\bibinfo {year} {1993})}\BibitemShut {NoStop}%
\bibitem [{\citenamefont {Matarrese}\ \emph {et~al.}(1994)\citenamefont
  {Matarrese}, \citenamefont {Pantano},\ and\ \citenamefont
  {Saez}}]{Matarrese:1993zf}%
  \BibitemOpen
  \bibfield  {author} {\bibinfo {author} {\bibfnamefont {S.}~\bibnamefont
  {Matarrese}}, \bibinfo {author} {\bibfnamefont {O.}~\bibnamefont {Pantano}},
  \ and\ \bibinfo {author} {\bibfnamefont {D.}~\bibnamefont {Saez}},\ }\href
  {\doibase 10.1103/PhysRevLett.72.320} {\bibfield  {journal} {\bibinfo
  {journal} {Phys. Rev. Lett.}\ }\textbf {\bibinfo {volume} {72}},\ \bibinfo
  {pages} {320} (\bibinfo {year} {1994})},\ \Eprint
  {http://arxiv.org/abs/astro-ph/9310036} {arXiv:astro-ph/9310036} \BibitemShut
  {NoStop}%
\bibitem [{\citenamefont {Matarrese}\ \emph {et~al.}(1998)\citenamefont
  {Matarrese}, \citenamefont {Mollerach},\ and\ \citenamefont
  {Bruni}}]{Matarrese:1997ay}%
  \BibitemOpen
  \bibfield  {author} {\bibinfo {author} {\bibfnamefont {S.}~\bibnamefont
  {Matarrese}}, \bibinfo {author} {\bibfnamefont {S.}~\bibnamefont
  {Mollerach}}, \ and\ \bibinfo {author} {\bibfnamefont {M.}~\bibnamefont
  {Bruni}},\ }\href {\doibase 10.1103/PhysRevD.58.043504} {\bibfield  {journal}
  {\bibinfo  {journal} {Phys. Rev. D}\ }\textbf {\bibinfo {volume} {58}},\
  \bibinfo {pages} {043504} (\bibinfo {year} {1998})},\ \Eprint
  {http://arxiv.org/abs/astro-ph/9707278} {arXiv:astro-ph/9707278} \BibitemShut
  {NoStop}%
\bibitem [{\citenamefont {Noh}\ and\ \citenamefont {Hwang}(2004)}]{Noh:2004bc}%
  \BibitemOpen
  \bibfield  {author} {\bibinfo {author} {\bibfnamefont {H.}~\bibnamefont
  {Noh}}\ and\ \bibinfo {author} {\bibfnamefont {J.-c.}\ \bibnamefont
  {Hwang}},\ }\href {\doibase 10.1103/PhysRevD.69.104011} {\bibfield  {journal}
  {\bibinfo  {journal} {Phys. Rev. D}\ }\textbf {\bibinfo {volume} {69}},\
  \bibinfo {pages} {104011} (\bibinfo {year} {2004})}\BibitemShut {NoStop}%
\bibitem [{\citenamefont {Carbone}\ and\ \citenamefont
  {Matarrese}(2005)}]{Carbone:2004iv}%
  \BibitemOpen
  \bibfield  {author} {\bibinfo {author} {\bibfnamefont {C.}~\bibnamefont
  {Carbone}}\ and\ \bibinfo {author} {\bibfnamefont {S.}~\bibnamefont
  {Matarrese}},\ }\href {\doibase 10.1103/PhysRevD.71.043508} {\bibfield
  {journal} {\bibinfo  {journal} {Phys. Rev. D}\ }\textbf {\bibinfo {volume}
  {71}},\ \bibinfo {pages} {043508} (\bibinfo {year} {2005})},\ \Eprint
  {http://arxiv.org/abs/astro-ph/0407611} {arXiv:astro-ph/0407611} \BibitemShut
  {NoStop}%
\bibitem [{\citenamefont {Nakamura}(2007)}]{Nakamura:2004rm}%
  \BibitemOpen
  \bibfield  {author} {\bibinfo {author} {\bibfnamefont {K.}~\bibnamefont
  {Nakamura}},\ }\href {\doibase 10.1143/PTP.117.17} {\bibfield  {journal}
  {\bibinfo  {journal} {Prog. Theor. Phys.}\ }\textbf {\bibinfo {volume}
  {117}},\ \bibinfo {pages} {17} (\bibinfo {year} {2007})},\ \Eprint
  {http://arxiv.org/abs/gr-qc/0605108} {arXiv:gr-qc/0605108} \BibitemShut
  {NoStop}%
\bibitem [{\citenamefont {Baumann}\ \emph {et~al.}(2007)\citenamefont
  {Baumann}, \citenamefont {Steinhardt}, \citenamefont {Takahashi},\ and\
  \citenamefont {Ichiki}}]{Baumann:2007zm}%
  \BibitemOpen
  \bibfield  {author} {\bibinfo {author} {\bibfnamefont {D.}~\bibnamefont
  {Baumann}}, \bibinfo {author} {\bibfnamefont {P.~J.}\ \bibnamefont
  {Steinhardt}}, \bibinfo {author} {\bibfnamefont {K.}~\bibnamefont
  {Takahashi}}, \ and\ \bibinfo {author} {\bibfnamefont {K.}~\bibnamefont
  {Ichiki}},\ }\href {\doibase 10.1103/PhysRevD.76.084019} {\bibfield
  {journal} {\bibinfo  {journal} {Phys. Rev. D}\ }\textbf {\bibinfo {volume}
  {76}},\ \bibinfo {pages} {084019} (\bibinfo {year} {2007})},\ \Eprint
  {http://arxiv.org/abs/hep-th/0703290} {arXiv:hep-th/0703290} \BibitemShut
  {NoStop}%
\bibitem [{\citenamefont {Nakama}\ \emph {et~al.}(2017)\citenamefont {Nakama},
  \citenamefont {Silk},\ and\ \citenamefont {Kamionkowski}}]{Nakama:2016gzw}%
  \BibitemOpen
  \bibfield  {author} {\bibinfo {author} {\bibfnamefont {T.}~\bibnamefont
  {Nakama}}, \bibinfo {author} {\bibfnamefont {J.}~\bibnamefont {Silk}}, \ and\
  \bibinfo {author} {\bibfnamefont {M.}~\bibnamefont {Kamionkowski}},\ }\href
  {\doibase 10.1103/PhysRevD.95.043511} {\bibfield  {journal} {\bibinfo
  {journal} {Phys. Rev. D}\ }\textbf {\bibinfo {volume} {95}},\ \bibinfo
  {pages} {043511} (\bibinfo {year} {2017})},\ \Eprint
  {http://arxiv.org/abs/1612.06264} {arXiv:1612.06264 [astro-ph.CO]}
  \BibitemShut {NoStop}%
\bibitem [{\citenamefont {Dom\`enech}(2020)}]{Domenech:2019quo}%
  \BibitemOpen
  \bibfield  {author} {\bibinfo {author} {\bibfnamefont {G.}~\bibnamefont
  {Dom\`enech}},\ }\href {\doibase 10.1142/S0218271820500285} {\bibfield
  {journal} {\bibinfo  {journal} {Int. J. Mod. Phys. D}\ }\textbf {\bibinfo
  {volume} {29}},\ \bibinfo {pages} {2050028} (\bibinfo {year} {2020})},\
  \Eprint {http://arxiv.org/abs/1912.05583} {arXiv:1912.05583 [gr-qc]}
  \BibitemShut {NoStop}%
\bibitem [{\citenamefont {Inomata}\ \emph {et~al.}(2020)\citenamefont
  {Inomata}, \citenamefont {Kawasaki}, \citenamefont {Mukaida}, \citenamefont
  {Terada},\ and\ \citenamefont {Yanagida}}]{Inomata:2020lmk}%
  \BibitemOpen
  \bibfield  {author} {\bibinfo {author} {\bibfnamefont {K.}~\bibnamefont
  {Inomata}}, \bibinfo {author} {\bibfnamefont {M.}~\bibnamefont {Kawasaki}},
  \bibinfo {author} {\bibfnamefont {K.}~\bibnamefont {Mukaida}}, \bibinfo
  {author} {\bibfnamefont {T.}~\bibnamefont {Terada}}, \ and\ \bibinfo {author}
  {\bibfnamefont {T.~T.}\ \bibnamefont {Yanagida}},\ }\href {\doibase
  10.1103/PhysRevD.101.123533} {\bibfield  {journal} {\bibinfo  {journal}
  {Phys. Rev. D}\ }\textbf {\bibinfo {volume} {101}},\ \bibinfo {pages}
  {123533} (\bibinfo {year} {2020})},\ \Eprint
  {http://arxiv.org/abs/2003.10455} {arXiv:2003.10455 [astro-ph.CO]}
  \BibitemShut {NoStop}%
\bibitem [{\citenamefont {Inomata}\ \emph
  {et~al.}(2019{\natexlab{a}})\citenamefont {Inomata}, \citenamefont {Kohri},
  \citenamefont {Nakama},\ and\ \citenamefont {Terada}}]{Inomata:2019ivs}%
  \BibitemOpen
  \bibfield  {author} {\bibinfo {author} {\bibfnamefont {K.}~\bibnamefont
  {Inomata}}, \bibinfo {author} {\bibfnamefont {K.}~\bibnamefont {Kohri}},
  \bibinfo {author} {\bibfnamefont {T.}~\bibnamefont {Nakama}}, \ and\ \bibinfo
  {author} {\bibfnamefont {T.}~\bibnamefont {Terada}},\ }\href {\doibase
  10.1103/PhysRevD.100.043532} {\bibfield  {journal} {\bibinfo  {journal}
  {Phys. Rev. D}\ }\textbf {\bibinfo {volume} {100}},\ \bibinfo {pages}
  {043532} (\bibinfo {year} {2019}{\natexlab{a}})},\ \Eprint
  {http://arxiv.org/abs/1904.12879} {arXiv:1904.12879 [astro-ph.CO]}
  \BibitemShut {NoStop}%
\bibitem [{\citenamefont {Inomata}\ \emph
  {et~al.}(2019{\natexlab{b}})\citenamefont {Inomata}, \citenamefont {Kohri},
  \citenamefont {Nakama},\ and\ \citenamefont {Terada}}]{Inomata:2019zqy}%
  \BibitemOpen
  \bibfield  {author} {\bibinfo {author} {\bibfnamefont {K.}~\bibnamefont
  {Inomata}}, \bibinfo {author} {\bibfnamefont {K.}~\bibnamefont {Kohri}},
  \bibinfo {author} {\bibfnamefont {T.}~\bibnamefont {Nakama}}, \ and\ \bibinfo
  {author} {\bibfnamefont {T.}~\bibnamefont {Terada}},\ }\href {\doibase
  10.1088/1475-7516/2019/10/071} {\bibfield  {journal} {\bibinfo  {journal}
  {JCAP}\ }\textbf {\bibinfo {volume} {10}},\ \bibinfo {pages} {071} (\bibinfo
  {year} {2019}{\natexlab{b}})},\ \Eprint {http://arxiv.org/abs/1904.12878}
  {arXiv:1904.12878 [astro-ph.CO]} \BibitemShut {NoStop}%
\bibitem [{\citenamefont {Ito}\ \emph {et~al.}(2020)\citenamefont {Ito},
  \citenamefont {Ikeda}, \citenamefont {Miuchi},\ and\ \citenamefont
  {Soda}}]{Ito:2019wcb}%
  \BibitemOpen
  \bibfield  {author} {\bibinfo {author} {\bibfnamefont {A.}~\bibnamefont
  {Ito}}, \bibinfo {author} {\bibfnamefont {T.}~\bibnamefont {Ikeda}}, \bibinfo
  {author} {\bibfnamefont {K.}~\bibnamefont {Miuchi}}, \ and\ \bibinfo {author}
  {\bibfnamefont {J.}~\bibnamefont {Soda}},\ }\href {\doibase
  10.1140/epjc/s10052-020-7735-y} {\bibfield  {journal} {\bibinfo  {journal}
  {Eur. Phys. J. C}\ }\textbf {\bibinfo {volume} {80}},\ \bibinfo {pages} {179}
  (\bibinfo {year} {2020})},\ \Eprint {http://arxiv.org/abs/1903.04843}
  {arXiv:1903.04843 [gr-qc]} \BibitemShut {NoStop}%
\bibitem [{\citenamefont {Ejlli}\ \emph {et~al.}(2019)\citenamefont {Ejlli},
  \citenamefont {Ejlli}, \citenamefont {Cruise}, \citenamefont {Pisano},\ and\
  \citenamefont {Grote}}]{Ejlli:2019bqj}%
  \BibitemOpen
  \bibfield  {author} {\bibinfo {author} {\bibfnamefont {A.}~\bibnamefont
  {Ejlli}}, \bibinfo {author} {\bibfnamefont {D.}~\bibnamefont {Ejlli}},
  \bibinfo {author} {\bibfnamefont {A.~M.}\ \bibnamefont {Cruise}}, \bibinfo
  {author} {\bibfnamefont {G.}~\bibnamefont {Pisano}}, \ and\ \bibinfo {author}
  {\bibfnamefont {H.}~\bibnamefont {Grote}},\ }\href {\doibase
  10.1140/epjc/s10052-019-7542-5} {\bibfield  {journal} {\bibinfo  {journal}
  {Eur. Phys. J. C}\ }\textbf {\bibinfo {volume} {79}},\ \bibinfo {pages}
  {1032} (\bibinfo {year} {2019})},\ \Eprint {http://arxiv.org/abs/1908.00232}
  {arXiv:1908.00232 [gr-qc]} \BibitemShut {NoStop}%
\bibitem [{\citenamefont {Berlin}\ \emph {et~al.}(2022)\citenamefont {Berlin},
  \citenamefont {Blas}, \citenamefont {Tito~D'Agnolo}, \citenamefont {Ellis},
  \citenamefont {Harnik}, \citenamefont {Kahn},\ and\ \citenamefont
  {Sch\"utte-Engel}}]{Berlin:2021txa}%
  \BibitemOpen
  \bibfield  {author} {\bibinfo {author} {\bibfnamefont {A.}~\bibnamefont
  {Berlin}}, \bibinfo {author} {\bibfnamefont {D.}~\bibnamefont {Blas}},
  \bibinfo {author} {\bibfnamefont {R.}~\bibnamefont {Tito~D'Agnolo}}, \bibinfo
  {author} {\bibfnamefont {S.~A.~R.}\ \bibnamefont {Ellis}}, \bibinfo {author}
  {\bibfnamefont {R.}~\bibnamefont {Harnik}}, \bibinfo {author} {\bibfnamefont
  {Y.}~\bibnamefont {Kahn}}, \ and\ \bibinfo {author} {\bibfnamefont
  {J.}~\bibnamefont {Sch\"utte-Engel}},\ }\href {\doibase
  10.1103/PhysRevD.105.116011} {\bibfield  {journal} {\bibinfo  {journal}
  {Phys. Rev. D}\ }\textbf {\bibinfo {volume} {105}},\ \bibinfo {pages}
  {116011} (\bibinfo {year} {2022})},\ \Eprint
  {http://arxiv.org/abs/2112.11465} {arXiv:2112.11465 [hep-ph]} \BibitemShut
  {NoStop}%
\bibitem [{\citenamefont {Domcke}\ \emph {et~al.}(2022)\citenamefont {Domcke},
  \citenamefont {Garcia-Cely},\ and\ \citenamefont {Rodd}}]{Domcke:2022rgu}%
  \BibitemOpen
  \bibfield  {author} {\bibinfo {author} {\bibfnamefont {V.}~\bibnamefont
  {Domcke}}, \bibinfo {author} {\bibfnamefont {C.}~\bibnamefont {Garcia-Cely}},
  \ and\ \bibinfo {author} {\bibfnamefont {N.~L.}\ \bibnamefont {Rodd}},\
  }\href {\doibase 10.1103/PhysRevLett.129.041101} {\bibfield  {journal}
  {\bibinfo  {journal} {Phys. Rev. Lett.}\ }\textbf {\bibinfo {volume} {129}},\
  \bibinfo {pages} {041101} (\bibinfo {year} {2022})},\ \Eprint
  {http://arxiv.org/abs/2202.00695} {arXiv:2202.00695 [hep-ph]} \BibitemShut
  {NoStop}%
\bibitem [{\citenamefont {Ito}\ and\ \citenamefont {Soda}(2023)}]{Ito:2022rxn}%
  \BibitemOpen
  \bibfield  {author} {\bibinfo {author} {\bibfnamefont {A.}~\bibnamefont
  {Ito}}\ and\ \bibinfo {author} {\bibfnamefont {J.}~\bibnamefont {Soda}},\
  }\href {\doibase 10.1140/epjc/s10052-023-11876-2} {\bibfield  {journal}
  {\bibinfo  {journal} {Eur. Phys. J. C}\ }\textbf {\bibinfo {volume} {83}},\
  \bibinfo {pages} {766} (\bibinfo {year} {2023})},\ \Eprint
  {http://arxiv.org/abs/2212.04094} {arXiv:2212.04094 [gr-qc]} \BibitemShut
  {NoStop}%
\bibitem [{\citenamefont {Inman}\ and\ \citenamefont
  {Ali-Ha\"\i{}moud}(2019)}]{Inman:2019wvr}%
  \BibitemOpen
  \bibfield  {author} {\bibinfo {author} {\bibfnamefont {D.}~\bibnamefont
  {Inman}}\ and\ \bibinfo {author} {\bibfnamefont {Y.}~\bibnamefont
  {Ali-Ha\"\i{}moud}},\ }\href {\doibase 10.1103/PhysRevD.100.083528}
  {\bibfield  {journal} {\bibinfo  {journal} {Phys. Rev. D}\ }\textbf {\bibinfo
  {volume} {100}},\ \bibinfo {pages} {083528} (\bibinfo {year} {2019})},\
  \Eprint {http://arxiv.org/abs/1907.08129} {arXiv:1907.08129 [astro-ph.CO]}
  \BibitemShut {NoStop}%
\bibitem [{\citenamefont {Kodama}\ and\ \citenamefont
  {Sasaki}(1986)}]{Kodama:1986fg}%
  \BibitemOpen
  \bibfield  {author} {\bibinfo {author} {\bibfnamefont {H.}~\bibnamefont
  {Kodama}}\ and\ \bibinfo {author} {\bibfnamefont {M.}~\bibnamefont
  {Sasaki}},\ }\href {\doibase 10.1142/S0217751X86000137} {\bibfield  {journal}
  {\bibinfo  {journal} {Int. J. Mod. Phys. A}\ }\textbf {\bibinfo {volume}
  {1}},\ \bibinfo {pages} {265} (\bibinfo {year} {1986})}\BibitemShut {NoStop}%
\bibitem [{\citenamefont {Kodama}\ and\ \citenamefont
  {Sasaki}(1987)}]{Kodama:1986ud}%
  \BibitemOpen
  \bibfield  {author} {\bibinfo {author} {\bibfnamefont {H.}~\bibnamefont
  {Kodama}}\ and\ \bibinfo {author} {\bibfnamefont {M.}~\bibnamefont
  {Sasaki}},\ }\href {\doibase 10.1142/S0217751X8700020X} {\bibfield  {journal}
  {\bibinfo  {journal} {Int. J. Mod. Phys. A}\ }\textbf {\bibinfo {volume}
  {2}},\ \bibinfo {pages} {491} (\bibinfo {year} {1987})}\BibitemShut {NoStop}%
\bibitem [{\citenamefont {Papanikolaou}\ \emph {et~al.}(2021)\citenamefont
  {Papanikolaou}, \citenamefont {Vennin},\ and\ \citenamefont
  {Langlois}}]{Papanikolaou:2020qtd}%
  \BibitemOpen
  \bibfield  {author} {\bibinfo {author} {\bibfnamefont {T.}~\bibnamefont
  {Papanikolaou}}, \bibinfo {author} {\bibfnamefont {V.}~\bibnamefont
  {Vennin}}, \ and\ \bibinfo {author} {\bibfnamefont {D.}~\bibnamefont
  {Langlois}},\ }\href {\doibase 10.1088/1475-7516/2021/03/053} {\bibfield
  {journal} {\bibinfo  {journal} {JCAP}\ }\textbf {\bibinfo {volume} {03}},\
  \bibinfo {pages} {053} (\bibinfo {year} {2021})},\ \Eprint
  {http://arxiv.org/abs/2010.11573} {arXiv:2010.11573 [astro-ph.CO]}
  \BibitemShut {NoStop}%
\bibitem [{\citenamefont {Dom\`enech}\ \emph {et~al.}(2021)\citenamefont
  {Dom\`enech}, \citenamefont {Lin},\ and\ \citenamefont
  {Sasaki}}]{Domenech:2020ssp}%
  \BibitemOpen
  \bibfield  {author} {\bibinfo {author} {\bibfnamefont {G.}~\bibnamefont
  {Dom\`enech}}, \bibinfo {author} {\bibfnamefont {C.}~\bibnamefont {Lin}}, \
  and\ \bibinfo {author} {\bibfnamefont {M.}~\bibnamefont {Sasaki}},\ }\href
  {\doibase 10.1088/1475-7516/2021/11/E01} {\bibfield  {journal} {\bibinfo
  {journal} {JCAP}\ }\textbf {\bibinfo {volume} {04}},\ \bibinfo {pages} {062}
  (\bibinfo {year} {2021})},\ \bibinfo {note} {[Erratum: JCAP 11, E01
  (2021)]},\ \Eprint {http://arxiv.org/abs/2012.08151} {arXiv:2012.08151
  [gr-qc]} \BibitemShut {NoStop}%
\bibitem [{\citenamefont {Kozaczuk}\ \emph {et~al.}(2022)\citenamefont
  {Kozaczuk}, \citenamefont {Lin},\ and\ \citenamefont
  {Villarama}}]{Kozaczuk:2021wcl}%
  \BibitemOpen
  \bibfield  {author} {\bibinfo {author} {\bibfnamefont {J.}~\bibnamefont
  {Kozaczuk}}, \bibinfo {author} {\bibfnamefont {T.}~\bibnamefont {Lin}}, \
  and\ \bibinfo {author} {\bibfnamefont {E.}~\bibnamefont {Villarama}},\ }\href
  {\doibase 10.1103/PhysRevD.105.123023} {\bibfield  {journal} {\bibinfo
  {journal} {Phys. Rev. D}\ }\textbf {\bibinfo {volume} {105}},\ \bibinfo
  {pages} {123023} (\bibinfo {year} {2022})},\ \Eprint
  {http://arxiv.org/abs/2108.12475} {arXiv:2108.12475 [astro-ph.CO]}
  \BibitemShut {NoStop}%
\bibitem [{\citenamefont {Papanikolaou}\ \emph {et~al.}(2022)\citenamefont
  {Papanikolaou}, \citenamefont {Tzerefos}, \citenamefont {Basilakos},\ and\
  \citenamefont {Saridakis}}]{Papanikolaou:2021uhe}%
  \BibitemOpen
  \bibfield  {author} {\bibinfo {author} {\bibfnamefont {T.}~\bibnamefont
  {Papanikolaou}}, \bibinfo {author} {\bibfnamefont {C.}~\bibnamefont
  {Tzerefos}}, \bibinfo {author} {\bibfnamefont {S.}~\bibnamefont {Basilakos}},
  \ and\ \bibinfo {author} {\bibfnamefont {E.~N.}\ \bibnamefont {Saridakis}},\
  }\href {\doibase 10.1088/1475-7516/2022/10/013} {\bibfield  {journal}
  {\bibinfo  {journal} {JCAP}\ }\textbf {\bibinfo {volume} {10}},\ \bibinfo
  {pages} {013} (\bibinfo {year} {2022})},\ \Eprint
  {http://arxiv.org/abs/2112.15059} {arXiv:2112.15059 [astro-ph.CO]}
  \BibitemShut {NoStop}%
\bibitem [{\citenamefont {Bhaumik}\ \emph {et~al.}(2022)\citenamefont
  {Bhaumik}, \citenamefont {Ghoshal},\ and\ \citenamefont
  {Lewicki}}]{Bhaumik:2022pil}%
  \BibitemOpen
  \bibfield  {author} {\bibinfo {author} {\bibfnamefont {N.}~\bibnamefont
  {Bhaumik}}, \bibinfo {author} {\bibfnamefont {A.}~\bibnamefont {Ghoshal}}, \
  and\ \bibinfo {author} {\bibfnamefont {M.}~\bibnamefont {Lewicki}},\ }\href
  {\doibase 10.1007/JHEP07(2022)130} {\bibfield  {journal} {\bibinfo  {journal}
  {JHEP}\ }\textbf {\bibinfo {volume} {07}},\ \bibinfo {pages} {130} (\bibinfo
  {year} {2022})},\ \Eprint {http://arxiv.org/abs/2205.06260} {arXiv:2205.06260
  [astro-ph.CO]} \BibitemShut {NoStop}%
\bibitem [{\citenamefont {Bhaumik}\ \emph {et~al.}(2023)\citenamefont
  {Bhaumik}, \citenamefont {Ghoshal}, \citenamefont {Jain},\ and\ \citenamefont
  {Lewicki}}]{Bhaumik:2022zdd}%
  \BibitemOpen
  \bibfield  {author} {\bibinfo {author} {\bibfnamefont {N.}~\bibnamefont
  {Bhaumik}}, \bibinfo {author} {\bibfnamefont {A.}~\bibnamefont {Ghoshal}},
  \bibinfo {author} {\bibfnamefont {R.~K.}\ \bibnamefont {Jain}}, \ and\
  \bibinfo {author} {\bibfnamefont {M.}~\bibnamefont {Lewicki}},\ }\href
  {\doibase 10.1007/JHEP05(2023)169} {\bibfield  {journal} {\bibinfo  {journal}
  {JHEP}\ }\textbf {\bibinfo {volume} {05}},\ \bibinfo {pages} {169} (\bibinfo
  {year} {2023})},\ \Eprint {http://arxiv.org/abs/2212.00775} {arXiv:2212.00775
  [astro-ph.CO]} \BibitemShut {NoStop}%
\bibitem [{\citenamefont {Papanikolaou}\ \emph {et~al.}(2023)\citenamefont
  {Papanikolaou}, \citenamefont {Tzerefos}, \citenamefont {Basilakos},\ and\
  \citenamefont {Saridakis}}]{Papanikolaou:2022hkg}%
  \BibitemOpen
  \bibfield  {author} {\bibinfo {author} {\bibfnamefont {T.}~\bibnamefont
  {Papanikolaou}}, \bibinfo {author} {\bibfnamefont {C.}~\bibnamefont
  {Tzerefos}}, \bibinfo {author} {\bibfnamefont {S.}~\bibnamefont {Basilakos}},
  \ and\ \bibinfo {author} {\bibfnamefont {E.~N.}\ \bibnamefont {Saridakis}},\
  }\href {\doibase 10.1140/epjc/s10052-022-11157-4} {\bibfield  {journal}
  {\bibinfo  {journal} {Eur. Phys. J. C}\ }\textbf {\bibinfo {volume} {83}},\
  \bibinfo {pages} {31} (\bibinfo {year} {2023})},\ \Eprint
  {http://arxiv.org/abs/2205.06094} {arXiv:2205.06094 [gr-qc]} \BibitemShut
  {NoStop}%
\bibitem [{\citenamefont {Papanikolaou}(2022)}]{Papanikolaou:2022chm}%
  \BibitemOpen
  \bibfield  {author} {\bibinfo {author} {\bibfnamefont {T.}~\bibnamefont
  {Papanikolaou}},\ }\href {\doibase 10.1088/1475-7516/2022/10/089} {\bibfield
  {journal} {\bibinfo  {journal} {JCAP}\ }\textbf {\bibinfo {volume} {10}},\
  \bibinfo {pages} {089} (\bibinfo {year} {2022})},\ \Eprint
  {http://arxiv.org/abs/2207.11041} {arXiv:2207.11041 [astro-ph.CO]}
  \BibitemShut {NoStop}%
\bibitem [{\citenamefont {Auffinger}(2023)}]{Auffinger:2022khh}%
  \BibitemOpen
  \bibfield  {author} {\bibinfo {author} {\bibfnamefont {J.}~\bibnamefont
  {Auffinger}},\ }\href {\doibase 10.1016/j.ppnp.2023.104040} {\bibfield
  {journal} {\bibinfo  {journal} {Prog. Part. Nucl. Phys.}\ }\textbf {\bibinfo
  {volume} {131}},\ \bibinfo {pages} {104040} (\bibinfo {year} {2023})},\
  \Eprint {http://arxiv.org/abs/2206.02672} {arXiv:2206.02672 [astro-ph.CO]}
  \BibitemShut {NoStop}%
\bibitem [{\citenamefont {Page}(1976{\natexlab{a}})}]{Page:1976df}%
  \BibitemOpen
  \bibfield  {author} {\bibinfo {author} {\bibfnamefont {D.~N.}\ \bibnamefont
  {Page}},\ }\href {\doibase 10.1103/PhysRevD.13.198} {\bibfield  {journal}
  {\bibinfo  {journal} {Phys. Rev. D}\ }\textbf {\bibinfo {volume} {13}},\
  \bibinfo {pages} {198} (\bibinfo {year} {1976}{\natexlab{a}})}\BibitemShut
  {NoStop}%
\bibitem [{\citenamefont {Page}(1976{\natexlab{b}})}]{Page:1976ki}%
  \BibitemOpen
  \bibfield  {author} {\bibinfo {author} {\bibfnamefont {D.~N.}\ \bibnamefont
  {Page}},\ }\href {\doibase 10.1103/PhysRevD.14.3260} {\bibfield  {journal}
  {\bibinfo  {journal} {Phys. Rev. D}\ }\textbf {\bibinfo {volume} {14}},\
  \bibinfo {pages} {3260} (\bibinfo {year} {1976}{\natexlab{b}})}\BibitemShut
  {NoStop}%
\bibitem [{\citenamefont {Page}(1977)}]{Page:1977um}%
  \BibitemOpen
  \bibfield  {author} {\bibinfo {author} {\bibfnamefont {D.~N.}\ \bibnamefont
  {Page}},\ }\href {\doibase 10.1103/PhysRevD.16.2402} {\bibfield  {journal}
  {\bibinfo  {journal} {Phys. Rev. D}\ }\textbf {\bibinfo {volume} {16}},\
  \bibinfo {pages} {2402} (\bibinfo {year} {1977})}\BibitemShut {NoStop}%
\bibitem [{\citenamefont {Hooper}\ \emph {et~al.}(2020)\citenamefont {Hooper},
  \citenamefont {Krnjaic}, \citenamefont {March-Russell}, \citenamefont
  {McDermott},\ and\ \citenamefont {Petrossian-Byrne}}]{Hooper:2020evu}%
  \BibitemOpen
  \bibfield  {author} {\bibinfo {author} {\bibfnamefont {D.}~\bibnamefont
  {Hooper}}, \bibinfo {author} {\bibfnamefont {G.}~\bibnamefont {Krnjaic}},
  \bibinfo {author} {\bibfnamefont {J.}~\bibnamefont {March-Russell}}, \bibinfo
  {author} {\bibfnamefont {S.~D.}\ \bibnamefont {McDermott}}, \ and\ \bibinfo
  {author} {\bibfnamefont {R.}~\bibnamefont {Petrossian-Byrne}},\ }\href@noop
  {} {\  (\bibinfo {year} {2020})},\ \Eprint {http://arxiv.org/abs/2004.00618}
  {arXiv:2004.00618 [astro-ph.CO]} \BibitemShut {NoStop}%
\bibitem [{\citenamefont {Abazajian}\ \emph {et~al.}(2016)\citenamefont
  {Abazajian} \emph {et~al.}}]{CMB-S4:2016ple}%
  \BibitemOpen
  \bibfield  {author} {\bibinfo {author} {\bibfnamefont {K.~N.}\ \bibnamefont
  {Abazajian}} \emph {et~al.} (\bibinfo {collaboration} {CMB-S4}),\ }\href@noop
  {} {\  (\bibinfo {year} {2016})},\ \Eprint {http://arxiv.org/abs/1610.02743}
  {arXiv:1610.02743 [astro-ph.CO]} \BibitemShut {NoStop}%
\bibitem [{\citenamefont {Aiola}\ \emph {et~al.}(2022)\citenamefont {Aiola}
  \emph {et~al.}}]{CMB-HD:2022bsz}%
  \BibitemOpen
  \bibfield  {author} {\bibinfo {author} {\bibfnamefont {S.}~\bibnamefont
  {Aiola}} \emph {et~al.} (\bibinfo {collaboration} {CMB-HD}),\ }\href@noop {}
  {\  (\bibinfo {year} {2022})},\ \Eprint {http://arxiv.org/abs/2203.05728}
  {arXiv:2203.05728 [astro-ph.CO]} \BibitemShut {NoStop}%
\bibitem [{\citenamefont {Matsas}\ \emph {et~al.}(1998)\citenamefont {Matsas},
  \citenamefont {Montero}, \citenamefont {Pleitez},\ and\ \citenamefont
  {Vanzella}}]{Matsas:1998zm}%
  \BibitemOpen
  \bibfield  {author} {\bibinfo {author} {\bibfnamefont {G.~E.~A.}\
  \bibnamefont {Matsas}}, \bibinfo {author} {\bibfnamefont {J.~C.}\
  \bibnamefont {Montero}}, \bibinfo {author} {\bibfnamefont {V.}~\bibnamefont
  {Pleitez}}, \ and\ \bibinfo {author} {\bibfnamefont {D.~A.~T.}\ \bibnamefont
  {Vanzella}},\ }in\ \href@noop {} {\emph {\bibinfo {booktitle} {{Conference on
  Topics in Theoretical Physics II: Festschrift for A.H. Zimerman}}}}\
  (\bibinfo {year} {1998})\ \Eprint {http://arxiv.org/abs/hep-ph/9810456}
  {arXiv:hep-ph/9810456} \BibitemShut {NoStop}%
\bibitem [{\citenamefont {Bell}\ and\ \citenamefont
  {Volkas}(1999)}]{Bell:1998jk}%
  \BibitemOpen
  \bibfield  {author} {\bibinfo {author} {\bibfnamefont {N.~F.}\ \bibnamefont
  {Bell}}\ and\ \bibinfo {author} {\bibfnamefont {R.~R.}\ \bibnamefont
  {Volkas}},\ }\href {\doibase 10.1103/PhysRevD.59.107301} {\bibfield
  {journal} {\bibinfo  {journal} {Phys. Rev.}\ }\textbf {\bibinfo {volume}
  {D59}},\ \bibinfo {pages} {107301} (\bibinfo {year} {1999})},\ \Eprint
  {http://arxiv.org/abs/astro-ph/9812301} {arXiv:astro-ph/9812301 [astro-ph]}
  \BibitemShut {NoStop}%
\bibitem [{\citenamefont {Fujita}\ \emph {et~al.}(2014)\citenamefont {Fujita},
  \citenamefont {Kawasaki}, \citenamefont {Harigaya},\ and\ \citenamefont
  {Matsuda}}]{Fujita:2014hha}%
  \BibitemOpen
  \bibfield  {author} {\bibinfo {author} {\bibfnamefont {T.}~\bibnamefont
  {Fujita}}, \bibinfo {author} {\bibfnamefont {M.}~\bibnamefont {Kawasaki}},
  \bibinfo {author} {\bibfnamefont {K.}~\bibnamefont {Harigaya}}, \ and\
  \bibinfo {author} {\bibfnamefont {R.}~\bibnamefont {Matsuda}},\ }\href
  {\doibase 10.1103/PhysRevD.89.103501} {\bibfield  {journal} {\bibinfo
  {journal} {Phys. Rev.}\ }\textbf {\bibinfo {volume} {D89}},\ \bibinfo {pages}
  {103501} (\bibinfo {year} {2014})},\ \Eprint {http://arxiv.org/abs/1401.1909}
  {arXiv:1401.1909 [astro-ph.CO]} \BibitemShut {NoStop}%
\bibitem [{\citenamefont {McDonald}\ \emph {et~al.}(2006)\citenamefont
  {McDonald} \emph {et~al.}}]{SDSS:2004kjl}%
  \BibitemOpen
  \bibfield  {author} {\bibinfo {author} {\bibfnamefont {P.}~\bibnamefont
  {McDonald}} \emph {et~al.} (\bibinfo {collaboration} {SDSS}),\ }\href
  {\doibase 10.1086/444361} {\bibfield  {journal} {\bibinfo  {journal}
  {Astrophys. J. Suppl.}\ }\textbf {\bibinfo {volume} {163}},\ \bibinfo {pages}
  {80} (\bibinfo {year} {2006})},\ \Eprint
  {http://arxiv.org/abs/astro-ph/0405013} {arXiv:astro-ph/0405013} \BibitemShut
  {NoStop}%
\bibitem [{\citenamefont {Becker}\ \emph {et~al.}(2007)\citenamefont {Becker},
  \citenamefont {Rauch},\ and\ \citenamefont {Sargent}}]{Becker:2006qj}%
  \BibitemOpen
  \bibfield  {author} {\bibinfo {author} {\bibfnamefont {G.~D.}\ \bibnamefont
  {Becker}}, \bibinfo {author} {\bibfnamefont {M.}~\bibnamefont {Rauch}}, \
  and\ \bibinfo {author} {\bibfnamefont {W.~L.~W.}\ \bibnamefont {Sargent}},\
  }\href {\doibase 10.1086/517866} {\bibfield  {journal} {\bibinfo  {journal}
  {Astrophys. J.}\ }\textbf {\bibinfo {volume} {662}},\ \bibinfo {pages} {72}
  (\bibinfo {year} {2007})},\ \Eprint {http://arxiv.org/abs/astro-ph/0607633}
  {arXiv:astro-ph/0607633} \BibitemShut {NoStop}%
\bibitem [{\citenamefont {Becker}\ \emph
  {et~al.}(2011{\natexlab{a}})\citenamefont {Becker}, \citenamefont {Bolton},
  \citenamefont {Haehnelt},\ and\ \citenamefont {Sargent}}]{Becker:2010cu}%
  \BibitemOpen
  \bibfield  {author} {\bibinfo {author} {\bibfnamefont {G.~D.}\ \bibnamefont
  {Becker}}, \bibinfo {author} {\bibfnamefont {J.~S.}\ \bibnamefont {Bolton}},
  \bibinfo {author} {\bibfnamefont {M.~G.}\ \bibnamefont {Haehnelt}}, \ and\
  \bibinfo {author} {\bibfnamefont {W.~L.~W.}\ \bibnamefont {Sargent}},\ }\href
  {\doibase 10.1111/j.1365-2966.2010.17507.x} {\bibfield  {journal} {\bibinfo
  {journal} {Mon. Not. Roy. Astron. Soc.}\ }\textbf {\bibinfo {volume} {410}},\
  \bibinfo {pages} {1096} (\bibinfo {year} {2011}{\natexlab{a}})},\ \Eprint
  {http://arxiv.org/abs/1008.2622} {arXiv:1008.2622 [astro-ph.CO]} \BibitemShut
  {NoStop}%
\bibitem [{\citenamefont {Calverley}\ \emph {et~al.}(2011)\citenamefont
  {Calverley}, \citenamefont {Becker}, \citenamefont {Haehnelt},\ and\
  \citenamefont {Bolton}}]{Calverley:2010tz}%
  \BibitemOpen
  \bibfield  {author} {\bibinfo {author} {\bibfnamefont {A.~P.}\ \bibnamefont
  {Calverley}}, \bibinfo {author} {\bibfnamefont {G.~D.}\ \bibnamefont
  {Becker}}, \bibinfo {author} {\bibfnamefont {M.~G.}\ \bibnamefont
  {Haehnelt}}, \ and\ \bibinfo {author} {\bibfnamefont {J.~S.}\ \bibnamefont
  {Bolton}},\ }\href {\doibase 10.1111/j.1365-2966.2010.18072.x} {\bibfield
  {journal} {\bibinfo  {journal} {Mon. Not. Roy. Astron. Soc.}\ }\textbf
  {\bibinfo {volume} {412}},\ \bibinfo {pages} {2543} (\bibinfo {year}
  {2011})},\ \Eprint {http://arxiv.org/abs/1011.5850} {arXiv:1011.5850
  [astro-ph.CO]} \BibitemShut {NoStop}%
\bibitem [{\citenamefont {Becker}\ \emph
  {et~al.}(2011{\natexlab{b}})\citenamefont {Becker}, \citenamefont {Sargent},
  \citenamefont {Rauch},\ and\ \citenamefont {Calverley}}]{Becker:2011ee}%
  \BibitemOpen
  \bibfield  {author} {\bibinfo {author} {\bibfnamefont {G.~D.}\ \bibnamefont
  {Becker}}, \bibinfo {author} {\bibfnamefont {W.~L.~W.}\ \bibnamefont
  {Sargent}}, \bibinfo {author} {\bibfnamefont {M.}~\bibnamefont {Rauch}}, \
  and\ \bibinfo {author} {\bibfnamefont {A.~P.}\ \bibnamefont {Calverley}},\
  }\href {\doibase 10.1088/0004-637X/735/2/93} {\bibfield  {journal} {\bibinfo
  {journal} {Astrophys. J.}\ }\textbf {\bibinfo {volume} {735}},\ \bibinfo
  {pages} {93} (\bibinfo {year} {2011}{\natexlab{b}})},\ \Eprint
  {http://arxiv.org/abs/1101.4399} {arXiv:1101.4399 [astro-ph.CO]} \BibitemShut
  {NoStop}%
\bibitem [{\citenamefont {Ahn}\ \emph {et~al.}(2014)\citenamefont {Ahn} \emph
  {et~al.}}]{SDSS:2013qvl}%
  \BibitemOpen
  \bibfield  {author} {\bibinfo {author} {\bibfnamefont {C.~P.}\ \bibnamefont
  {Ahn}} \emph {et~al.} (\bibinfo {collaboration} {SDSS}),\ }\href {\doibase
  10.1088/0067-0049/211/2/17} {\bibfield  {journal} {\bibinfo  {journal}
  {Astrophys. J. Suppl.}\ }\textbf {\bibinfo {volume} {211}},\ \bibinfo {pages}
  {17} (\bibinfo {year} {2014})},\ \Eprint {http://arxiv.org/abs/1307.7735}
  {arXiv:1307.7735 [astro-ph.IM]} \BibitemShut {NoStop}%
\bibitem [{\citenamefont {L\'opez}\ \emph {et~al.}(2016)\citenamefont {L\'opez}
  \emph {et~al.}}]{Lopez:2016}%
  \BibitemOpen
  \bibfield  {author} {\bibinfo {author} {\bibfnamefont {S.}~\bibnamefont
  {L\'opez}} \emph {et~al.},\ }\href {\doibase 10.1051/0004-6361/201628161}
  {\bibfield  {journal} {\bibinfo  {journal} {Astron. Astrophys.}\ }\textbf
  {\bibinfo {volume} {594}},\ \bibinfo {pages} {A91} (\bibinfo {year}
  {2016})},\ \Eprint {http://arxiv.org/abs/1607.08776} {arXiv:1607.08776
  [astro-ph.GA]} \BibitemShut {NoStop}%
\bibitem [{\citenamefont {Dienes}\ and\ \citenamefont
  {Thomas}(2012{\natexlab{a}})}]{Dienes:2011ja}%
  \BibitemOpen
  \bibfield  {author} {\bibinfo {author} {\bibfnamefont {K.~R.}\ \bibnamefont
  {Dienes}}\ and\ \bibinfo {author} {\bibfnamefont {B.}~\bibnamefont
  {Thomas}},\ }\href {\doibase 10.1103/PhysRevD.85.083523} {\bibfield
  {journal} {\bibinfo  {journal} {Phys. Rev. D}\ }\textbf {\bibinfo {volume}
  {85}},\ \bibinfo {pages} {083523} (\bibinfo {year} {2012}{\natexlab{a}})},\
  \Eprint {http://arxiv.org/abs/1106.4546} {arXiv:1106.4546 [hep-ph]}
  \BibitemShut {NoStop}%
\bibitem [{\citenamefont {Dienes}\ and\ \citenamefont
  {Thomas}(2012{\natexlab{b}})}]{Dienes:2011sa}%
  \BibitemOpen
  \bibfield  {author} {\bibinfo {author} {\bibfnamefont {K.~R.}\ \bibnamefont
  {Dienes}}\ and\ \bibinfo {author} {\bibfnamefont {B.}~\bibnamefont
  {Thomas}},\ }\href {\doibase 10.1103/PhysRevD.85.083524} {\bibfield
  {journal} {\bibinfo  {journal} {Phys. Rev. D}\ }\textbf {\bibinfo {volume}
  {85}},\ \bibinfo {pages} {083524} (\bibinfo {year} {2012}{\natexlab{b}})},\
  \Eprint {http://arxiv.org/abs/1107.0721} {arXiv:1107.0721 [hep-ph]}
  \BibitemShut {NoStop}%
\bibitem [{\citenamefont {Dienes}\ and\ \citenamefont
  {Thomas}(2012{\natexlab{c}})}]{Dienes:2012jb}%
  \BibitemOpen
  \bibfield  {author} {\bibinfo {author} {\bibfnamefont {K.~R.}\ \bibnamefont
  {Dienes}}\ and\ \bibinfo {author} {\bibfnamefont {B.}~\bibnamefont
  {Thomas}},\ }\href {\doibase 10.1103/PhysRevD.86.055013} {\bibfield
  {journal} {\bibinfo  {journal} {Phys. Rev. D}\ }\textbf {\bibinfo {volume}
  {86}},\ \bibinfo {pages} {055013} (\bibinfo {year} {2012}{\natexlab{c}})},\
  \Eprint {http://arxiv.org/abs/1203.1923} {arXiv:1203.1923 [hep-ph]}
  \BibitemShut {NoStop}%
\bibitem [{\citenamefont {Joyce}(1997)}]{Joyce:1996cp}%
  \BibitemOpen
  \bibfield  {author} {\bibinfo {author} {\bibfnamefont {M.}~\bibnamefont
  {Joyce}},\ }\href {\doibase 10.1103/PhysRevD.55.1875} {\bibfield  {journal}
  {\bibinfo  {journal} {Phys. Rev. D}\ }\textbf {\bibinfo {volume} {55}},\
  \bibinfo {pages} {1875} (\bibinfo {year} {1997})},\ \Eprint
  {http://arxiv.org/abs/hep-ph/9606223} {arXiv:hep-ph/9606223} \BibitemShut
  {NoStop}%
\bibitem [{\citenamefont {Joyce}\ and\ \citenamefont
  {Prokopec}(1998)}]{Joyce:1997fc}%
  \BibitemOpen
  \bibfield  {author} {\bibinfo {author} {\bibfnamefont {M.}~\bibnamefont
  {Joyce}}\ and\ \bibinfo {author} {\bibfnamefont {T.}~\bibnamefont
  {Prokopec}},\ }\href {\doibase 10.1103/PhysRevD.57.6022} {\bibfield
  {journal} {\bibinfo  {journal} {Phys. Rev. D}\ }\textbf {\bibinfo {volume}
  {57}},\ \bibinfo {pages} {6022} (\bibinfo {year} {1998})},\ \Eprint
  {http://arxiv.org/abs/hep-ph/9709320} {arXiv:hep-ph/9709320} \BibitemShut
  {NoStop}%
\bibitem [{\citenamefont {Servant}(2002)}]{Servant:2001jh}%
  \BibitemOpen
  \bibfield  {author} {\bibinfo {author} {\bibfnamefont {G.}~\bibnamefont
  {Servant}},\ }\href {\doibase 10.1088/1126-6708/2002/01/044} {\bibfield
  {journal} {\bibinfo  {journal} {JHEP}\ }\textbf {\bibinfo {volume} {01}},\
  \bibinfo {pages} {044} (\bibinfo {year} {2002})},\ \Eprint
  {http://arxiv.org/abs/hep-ph/0112209} {arXiv:hep-ph/0112209} \BibitemShut
  {NoStop}%
\bibitem [{\citenamefont {Barenboim}\ and\ \citenamefont
  {Rasero}(2012)}]{Barenboim:2012nh}%
  \BibitemOpen
  \bibfield  {author} {\bibinfo {author} {\bibfnamefont {G.}~\bibnamefont
  {Barenboim}}\ and\ \bibinfo {author} {\bibfnamefont {J.}~\bibnamefont
  {Rasero}},\ }\href {\doibase 10.1007/JHEP07(2012)028} {\bibfield  {journal}
  {\bibinfo  {journal} {JHEP}\ }\textbf {\bibinfo {volume} {07}},\ \bibinfo
  {pages} {028} (\bibinfo {year} {2012})},\ \Eprint
  {http://arxiv.org/abs/1202.6070} {arXiv:1202.6070 [hep-ph]} \BibitemShut
  {NoStop}%
\bibitem [{\citenamefont {Toussaint}\ \emph {et~al.}(1979)\citenamefont
  {Toussaint}, \citenamefont {Treiman}, \citenamefont {Wilczek},\ and\
  \citenamefont {Zee}}]{Toussaint:1978br}%
  \BibitemOpen
  \bibfield  {author} {\bibinfo {author} {\bibfnamefont {D.}~\bibnamefont
  {Toussaint}}, \bibinfo {author} {\bibfnamefont {S.~B.}\ \bibnamefont
  {Treiman}}, \bibinfo {author} {\bibfnamefont {F.}~\bibnamefont {Wilczek}}, \
  and\ \bibinfo {author} {\bibfnamefont {A.}~\bibnamefont {Zee}},\ }\href
  {\doibase 10.1103/PhysRevD.19.1036} {\bibfield  {journal} {\bibinfo
  {journal} {Phys. Rev. D}\ }\textbf {\bibinfo {volume} {19}},\ \bibinfo
  {pages} {1036} (\bibinfo {year} {1979})}\BibitemShut {NoStop}%
\bibitem [{\citenamefont {Turner}(1979)}]{Turner:1979bt}%
  \BibitemOpen
  \bibfield  {author} {\bibinfo {author} {\bibfnamefont {M.~S.}\ \bibnamefont
  {Turner}},\ }\href {\doibase 10.1016/0370-2693(79)90095-9} {\bibfield
  {journal} {\bibinfo  {journal} {Phys. Lett. B}\ }\textbf {\bibinfo {volume}
  {89}},\ \bibinfo {pages} {155} (\bibinfo {year} {1979})}\BibitemShut
  {NoStop}%
\bibitem [{\citenamefont {Grillo}(1980)}]{Grillo:1980rt}%
  \BibitemOpen
  \bibfield  {author} {\bibinfo {author} {\bibfnamefont {A.~F.}\ \bibnamefont
  {Grillo}},\ }\href {\doibase 10.1016/0370-2693(80)90897-7} {\bibfield
  {journal} {\bibinfo  {journal} {Phys. Lett. B}\ }\textbf {\bibinfo {volume}
  {94}},\ \bibinfo {pages} {364} (\bibinfo {year} {1980})}\BibitemShut
  {NoStop}%
\bibitem [{\citenamefont {Hook}(2014)}]{Hook:2014mla}%
  \BibitemOpen
  \bibfield  {author} {\bibinfo {author} {\bibfnamefont {A.}~\bibnamefont
  {Hook}},\ }\href {\doibase 10.1103/PhysRevD.90.083535} {\bibfield  {journal}
  {\bibinfo  {journal} {Phys. Rev. D}\ }\textbf {\bibinfo {volume} {90}},\
  \bibinfo {pages} {083535} (\bibinfo {year} {2014})},\ \Eprint
  {http://arxiv.org/abs/1404.0113} {arXiv:1404.0113 [hep-ph]} \BibitemShut
  {NoStop}%
\bibitem [{\citenamefont {Granelli}\ \emph {et~al.}(2021)\citenamefont
  {Granelli}, \citenamefont {Moffat}, \citenamefont {Perez-Gonzalez},
  \citenamefont {Schulz},\ and\ \citenamefont {Turner}}]{Granelli:2020pim}%
  \BibitemOpen
  \bibfield  {author} {\bibinfo {author} {\bibfnamefont {A.}~\bibnamefont
  {Granelli}}, \bibinfo {author} {\bibfnamefont {K.}~\bibnamefont {Moffat}},
  \bibinfo {author} {\bibfnamefont {Y.~F.}\ \bibnamefont {Perez-Gonzalez}},
  \bibinfo {author} {\bibfnamefont {H.}~\bibnamefont {Schulz}}, \ and\ \bibinfo
  {author} {\bibfnamefont {J.}~\bibnamefont {Turner}},\ }\href {\doibase
  10.1016/j.cpc.2020.107813} {\bibfield  {journal} {\bibinfo  {journal}
  {Comput. Phys. Commun.}\ }\textbf {\bibinfo {volume} {262}},\ \bibinfo
  {pages} {107813} (\bibinfo {year} {2021})},\ \Eprint
  {http://arxiv.org/abs/2007.09150} {arXiv:2007.09150 [hep-ph]} \BibitemShut
  {NoStop}%
\bibitem [{\citenamefont {Chaudhuri}\ and\ \citenamefont
  {Dolgov}(2021)}]{Chaudhuri:2020wjo}%
  \BibitemOpen
  \bibfield  {author} {\bibinfo {author} {\bibfnamefont {A.}~\bibnamefont
  {Chaudhuri}}\ and\ \bibinfo {author} {\bibfnamefont {A.}~\bibnamefont
  {Dolgov}},\ }\href {\doibase 10.1134/S1063776121110078} {\bibfield  {journal}
  {\bibinfo  {journal} {J. Exp. Theor. Phys.}\ }\textbf {\bibinfo {volume}
  {133}},\ \bibinfo {pages} {552} (\bibinfo {year} {2021})},\ \Eprint
  {http://arxiv.org/abs/2001.11219} {arXiv:2001.11219 [astro-ph.CO]}
  \BibitemShut {NoStop}%
\bibitem [{\citenamefont {Bernal}\ \emph
  {et~al.}(2022{\natexlab{b}})\citenamefont {Bernal}, \citenamefont {Fong},
  \citenamefont {Perez-Gonzalez},\ and\ \citenamefont
  {Turner}}]{Bernal:2022pue}%
  \BibitemOpen
  \bibfield  {author} {\bibinfo {author} {\bibfnamefont {N.}~\bibnamefont
  {Bernal}}, \bibinfo {author} {\bibfnamefont {C.~S.}\ \bibnamefont {Fong}},
  \bibinfo {author} {\bibfnamefont {Y.~F.}\ \bibnamefont {Perez-Gonzalez}}, \
  and\ \bibinfo {author} {\bibfnamefont {J.}~\bibnamefont {Turner}},\ }\href
  {\doibase 10.1103/PhysRevD.106.035019} {\bibfield  {journal} {\bibinfo
  {journal} {Phys. Rev. D}\ }\textbf {\bibinfo {volume} {106}},\ \bibinfo
  {pages} {035019} (\bibinfo {year} {2022}{\natexlab{b}})},\ \Eprint
  {http://arxiv.org/abs/2203.08823} {arXiv:2203.08823 [hep-ph]} \BibitemShut
  {NoStop}%
\bibitem [{\citenamefont {Dienes}\ \emph {et~al.}(2025)\citenamefont {Dienes},
  \citenamefont {Heurtier}, \citenamefont {Huang}, \citenamefont {Tait},\ and\
  \citenamefont {Thomas}}]{Dienes:2025qdw}%
  \BibitemOpen
  \bibfield  {author} {\bibinfo {author} {\bibfnamefont {K.~R.}\ \bibnamefont
  {Dienes}}, \bibinfo {author} {\bibfnamefont {L.}~\bibnamefont {Heurtier}},
  \bibinfo {author} {\bibfnamefont {F.}~\bibnamefont {Huang}}, \bibinfo
  {author} {\bibfnamefont {T.~M.~P.}\ \bibnamefont {Tait}}, \ and\ \bibinfo
  {author} {\bibfnamefont {B.}~\bibnamefont {Thomas}},\ }\href@noop {} {\
  (\bibinfo {year} {2025})},\ \Eprint {http://arxiv.org/abs/2510.06551}
  {arXiv:2510.06551 [astro-ph.CO]} \BibitemShut {NoStop}%
\bibitem [{\citenamefont {Erickcek}\ and\ \citenamefont
  {Sigurdson}(2011)}]{Erickcek:2011us}%
  \BibitemOpen
  \bibfield  {author} {\bibinfo {author} {\bibfnamefont {A.~L.}\ \bibnamefont
  {Erickcek}}\ and\ \bibinfo {author} {\bibfnamefont {K.}~\bibnamefont
  {Sigurdson}},\ }\href {\doibase 10.1103/PhysRevD.84.083503} {\bibfield
  {journal} {\bibinfo  {journal} {Phys. Rev. D}\ }\textbf {\bibinfo {volume}
  {84}},\ \bibinfo {pages} {083503} (\bibinfo {year} {2011})},\ \Eprint
  {http://arxiv.org/abs/1106.0536} {arXiv:1106.0536 [astro-ph.CO]} \BibitemShut
  {NoStop}%
\bibitem [{\citenamefont {Fan}\ \emph {et~al.}(2014)\citenamefont {Fan},
  \citenamefont {\"Ozsoy},\ and\ \citenamefont {Watson}}]{Fan:2014zua}%
  \BibitemOpen
  \bibfield  {author} {\bibinfo {author} {\bibfnamefont {J.}~\bibnamefont
  {Fan}}, \bibinfo {author} {\bibfnamefont {O.}~\bibnamefont {\"Ozsoy}}, \ and\
  \bibinfo {author} {\bibfnamefont {S.}~\bibnamefont {Watson}},\ }\href
  {\doibase 10.1103/PhysRevD.90.043536} {\bibfield  {journal} {\bibinfo
  {journal} {Phys. Rev. D}\ }\textbf {\bibinfo {volume} {90}},\ \bibinfo
  {pages} {043536} (\bibinfo {year} {2014})},\ \Eprint
  {http://arxiv.org/abs/1405.7373} {arXiv:1405.7373 [hep-ph]} \BibitemShut
  {NoStop}%
\bibitem [{\citenamefont {Georg}\ \emph {et~al.}(2016)\citenamefont {Georg},
  \citenamefont {\c{S}eng\"or},\ and\ \citenamefont {Watson}}]{Georg:2016yxa}%
  \BibitemOpen
  \bibfield  {author} {\bibinfo {author} {\bibfnamefont {J.}~\bibnamefont
  {Georg}}, \bibinfo {author} {\bibfnamefont {G.}~\bibnamefont {\c{S}eng\"or}},
  \ and\ \bibinfo {author} {\bibfnamefont {S.}~\bibnamefont {Watson}},\ }\href
  {\doibase 10.1103/PhysRevD.93.123523} {\bibfield  {journal} {\bibinfo
  {journal} {Phys. Rev. D}\ }\textbf {\bibinfo {volume} {93}},\ \bibinfo
  {pages} {123523} (\bibinfo {year} {2016})},\ \Eprint
  {http://arxiv.org/abs/1603.00023} {arXiv:1603.00023 [hep-ph]} \BibitemShut
  {NoStop}%
\end{thebibliography}%
\end{document}